\documentclass[
aps,
prb,
reprint,
amsmath,
amssymb,
superscriptaddress,
longbibliography,
]{revtex4-2}

\usepackage{color}
\usepackage{xcolor}
\usepackage{graphicx}
\usepackage{overpic}
\usepackage{amsmath}
\usepackage[normalem]{ulem}
\usepackage{dcolumn}
\usepackage{bm}
\usepackage{hyperref}
\hypersetup{
    colorlinks=true,
    linkcolor=blue,
    citecolor=blue,
    urlcolor=blue
}

\begin{document}

\preprint{APS/123-QED}
\title{Displacement-Field-Driven Semimetal-Superconductor Transition in Magic-Angle Twisted Trilayer Graphene}
\author{Bokai Liang}
\affiliation{Department of Physics and Astronomy, Johns Hopkins University, Baltimore, Maryland 21218, USA}

\author{Jing-Yu Zhao}
\affiliation{Department of Physics and Astronomy, Johns Hopkins University, Baltimore, Maryland 21218, USA}

\author{Ya-Hui Zhang} 
\affiliation{Department of Physics and Astronomy, Johns Hopkins University, Baltimore, Maryland 21218, USA}

\date{\today}

\begin{abstract}
Magic-angle twisted trilayer graphene(MATTG) hosts versatile displacement-field-tuned correlated phenomena. MATTG consists of a dispersive Dirac cone which hybridizes with the flat band from a twisted bilayer graphene (TBG) sector. The hybridization strength increases with the displacement field $D$ and naively one may expect D-driven heavy fermion physics. However, the TBG Hubbard bands have a momentum-selective Mott gap, which is small at the $\Gamma$ point due to the band topology, and a rigid local moment description as in the familiar Kondo lattice model is invalid.  Here we show that the dominant effect of the displacement field is to induce an energy shift of the Dirac cone and self-doping into the TBG sector. We illustrate this picture in a concrete calculation using a slave-particle theory at the filling $\nu=\pm 2$. We find that increasing $D$ drives a transition from a semimetal into  a superconducting state.
We also discuss the enhancement of the superconductivity by $D$ near $\nu=\pm2$ and the particle-hole asymmetry of the phase diagram. 
Our results provide a unified picture for electric-field-tunable superconductivity, Mottness, and heavy-fermion-like behavior in MATTG.
\end{abstract}

\maketitle

\textit{Introduction---}
Since the discovery of correlated insulating states and superconductivity in magic-angle twisted bilayer graphene (MATBG)\cite{cao2018unconventional, cao2018correlated, yankowitz2019tuning, lu2019superconductors, stepanov2020untying, arora2020superconductivity, cao2021nematicity, liu2021tuning}, graphene-based moir\'{e} superlattices have attracted wide interest as a tunable platform for strongly correlated physics\cite{andrei2020graphene,andrei2021marvels,nuckolls2024microscopic}. The emergence of nearly flat bands intertwined with nontrivial band topology enables a wide range of interaction-driven phases\cite{po2018origin,ahn2019failure,cao2020strange,ledwith2021strong,song2021twisted} that can be continuously tuned into one another by varying the carrier filling, temperature, and pressure.

Compared with MATBG, magic-angle twisted trilayer graphene (MATTG) exhibits a more versatile phase diagram\cite{park2021tunable,cao2021pauli, hao2021electric, park2022robust, zhang2022promotion, mukherjee2025superconducting}, owing in particular to an additional control knob: the out-of-plane displacement field. Experiments have observed electric-field-tunable superconductivity and various symmetry-breaking states\cite{park2021tunable,hao2021electric,cao2021pauli,kim2022evidence,shen2023dirac,zhang2022promotion,pierce2025tunable}.
Signatures of unconventional superconductivity have also been widely discussed both experimentally and theoretically\cite{oh2021evidence,kim2022evidence,park2026experimental,kim2026resolving}. More recently, an experiment reported a displacement-field-tuned transition from an antiferromagnetic semimetal to a heavy fermion liquid(HFL) at $\nu=3$ with strong quantum critical behavior, at relatively high temperature \cite{zhang2025electrically}. Understanding the displacement-field-tuned correlated physics in TTG is therefore a central theoretical problem.

At the single-particle level, the role of the displacement field in TTG has been well understood. 
At low energies, MATTG can be approximately decomposed into a MATBG-like flat-band sector and a monolayer graphene Dirac cone. They decouple at zero displacement field $D$ due to mirror symmetry.
The displacement field breaks mirror symmetry, hybridizes these two sectors, reshapes the Fermi surface, and tunes nearby symmetry-breaking instabilities~\cite{khalaf2019magic,cualuguaru2021twisted,lei2021mirror,xie2021twisted,christos2022correlated,fischer2022unconventional,pierce2025tunable}. However, now it is appreciated that MATBG and MATTG are in the strongly correlated regime and the interaction changes the band structure qualitatively. For example, recent works  \cite{wong2020cascade,song2022magic,christos2022correlated,yu2023magic, nuckolls2023quantum,herzog2024topological,rai2024dynamical, youn2024hundness,zhao2025ancilla, zhao2025mixed, zhao2025resonating,xiao2026imaging} have suggested that Mott physics plays an important role, especially near the integer fillings $\nu=0,\pm2$. As a result, there is a large Mott gap for the TBG sector at order of the Hubbard $U$ at the $K_-$ point of the mini Brillouin zone (mBZ). 
Due to the Mott gap, the Dirac cone at $K_-$ remains robust against hybridization with the TBG sector.
In this work, we  ask how the displacement field acts on such Mott-reconstructed states, rather than treating it merely as a perturbation to the noninteracting band structure at $D=0$.

In a naive picture, if the TBG band is topologically trivial, we can map the problem to a Kondo lattice model with Dirac cone couples to the localized spin moments from the TBG sector through a Kondo coupling $J\sim \frac{D^2}{U}$.  However, this simple heavy fermion picture does not work for MATTG because of the band topology of the TBG sector\cite{po2018origin,kang2018symmetry, song2019all,ahn2019failure, song2021twisted}. It turns out that the Mott state of TBG is intrinsically of a mixed-valence nature\cite{zhao2025mixed}. For example, at the filling $\nu_{\mathrm{TBG}}=-2$, there are around $2.3$ electrons on the AA site and  $0.3$ holes away from the AA site. One consequence is that the Mott gap becomes significantly smaller than $U$  at the $\Gamma$ point. In this work, we will show that the dominant effect of $D$ is to induce an energy shift of the Dirac cone at order $\frac{D^2}{U}$ and self-dope holes into the TBG lower Hubbard band at the $\Gamma$ point.

We capture this effect using a slave-particle theory for the topological heavy fermion model of TTG \cite{yu2023magic} near $\nu=\pm2$, following Ref.~\onlinecite{zhao2025resonating}, to capture both the strong on-site interaction and the hybridization with local moments\cite{wong2020cascade,rozen2021entropic,saito2021isospin,nuckolls2023quantum,ghosh2025thermopower}. 
At $\nu=-2$ and $D=0$, we identify a symmetric  semimetal phase. Note that the ground state is actually in a symmetry breaking phase\cite{po2018origin,bultinck2020ground,parker2021strain,kwan2021kekule,wagner2022global,christos2022correlated}, but in this work we restrict to a symmetric ansatz.
The MATBG-like flat-band sector remains decoupled from the Dirac-cone sector.
The former forms a correlated insulating state, in which the original flat bands split into upper and lower Hubbard bands with a momentum-selective Mott gap that is small near the $\Gamma$ point, as discussed in \cite{zhao2025mixed}.
The latter remains an essentially intact semimetal.
Upon increasing $D$, the displacement field hybridizes the Dirac cone with these Hubbard bands.
In contrast to the noninteracting band picture, the Mott gap at $K_-$ suppresses the effective hybridization and the Dirac cone remains stable.
Instead, its energy is pushed downward by level repulsion with the Hubbard bands, 
leading to self-hole-doping of the lower Hubbard band once $D$ becomes sufficiently large. 
These doped holes strongly hybridize with the local $f$ moments, allowing the local pairing correlations of the $f$ sector to acquire coherent spectral weight on the emergent Fermi surface, which leads to the phase transition into a superconducting (SC) phase.

Furthermore, we show that an energy shift of the Dirac cone---a feature widely recognized in previous studies---is essential for realizing the displacement-field-driven transition. 
This observation provides a natural perspective on the pronounced particle-hole asymmetry in TTG. We further show that a large displacement field substantially enhances the superconducting pairing near $\nu=\pm2$ and identify a displacement-field-tuned two-gap structure in the SC phase.
Overall, our work establishes a unified theoretical framework for the electric-field-tunable phase transition and superconductivity in TTG.

\textit{Model---}We employ the $f$-$c$-$d$ model framework \cite{yu2023magic}, which describes TTG in a manner analogous to the topological heavy fermion model (THFM) for twisted bilayer graphene (TBG) \cite{song2022magic}. The full Hamiltonian is given by
\begin{equation}
H_0 = H_0^{d} + H_0^{c} + H_0^{fd} + H_0^{fc} - \mu N + wN_d+H_{\text{int}}^{(f)}.
\label{original_Hamiltonian}
\end{equation}
The Hamiltonian consists of localized $f$ electron without dispersion, two dispersive conduction bands ($c_1$ and $c_2$) governed by $H_0^{c}$ and in MATTG, an additional Dirac band $d$ governed by $H_0^d$. 
With hybridization $H_0^{fc}$, the combined $f$ and $c$ sectors reproduce the flat bands of MATBG~\cite{song2022magic}.
In MATTG, a finite displacement field $D$ further hybridizes the $f$ orbital with the Dirac band via $H_0^{fd}\propto D$.

The hybridization terms are
\begin{equation}
    H_{0}^{fc} = \sum_{\bm{k}, \bm{G}} f_{\bm{k}}^{\dagger} \gamma(\bm{k}+\bm{G})c_{\bm{k}+\bm{G}} + \text{h.c.},
\end{equation}
\begin{equation}
    H_{0}^{fd} = \sum_{\bm{k} \in \text{MBZ}}\sum_{\bm{p}}^{|\bm{p}| < \Lambda_d} f_{\bm{k}}^{\dagger} M_1D\tilde{h}(\bm{k},{\bm{p}})d_{\bm{p}} + \text{h.c.},
\end{equation}
where $f_{\bm{k}} = \{f_{\bm{k};\alpha}\}$,  $c_{\bm{k}} = \{c_{1;\bm{k};\alpha}, c_{2;\bm{k};\alpha}\}$, and $d_{\bm{k}} = \{d_{\bm{k};\alpha}\}$. Here, $\alpha = (a, \eta, s)$ is a composite index encompassing the orbital $a$, valley $\eta \in \{K,K'\}$, and spin $s \in \{\uparrow, \downarrow\}$. 
$M_1=-0.1397$ is a constant. Readers can find details in the appendix.

We also include an energy offset $w$ for the Dirac cone electron $d$ in Eq.~\eqref{original_Hamiltonian}, 
which arises from the inequivalent electrostatic environments of the inner and outer layers~\cite{wu2021lattice,fischer2022unconventional} and is shown later to be important for the self-doping mechanism and to explain the particle-hole asymmetry.

The flat $f$-orbital is split by a strong on-site interaction: 
\begin{equation}
H_{\text{int}}^{(f)} = \frac{U}{2} \sum_i \left(n_{i ; f} - 4 - \kappa \nu\right)^2 + \sum_i h_{i ; J}^{(f)}.
\end{equation}
which consists of a large Hubbard interaction $U$ and spin interaction $J$.
Here, $h_{i;J}^{(f)}$ captures effective local-moment interactions generated by electron-electron and electron-phonon processes\cite{chen2024strong,wang2024molecular,wang2025electron}, which favors inter-valley spin-singlet pairs. 
An additional phenomenological parameter $\kappa=0.8$ is introduced to fit the Hartree term of the Coulomb interaction.

\begin{figure}[t]
    \centering
    \includegraphics[width=\columnwidth]{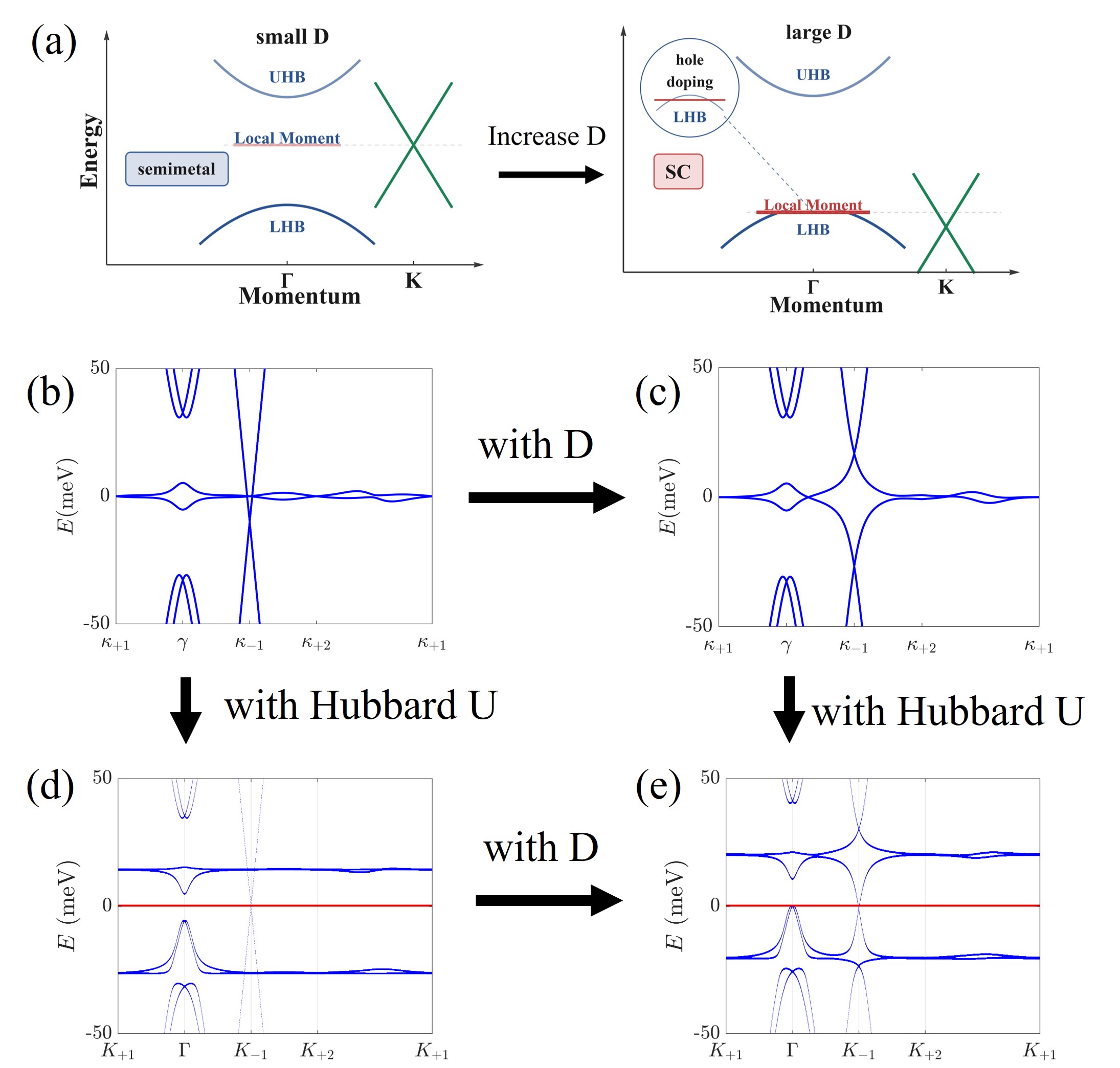}
    \caption{(a)Schematic interpretation of the central mechanism governing the electrically tunable phase transition.
    At integer filling $\nu=-2$, the original flat bands are split into upper and lower Hubbard bands, whereas the additional Dirac cone at $K$ remains unaffected. 
    Upon increasing the displacement field $D$ or decreasing the energy offset $w$, the Dirac cone is pushed downward in energy. 
    This gives rise to self-doped electrons into the Dirac cone and holes into the lower Hubbard band, and triggers the phase transition.
    (b) and (c) Noninteracting band structures of Eq.~\eqref{original_Hamiltonian} for $D=0$ and $D=120\,\mathrm{meV}$, respectively, with an energy offset $w=-10\,\mathrm{meV}$ 
    (d) and (e) Corresponding interacting mean-field spectra of Eq.~\eqref{mean field Hamiltonian} with $U=40\mathrm{meV}$ and $J=0.3\mathrm{meV}$, where (d) corresponds to (b) and (e) to (c). The spinon $\psi'$ bands are marked with red.
    }
    \label{figure1}	
\end{figure}

\textit{Parton construction---} 
In the limit of large on-site $U$, different local valence sectors of the $f$ orbital are well separated.  
In the vicinity of $\nu=-2$, this separation allows us to project onto the states with occupations $n_{i;f}=1,2,3$ at each site $i$. 
The local states are referred to as singlons, doublons, and triplons, respectively. 
The restricted Hilbert space is therefore spanned by 8 singlon states $|i;\alpha\rangle_s=f_{i;\alpha}^{\dagger}|0\rangle$, 28 doublon states $|i;\alpha\beta\rangle_d=f_{i;\alpha}^{\dagger}f_{i;\beta}^{\dagger}|0\rangle$ with $\alpha<\beta$, and 56 triplon states $|i;\alpha\beta\gamma\rangle_t=f_{i;\alpha}^{\dagger}f_{i;\beta}^{\dagger}f_{i;\gamma}^{\dagger}|0\rangle$ with $\alpha<\beta<\gamma$.
We introduce a set of parton fermions in the restricted Hilbert space.
The singlon and triplon are represented as $|i ; \alpha\rangle_s=s_{i ; \alpha}^{\dagger}|0\rangle$ and $|i ; \alpha \beta \gamma\rangle_t=t_{i ; \alpha \beta \gamma}^{\dagger}|0\rangle$, respectively, 
while the doublon states are represented by spinon bilinears, $|i ; \alpha \beta\rangle_d=\psi_{i ; \alpha}^{\prime \dagger} \psi_{i ; \beta}^{\prime \dagger}|0\rangle$. 
The $s$, $t$, and $\psi'$ are fermionic operators satisfying the local constraint  
\begin{equation} \label{constraint}
    n_{i ; s} + \frac{1}{2} n_{i ; \psi^{\prime}} + n_{i ; t} = 1.  
\end{equation}
where $n_{i;s}$, $n_{i;\psi'}$, and $n_{i;t}$ denote the corresponding parton number operators.
Because of the parton gauge redundancy, the physical charge can be assigned relative to the doublon background. 
We choose the singlon, triplon, and spinon $\psi'$ to carry physical charges $-1$, $+1$, and $0$, respectively. 
In this convention, $\psi'$ is a neutral spinon describing the local moment, while $s$ and $t$ describe charge fluctuations below and above the doublon sector. 
The total charge density is therefore
\begin{equation}
    n_{\mathrm{tot}}-2 = \nu+2=n_c+n_d+n_t-n_s ,
\end{equation}
where $n_c$ and $n_d$ denote the charge densities of the conduction and Dirac bands, respectively.

Within the projected Hilbert space, the interaction term can be rewritten as
\begin{equation}
P_G H_{\mathrm{int}}^{(f)} P_G = \sum_i \left(E_s n_{i ; s} + E_t n_{i ; t} + h_{i ; J}^{(f)} + \text{const.} \right),
\end{equation}
where $E_s = U / 2 + U(2+\kappa \nu)$ and $E_t = U / 2 - U(2+\kappa \nu)$ represent the on-site energies of the singlon and triplon with reference to the doublon state.
The physical $f$-electron operator is decomposed as:
\begin{equation} \label{eq:fst}
    P_Gf_{i ; \alpha}P_G = \sum_\beta s_{i ; \beta}^{\dagger} \psi_{i ; \beta}^{\prime} \psi_{i ; \alpha}^{\prime} + \sum_{\beta < \gamma} t_{i ; \alpha \beta \gamma} \psi_{i ; \beta}^{\prime \dagger} \psi_{i ; \gamma}^{\prime \dagger}.
\end{equation}

\textit{Mean field theory---}
Eq.~\eqref{original_Hamiltonian} can be in principle solved in the projected Hilbert space. 
To get an intuitive understanding of the phase diagram, here we develop a parton mean-field theory following Ref.~\cite{zhao2025resonating}. 
There are two natural mean-field channels for the projected fermion $f$ in Eq.~\eqref{eq:fst}.  First, when the slave boson $B_s\sim s^\dagger \psi'$ condenses, the physical $f$ fermion is effectively reconstructed as $f\sim B\psi'$. 
The spinon $\psi'$ now has a finite overlap with the physical electrons. On the other hand, when the $\psi'$ fermions are paired $\Delta\sim \sum_{a\eta s}s\langle \psi'_{i;a\bar\eta\bar s}\psi'_{i;a\eta s}\rangle \neq 0$, 
the physical electron has the form $f\sim \Delta s^\dagger + t\Delta^*$. 

When $B=0$ and $\Delta\neq0$, the original $f$ flat band is split into the lower and upper Hubbard bands, represented by $s$ and $t$, respectively.
Hybridization with the $c$ electrons further opens a momentum-selective Mott gap in the TBG sector, as shown in Fig.~\ref{figure1}(d).
This gap is strongly suppressed near the $\Gamma$ point but remains large away from $\Gamma$, due to the intrinsic topological property of TBG.
The resulting state is insulating at $\nu=-2$ and evolves into a small-Fermi-liquid phase upon hole doping to $\nu=-2-x$ with $x>0$~\cite{zhao2025resonating}. When $B$ also condenses while $\Delta$ remains finite, the local $d$-wave spin-singlet pairing of $\psi^\prime$ is transferred to the pairing instability of the Fermi surface, yielding a coherent SC phase.
If $\Delta$ vanishes while $B \neq 0$, we get a HFL phase. 

 The above idea can be captured more precisely by the full mean-field Hamiltonian
\begin{equation}
\begin{aligned}
H_{\mathrm{MF}}= & H_0^{c}+H_0^{d}+H_{\mathrm{MF}}^{\Delta}+H_{\mathrm{MF}}^B+\sum_i\left(E_s-\lambda+\mu\right) n_{i ; s} \\
& +\sum_i\left(E_t-\lambda-\mu\right) n_{i ; t}-\mu N_c \\
H_{\mathrm{MF}}^{\Delta}= & \sum_{\bm{k}, \bm{G}} \left(\Delta^* s_{-\bm{k}}+\sqrt{3} \Delta t_{\bm{k}}^{\dagger}\right) \gamma(\bm{k}+\bm{G}) c_{\bm{k}+\bm{G}}+\text { h.c. } \\
&+\sum_{\bm{p}}^{|\bm{p}|<\Lambda_d} \left(\Delta^* s_{-\bm{k}}+\sqrt{3} \Delta t_{\bm{k}}^{\dagger}\right) \tilde{h}(\bm{p},\bm{k})d_{\bm{p}} +\text { h.c. } \\
& -2 J \Delta^* \psi_{-\bm{k}}^{\prime} \psi_{\bm{k}}^{\prime}+\text { h.c. } \\
H_{\mathrm{MF}}^B= & \sum_{\bm{k}, \bm{G}} \left(B_s^* \psi_{\bm{k}}^{\prime \dagger}+B_t \psi_{-\bm{k}}^{\prime}\right) \gamma(\bm{k}+\bm{G}) c_{\bm{k}+\bm{G}}+\text { h.c. } \\
& +\sum_{\bm{p}}^{|\bm{p}|<\Lambda_d} \left(B_s^* \psi_{\bm{k}}^{\prime \dagger}+B_t \psi_{-\bm{k}}^{\prime}\right) \tilde{h}(\bm{p},\bm{k})d_{\bm{p}} +\text { h.c. } \\
& +\sum_{\bm{k}} B_{c}^{\prime*} s_{\bm{k}}^{\dagger} \psi_{\bm{k}}^{\prime}+\Delta_{c}^{\prime} t_{\bm{k}}^{\dagger} \psi_{\bm{k}}^{\prime}+\text { h.c. } \\
& +\sum_{\bm{k}} B_{d}^{\prime*} s_{\bm{k}}^{\dagger} \psi_{\bm{k}}^{\prime}+\Delta_{d}^{\prime} t_{\bm{k}}^{\dagger} \psi_{\bm{k}}^{\prime}+\text { h.c. }
\end{aligned}
\label{mean field Hamiltonian}
\end{equation}
Here $s_{-\bm{k}}, t_{\bm{k}}$ are eight-component operators considering the normalization and the pairing channel restriction, $s_{-\bm{k}}= \left\{\alpha s_{-\bm{k} ; \bar{\alpha}}\right\}, t_{\bm{k}}^{\dagger}=\left\{t_{\bm{k} ; \alpha}^{\dagger}\right\}$, where $t_{\bm{k} ; \alpha}=\frac{1}{2 \sqrt{3}} \sum_\beta \beta t_{\bm{k} ; \alpha \beta \bar{\beta}}$. And similar for $\psi_{\bm{k}}^{\prime}=\left\{\psi_{\bm{k} ; \alpha}^{\prime}\right\}$ and $\psi_{-\bm{k}}^{\prime}=\left\{\alpha \psi_{-\bar{\alpha}}^{\prime}\right\}$. 
We adopt the time-reversal invariant $d$-wave ansatz characterized by $H_{J, \mathrm{MF}}=-2 J \Delta^* \sum_{i, a \eta s} s \psi_{i ; a \bar{\eta} \bar{s}}^{\prime} \psi_{i ; a \eta s}^{\prime}$ for the spin interaction, where $s=\pm$ denotes spin up and down. The index $\bar{\alpha}=a\bar{\eta}\bar{s}$ represents the pairing partner of $\alpha=a \eta s$. 
Here the order parameters $\Delta \propto \alpha\left\langle\psi_{\bar{\alpha}}^{\prime} \psi_\alpha^{\prime}\right\rangle$, $B_s \propto\left\langle s_\beta^{\dagger} \psi_\beta^{\prime}\right\rangle, B_t \propto\left\langle t_\beta^{\dagger} \psi_\beta^{\prime}\right\rangle$, $B_{c}^{\prime} \propto \left\langle \psi_\alpha^{\prime\dagger} \gamma_{\alpha \beta} c_\beta \right\rangle$, 
$\Delta_{c}^{\prime} \propto \alpha \left\langle \psi_{\bar{\alpha}}^{\prime } \gamma_{\alpha \beta} c_\beta \right\rangle$, 
$B_{d}^{\prime} \propto \left\langle \psi_\alpha^{\prime \dagger} \tilde{h}_{\alpha \beta} d_\beta \right\rangle$ 
and $\Delta_{d}^{\prime} \propto \alpha \left\langle \psi_{\bar{\alpha}}^{\prime } \tilde{h}_{\alpha \beta} d_\beta \right\rangle$  are determined self-consistently. $\lambda$ and $\mu$ are introduced to satisfy the constraints in Eq.~\eqref{constraint}. 
Here $\Delta$ describes neutral local-moment pairing, while the remaining order parameters are slave-boson condensates that encode the hybridization between local moments and physical electrons.
A small temperature $T=0.001\,\mathrm{meV}$ is introduced in solving the mean-field equation for numerical stabilization.

\begin{figure}[t]
    \centering
    \includegraphics[width=\linewidth]{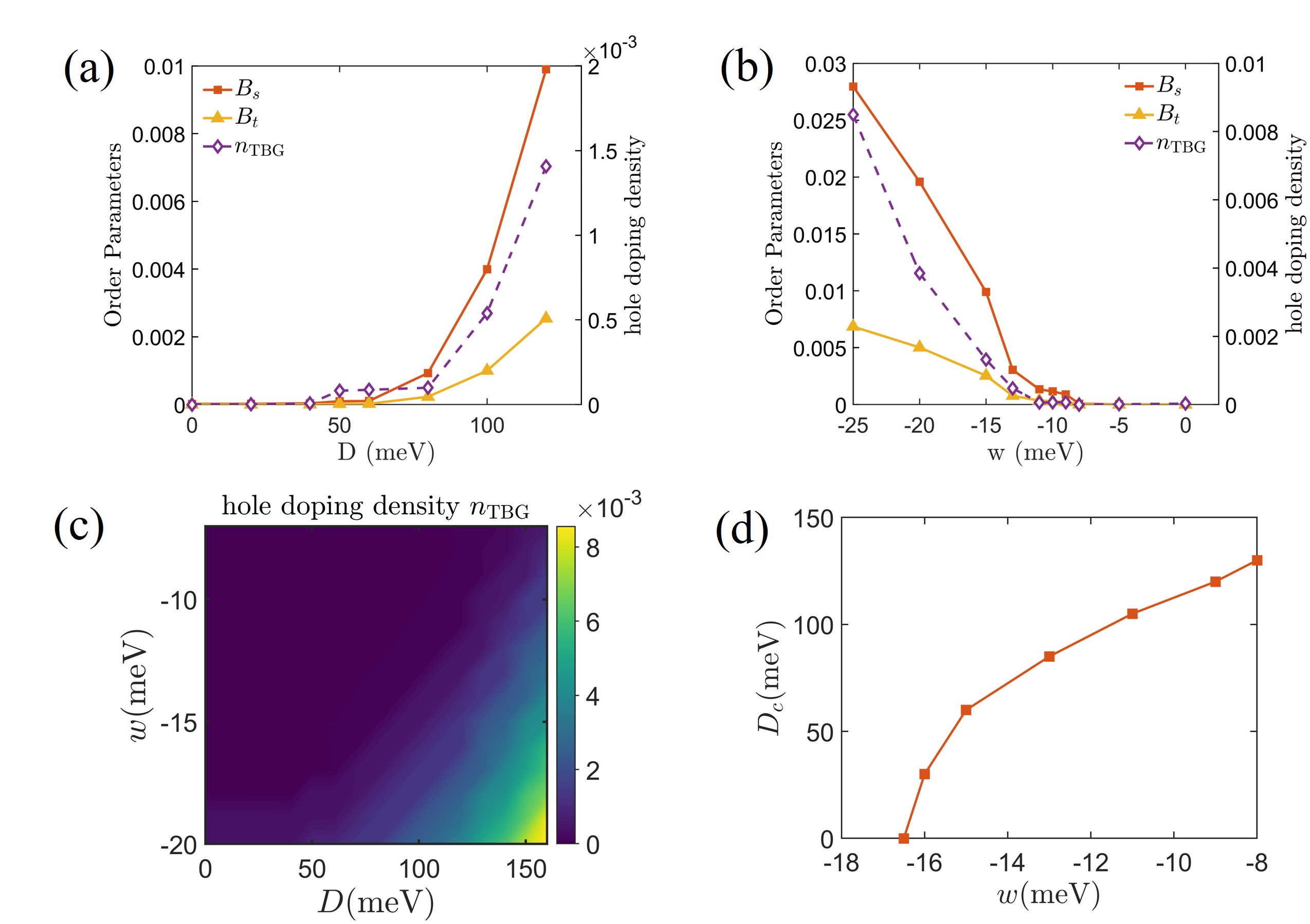}
    \caption{(a-b)Order parameters and hole doping density $n_{\rm TBG}$ at $\nu=-2$ (a) as a function of the displacement field $D$ with a fixed $w=-15\,\mathrm{meV}$, 
    and (b) as a function of the $d$-band energy offset $w$, with a fixed $D=120\,\mathrm{meV}$.
    (c) Hole doping density $n_{\rm TBG}$ in the $(D,w)$ plane. 
    (d) Critical displacement field $D_c$ for phase transition as a function of $w$.
    Here $n_{\rm TBG}$ denotes the unoccupied LHB spectral weight of $c$ and $f$ electrons
    integrated over the \(\Gamma\)-centered pocket.
    Other parameters used across all panels are $J=0.3\,\mathrm{meV}$, $\kappa=0.8$, and $U=40\,\mathrm{meV}$.
     }
    \label{figure2}
\end{figure}

\textit{Mott band and the $D$-tuned self-doping ---}
At small displacement field $D$, the TBG sector of MATTG forms a Mott insulator as in MATBG, with the flat bands split into upper and lower Hubbard bands (UHB and LHB)~\cite{zhao2025mixed}; this phase has $\Delta\neq 0$ and all other order parameters vanishing. The additional Dirac cone around $K_-$ [Fig.~\ref{figure1}(d)] renders the full system a semimetal.

At finite $D$, the large Mott gap protects this Dirac cone, which stays intact rather than splitting into two branches as in the noninteracting case [Fig.~\ref{figure1}(c)]. 
The dominant effect of $D$ is instead to lower the energy of the Dirac cone: since the UHB carries more spectral weight at $\nu=-2$, band repulsion pushes it downward, leading to a second-order shift $\delta E_D\sim \nu (M_1D)^2/U$ at small $D$ (see the appendix). 
While the middle Dirac cone stays within the Mott gap, the system remains semimetallic. 
On the other hand, once it drops below the top of the LHB, the system becomes self-doped, with hole pockets around $\Gamma$ and electron pockets around $K$ [Fig.~\ref{figure1}(a) and (e)]; the same self-doping follows from directly lowering the Dirac-cone energy offset $w$.

\textit{Phase transition driven by $D$ and $w$---}
The finite low-energy density of states generated by self-doping enhances the hybridization between the local moments and the physical electrons.
This is confirmed by the mean-field calculations in Figs.~\ref{figure2}(a) and \ref{figure2}(b), where the transition is driven by tuning $D$ and $w$, respectively.
In both cases, we find a finite $\Gamma$-centered TBG-sector hole density, defined as $n_{\rm TBG}\equiv n_{\Gamma}^{0}-n_{\Gamma}(D,w)$, where $n_{\Gamma}(D,w)=\sum_{|\mathbf{k}|<0.2\pi/a}(n_{c_1;\mathbf{k}}+n_{c_2;\mathbf{k}}+n_{f;\mathbf{k}})$ and $n_{\Gamma}^{0}$ is the corresponding value before self-doping.
For relatively weak spin interaction $J$, once $n_{\rm TBG}$ exceeds a small threshold, the slave bosons condense while $\Delta$ remains finite, driving the system into a SC state even at the integer filling $\nu=-2$.

\begin{figure}[t]
    \centering
    \includegraphics[width=\linewidth]{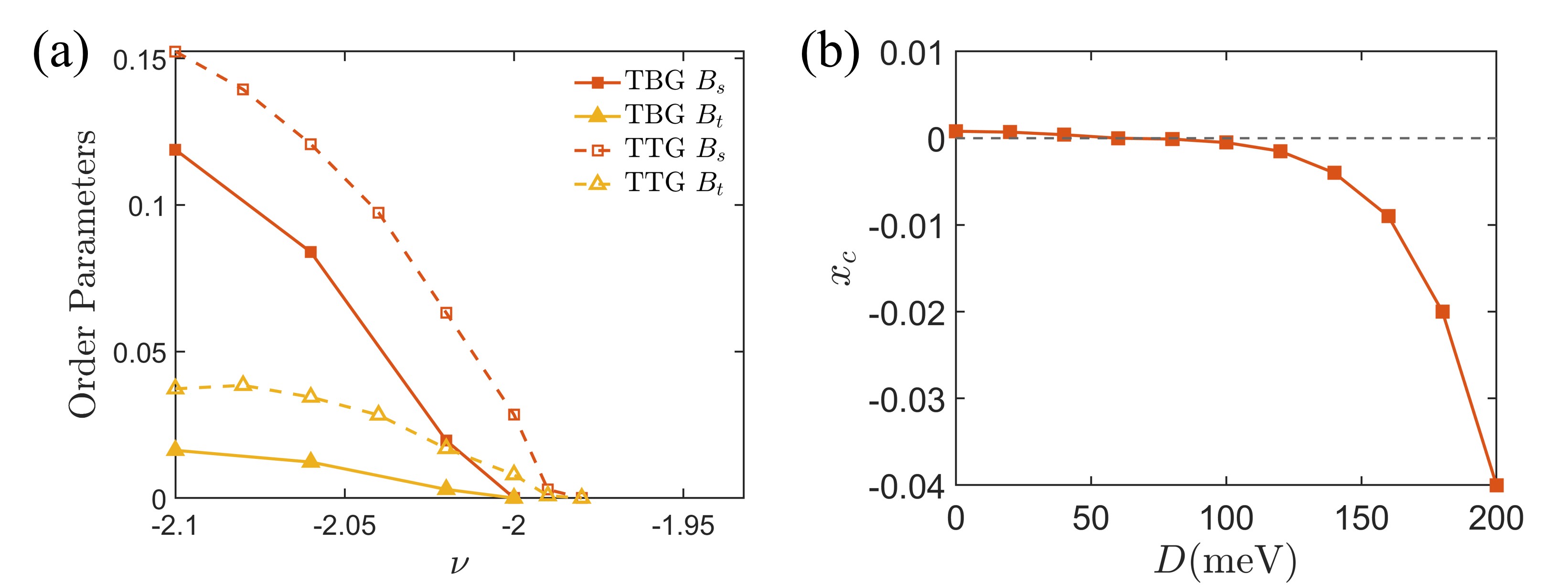}
    \caption{Order parameters as a function of filling for TBG and TTG under a finite displacement field. 
    The TTG curves use $D=160\,\mathrm{meV}$, while the TBG curves set $D=0$ and remove the $d$ band. 
    (b) Critical doping $x_c$ where SC emerges versus displacement field $D$.
    Other parameters are $w=-15\,\mathrm{meV}$ and $J=0.3\,\mathrm{meV}$.
    }
    \label{figure4}
\end{figure}

Importantly, this transition is not driven by a direct enhancement of the bare coupling to the monolayer $d$ electrons. Although increasing $D$ also enhances the $d$-$f$ hybridization matrix element, no transition occurs when only the Dirac cone is doped; rather, $D$ acts indirectly, by reshaping the low-energy band structure and self-doping the TBG sector.


A related question is why no transition occurs at $\nu=-2$ when increasing $w$ to drive self-electron-doping in the UHB.
The reason is that, in the $f$-$c$-$d$ model, the $f$ orbital couples predominantly to $c_1$ and negligibly to $c_2$ because they carry different angular momentum. As detailed in the appendix, the LHB states carry mainly $c_1$ weight, whereas the UHB states are predominantly $c_2$-like, so proximity to the UHB does not strongly enhance the hybridization, and the SC phase requires self-doping in the $c_1$-dominated LHB.

\textit{Enhancement of superconductivity---}
Self-doping also accounts for the displacement-field enhancement of superconductivity at the filling $\nu=-2-x$. In Fig.~\ref{figure4} we compare MATTG with a MATBG reference (setting $D=0$ and removing the $d$ bands from Eq.~\eqref{original_Hamiltonian}): the finite-$D$ MATTG system develops superconductivity at smaller or even negative hole doping $x$ and with larger hybridization order parameters.
We also track the critical doping $x_c$ where SC emerges as a function of displacement field $D$. As shown in Fig.~\ref{figure4}(b), in the large $D$ regime, $x_c$ decreases significantly as $D$ increases and even evolves into a negative value. The negative $x_c$ is possible because self-doping keeps the TBG sector hole doped even for $x<0$.
This is consistent with experiments, where a strong displacement field allows superconductivity to appear over an enlarged filling range with $|\nu|<2$~\cite{park2021tunable,cao2021pauli,hao2021electric,zhang2022promotion}.

\begin{figure}[t]
    \centering
    
    \includegraphics[width=\linewidth]{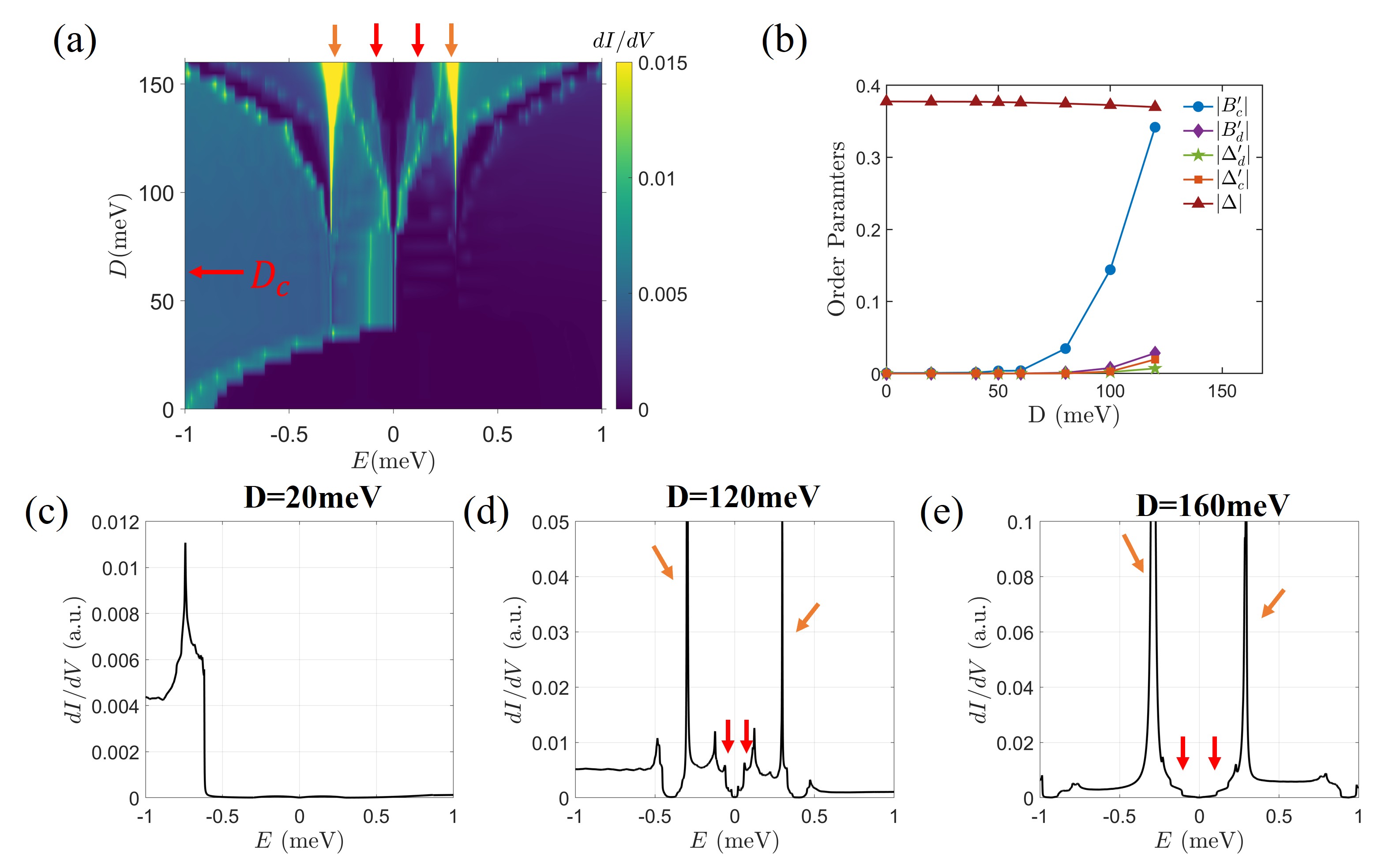}
    \caption{(a) Color map of the STM signal as a function of energy $E$ and
    displacement field $D$. The critical displacement field $D_c$ for the phase transition is marked by red arrow. 
    (b) Order parameters as a function of $D$.
    (c-e) Representative STM spectra at $x=0$ for $D=20$meV, $D=120$ meV, $D=160$meV, respectively. 
    The pseudogap and SC gap are marked with orange and red arrows.
    In all the calculations, we choose parameters $w=-15\,{\rm meV}$ and $J=0.3\,{\rm meV}$.
    }
    \label{fig:STM}
\end{figure}

\textit{Two-gap structure and its evolution---} 
The superconducting phase displays two distinct energy scales in the STM
spectrum, as shown in Fig.~\ref{fig:STM}. The outer scale is a broad
pseudogap, which originates from the spinon pairing $\Delta$ and is at the energy scale of $J$. The inner scale is the superconducting gap induced for the small hole pocket. Such two-gap phenomenology has also been reported in recent tunneling measurements in magic-angle graphene
systems at finite filling\cite{oh2021evidence,park2026experimental,kim2026resolving}.

For $D$ below the phase transition, the system is in the semimetal phase, with a large charge transfer gap in the STM spectrum[Fig.~\ref{fig:STM}(c)]. 
When $D$ approaches the critical value $D_c$, 
self doping happens, but the hybridization order parameters are still weak, resulting in a nearly negligible superconducting gap in the physical-electron sector.
At intermediate $D$, the low-energy physical-electron spectral weight is further depleted from the inner-gap region and a clear two-gap structure emerges [Fig.~\ref{fig:STM}(d)]. At larger $D$, the hybridization $B$ becomes strong enough that the two-gap structure evolves into a U-shaped gap[Fig.~\ref{fig:STM}(e)].

\textit{Particle-hole asymmetry in the TTG phase diagram---}
A striking yet poorly understood feature of TTG is its pronounced particle-hole (PH) asymmetry in the phase diagram.
At positive integer fillings $(\nu=2,3)$, experiments observe not only a $D$-tuned transition, but also a broad high-resistance regime emanating from the critical displacement field at finite temperature~\cite{zhang2025electrically,zhang2022promotion,park2021tunable}, which is naturally interpreted as a quantum-critical regime associated with the onset of local-moment hybridization~\cite{zhang2025electrically}. In contrast, no analogous transition or high-resistance quantum-critical behavior is observed at the corresponding negative fillings $\nu=-2,-3$.
We attribute this asymmetry to an intrinsic energy offset of the Dirac cone at $D=0$, which controls how hard $D$ can drive self-doping and hence the semimetal-superconductor transition.
As discussed above, this energy offset originates from the intrinsic layer asymmetry, which can be incorporated through a middle-layer onsite potential, $E_{\rm sh}\phi_{\mathbf r,2,\eta s}^{\dagger}\sigma_0\phi_{\mathbf r,2,\eta s}$~\cite{fischer2022unconventional}, where $\phi_{\mathbf r,2,\eta s}$ is the wave function of the middle layer. We choose $E_{\rm sh}=-30{\rm meV}$, close to the estimate in Ref.~\cite{fischer2022unconventional}. As shown in the appendix, this on-site potential effectively acts as a positive Dirac-cone offset and gives $w\simeq 15{\rm meV}$.

\begin{figure}[t]
    \centering
    \includegraphics[width=\linewidth]{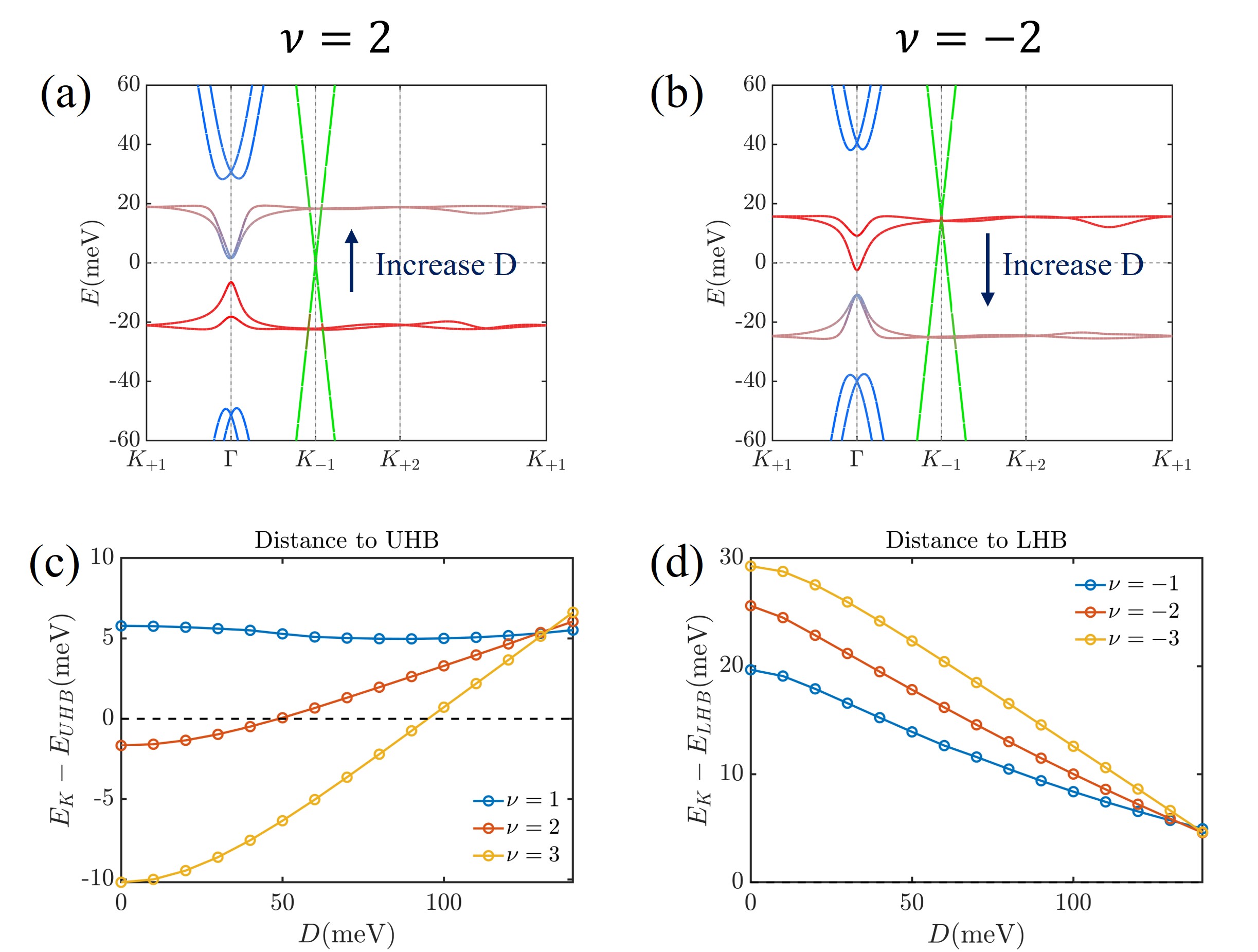}
    \caption{(a-b)Representative ancilla bands illustrating particle-hole asymmetry. The red, purple, green, blue colors of the dispersion line represent the weight of flat bands, Dirac cone, remote bands, ancilla fermion $\psi$. The black arrows mark the direction of the Dirac cone shift when increasing $D$.
    (c-d) Distance of the middle Dirac cone to UHB for positive fillings(c) and LHB for negative fillings(d). According to our theory, a sign change of this distance signals the onset of Hubbard-band doping and marks the transition.
    $E_{sh}=-30\,\mathrm{meV}$.
    }
    \label{figure5}
\end{figure}
The results at $\nu=-2$ can be mapped to $\nu=2$ by a particle-hole transformation with $w \rightarrow -w$. Therefore we conclude that there is a semimetal to superconductor transition at $\nu=+2$ for the realistive value of $w=15$ meV. 
By comparison, at $\nu=-2$ with $w>0$, no self-doping occurs and we do not expect such a transition at moderate $D$. 
The same PH asymmetry also applies to odd fillings, where the spinon cannot be paired and the above slave particle theory does not work anymore. However, we still expect similar self-doping effect and a transition to close the Mott gap. 
The difference is that the resulting phase is metallic instead of superconducting. 
To describe this regime, we instead use the ancilla theory, which is equivalent to the slave-particle model at $\nu=\pm2$~\cite{zhao2025mixed,zhao2025resonating} and captures the physics of Mottness and local moments~\cite{zhang2020pseudogap,mascot2022electronic,zhou2025variational} also at odd fillings. 
We illustrate the PH-asymmetry mechanism in Fig.~\ref{figure5} using Mott band structure without coupling to the local moment \cite{zhou2025variational}.
At positive fillings $\nu=2,3$, the UHB lies close to zero energy, so increasing the displacement field can readily shift the Dirac cone upward to UHB, which leads to self-doping into the UHB and triggers the transition [Figs.~\ref{figure5}(a) and \ref{figure5}(c)].
In contrast, at negative fillings with $\nu<0$, the intrinsic Dirac-cone offset places the LHB far from zero energy.
This large energy separation prevents $D$ from inducing LHB self-doping, thereby suppressing the corresponding displacement-field-driven transition [Figs.~\ref{figure5}(b) and \ref{figure5}(d)].
Further details are provided in the appendix.  

\textit{Discussion---}
The character of the $D$-driven phase is set by the filling parity. At even fillings $|\nu|=2$, the spinon $\psi^\prime$ forms intervalley spin-singlet pairs ($\Delta\neq0$) already in the semimetal; once self-doping condenses the slave boson $B$, this local pairing is transferred to the physical electrons and leads to  a superconducting phase. Around the filling $|\nu|=2$, there are actually two different symmetric Fermi liquids\cite{zhang2020spin,zhao2025ancilla,zhao2025resonating,zhao2025mixed}: a Fermi liquid with large Fermi surface and a second Fermi liquid (sFL) with small $\Gamma$-centered hole pockets. At small self doping across $D_c$, the normal state is a sFL phase with small hole pockets and a conventional Fermi liquid may be reached only at much larger $D$.  At odd fillings $|\nu|=1,3$, the odd electron number per site precludes onsite spinon pairing, so $\psi^\prime$ stays gapless and both superconductor and sFL are not expected.  In this case, self-doping then condenses $B$ and the system instead becomes a heavy Fermi liquid (HFL), consistent with the $D$-tuned semimetal-to-HFL transition observed at $\nu=3$~\cite{zhang2025electrically}. The present slave-particle construction is tailored to $\nu=\pm2$ and does not work for the odd fillings, but the same self-doping mechanism should drive the transition there. We need to emphasize that the HFL here is not driven by Kondo physics, instead it originates from slightly doped Mott insulator in the TBG sector.

In our calculations for $|\nu|=2$ the  on-site spin interaction is small, $J=0.3\,{\rm meV}$, so a small self-doped hole density already triggers the transition; larger $J$ stabilizes the pairing of the spinon $\psi'$ and requires stronger self-doping, pushing the transition to larger $D$ or to $|\nu|>2$. In reality there are also  spin interactions that can act across sites, valleys, and orbitals~\cite{hu2023kondo,chou2023kondo}, which lead to symmetry breaking phases.  In this work we restrict to a symmetric Mott state before the transition and leave it to future work to study how symmetry breaking intertwines with the transition.

The displacement-field enhancement of superconductivity is restricted to fillings near $\nu=\pm2$; farther from $|\nu|=2$, large $D$ instead suppresses superconductivity through a strong reduction of the spinon pairing $\Delta$, consistent with the experimental results~\cite{park2021tunable,hao2021electric,zhang2022promotion} (see the appendix). Finally, our conclusions rest on a mean-field treatment; a quantitative account of the transition and the finite-temperature regime would require methods that capture dynamical correlations and fluctuations beyond mean field, such as variational Monte Carlo\cite{shackleton2024emergent, muller2025polaronic} or dynamical mean-field theory~\cite{zhao2026pseudogap}.

\textit{Conclusion---}We have shown that, in magic-angle twisted trilayer graphene, the displacement field acts not through a direct Kondo-like coupling but by shifting the Dirac cone and self-doping the Mott-reconstructed TBG sector. A slave-particle calculation at $\nu=\pm2$ shows that this self-doping condenses the slave boson and turns the semimetal into a superconductor, and that the same mechanism underlies the displacement-field enhancement of pairing, the two-gap tunneling spectrum, and the particle-hole asymmetry of the phase diagram. Which phase emerges is set by the filling parity: paired spinons give superconductivity at even fillings, whereas unpaired spinons give a heavy Fermi liquid at odd fillings. Together these results provide a unified picture of the electric-field-tunable superconductivity, Mottness, and heavy-fermion behavior of MATTG, and motivate treatments beyond mean field to capture the finite-temperature quantum-critical physics.

\textit{Acknowledgement---} This work was supported by the  National Science Foundation under Grant No. DMR-2237031.

\bibliography{ref}

@article{cao2018unconventional,
  title={Unconventional superconductivity in magic-angle graphene superlattices},
  author={Cao, Yuan and Fatemi, Valla and Fang, Shiang and Watanabe, Kenji and Taniguchi, Takashi and Kaxiras, Efthimios and Jarillo-Herrero, Pablo},
  journal={Nature},
  volume={556},
  number={7699},
  pages={43--50},
  year={2018},
  publisher={Nature Publishing Group UK London},
  url={https://doi.org/10.1038/nature26160}
}

@article{zhang2020spin,
  title={Spin liquids and pseudogap metals in the SU (4) Hubbard model in a moir{\'e} superlattice},
  author={Zhang, Ya-Hui and Mao, Dan},
  journal={Physical Review B},
  volume={101},
  number={3},
  pages={035122},
  year={2020},
  publisher={APS}
}

@article{cao2018correlated,
  title={Correlated insulator behaviour at half-filling in magic-angle graphene superlattices},
  author={Cao, Yuan and Fatemi, Valla and Demir, Ahmet and Fang, Shiang and Tomarken, Spencer L and Luo, Jason Y and Sanchez-Yamagishi, Javier D and Watanabe, Kenji and Taniguchi, Takashi and Kaxiras, Efthimios and others},
  journal={Nature},
  volume={556},
  number={7699},
  pages={80--84},
  year={2018},
  publisher={Nature Publishing Group UK London},
  url={https://doi.org/10.1038/nature26154}
}

@article{yankowitz2019tuning,
  title={Tuning superconductivity in twisted bilayer graphene},
  author={Yankowitz, Matthew and Chen, Shaowen and Polshyn, Hryhoriy and Zhang, Yuxuan and Watanabe, Kenji and Taniguchi, Takashi and Graf, David and Young, Andrea F and Dean, Cory R},
  journal={Science},
  volume={363},
  number={6431},
  pages={1059--1064},
  year={2019},
  publisher={American Association for the Advancement of Science},
  URL = {https://www.science.org/doi/abs/10.1126/science.aav1910}
}

@article{lu2019superconductors,
  title={Superconductors, orbital magnets and correlated states in magic-angle bilayer graphene},
  author={Lu, Xiaobo and Stepanov, Petr and Yang, Wei and Xie, Ming and Aamir, Mohammed Ali and Das, Ipsita and Urgell, Carles and Watanabe, Kenji and Taniguchi, Takashi and Zhang, Guangyu and others},
  journal={Nature},
  volume={574},
  number={7780},
  pages={653--657},
  year={2019},
  publisher={Nature Publishing Group UK London},
  url={https://doi.org/10.1038/s41586-019-1695-0}
}

@article{stepanov2020untying,
  title={Untying the insulating and superconducting orders in magic-angle graphene},
  author={Stepanov, Petr and Das, Ipsita and Lu, Xiaobo and Fahimniya, Ali and Watanabe, Kenji and Taniguchi, Takashi and Koppens, Frank HL and Lischner, Johannes and Levitov, Leonid and Efetov, Dmitri K},
  journal={Nature},
  volume={583},
  number={7816},
  pages={375--378},
  year={2020},
  publisher={Nature Publishing Group UK London},
  url={ https://doi.org/10.1038/s41586-020-2459}
}

@article{cao2021nematicity,
  title={Nematicity and competing orders in superconducting magic-angle graphene},
  author={Cao, Yuan and Rodan-Legrain, Daniel and Park, Jeong Min and Yuan, Noah FQ and Watanabe, Kenji and Taniguchi, Takashi and Fernandes, Rafael M and Fu, Liang and Jarillo-Herrero, Pablo},
  journal={science},
  volume={372},
  number={6539},
  pages={264--271},
  year={2021},
  publisher={American Association for the Advancement of Science},
  URL = {https://www.science.org/doi/abs/10.1126/science.abc2836}
}

@article{liu2021tuning,
  title={Tuning electron correlation in magic-angle twisted bilayer graphene using Coulomb screening},
  author={Liu, Xiaoxue and Wang, Zhi and Watanabe, Kenji and Taniguchi, Takashi and Vafek, Oskar and Li, JIA},
  journal={Science},
  volume={371},
  number={6535},
  pages={1261--1265},
  year={2021},
  publisher={American Association for the Advancement of Science},
  URL = {https://www.science.org/doi/abs/10.1126/science.abb8754}
}

@article{arora2020superconductivity,
  title={Superconductivity in metallic twisted bilayer graphene stabilized by WSe2},
  author={Arora, Harpreet Singh and Polski, Robert and Zhang, Yiran and Thomson, Alex and Choi, Youngjoon and Kim, Hyunjin and Lin, Zhong and Wilson, Ilham Zaky and Xu, Xiaodong and Chu, Jiun-Haw and others},
  journal={Nature},
  volume={583},
  number={7816},
  pages={379--384},
  year={2020},
  publisher={Nature Publishing Group UK London},
  url={https://doi.org/10.1038/s41586-020-2473-8}
}

@article{cao2020strange,
  title={Strange metal in magic-angle graphene with near Planckian dissipation},
  author={Cao, Yuan and Chowdhury, Debanjan and Rodan-Legrain, Daniel and Rubies-Bigorda, Oriol and Watanabe, Kenji and Taniguchi, Takashi and Senthil, T and Jarillo-Herrero, Pablo},
  journal={Physical review letters},
  volume={124},
  number={7},
  pages={076801},
  year={2020},
  publisher={APS},
  url = {https://link.aps.org/doi/10.1103/PhysRevLett.124.076801}
}

@article{po2018origin,
  title={Origin of Mott insulating behavior and superconductivity in twisted bilayer graphene},
  author={Po, Hoi Chun and Zou, Liujun and Vishwanath, Ashvin and Senthil, T},
  journal={Physical Review X},
  volume={8},
  number={3},
  pages={031089},
  year={2018},
  publisher={APS},
  url = {https://link.aps.org/doi/10.1103/PhysRevX.8.031089}
}

@article{ahn2019failure,
  title={Failure of Nielsen-Ninomiya theorem and fragile topology in two-dimensional systems with space-time inversion symmetry: application to twisted bilayer graphene at magic angle},
  author={Ahn, Junyeong and Park, Sungjoon and Yang, Bohm-Jung},
  journal={Physical Review X},
  volume={9},
  number={2},
  pages={021013},
  year={2019},
  publisher={APS},
  url = {https://link.aps.org/doi/10.1103/PhysRevX.9.021013}
}

@article{ledwith2021strong,
  title={Strong coupling theory of magic-angle graphene: A pedagogical introduction},
  author={Ledwith, Patrick J and Khalaf, Eslam and Vishwanath, Ashvin},
  journal={Annals of Physics},
  volume={435},
  pages={168646},
  year={2021},
  publisher={Elsevier},
  url={https://doi.org/10.1016/j.aop.2021.168646}
}

@article{song2021twisted,
  title={Twisted bilayer graphene. II. Stable symmetry anomaly},
  author={Song, Zhi-Da and Lian, Biao and Regnault, Nicolas and Bernevig, B Andrei},
  journal={Physical Review B},
  volume={103},
  number={20},
  pages={205412},
  year={2021},
  publisher={APS},
  url = {https://link.aps.org/doi/10.1103/PhysRevB.103.205412}
}

@article{park2021tunable,
  title={Tunable strongly coupled superconductivity in magic-angle twisted trilayer graphene},
  author={Park, Jeong Min and Cao, Yuan and Watanabe, Kenji and Taniguchi, Takashi and Jarillo-Herrero, Pablo},
  journal={Nature},
  volume={590},
  number={7845},
  pages={249--255},
  year={2021},
  publisher={Nature Publishing Group UK London},
  url={https://doi.org/10.1038/s41586-021-03192-0}
}

@article{cao2021pauli,
  title={Pauli-limit violation and re-entrant superconductivity in moir{\'e} graphene},
  author={Cao, Yuan and Park, Jeong Min and Watanabe, Kenji and Taniguchi, Takashi and Jarillo-Herrero, Pablo},
  journal={Nature},
  volume={595},
  number={7868},
  pages={526--531},
  year={2021},
  publisher={Nature Publishing Group UK London},
  url={https://doi.org/10.1038/s41586-021-03685-y}
}

@article{hao2021electric,
  title={Electric field--tunable superconductivity in alternating-twist magic-angle trilayer graphene},
  author={Hao, Zeyu and Zimmerman, AM and Ledwith, Patrick and Khalaf, Eslam and Najafabadi, Danial Haie and Watanabe, Kenji and Taniguchi, Takashi and Vishwanath, Ashvin and Kim, Philip},
  journal={Science},
  volume={371},
  number={6534},
  pages={1133--1138},
  year={2021},
  publisher={American Association for the Advancement of Science},
  url={https://www.science.org/doi/10.1126/science.abg0399}
}

@article{park2022robust,
  title={Robust superconductivity in magic-angle multilayer graphene family},
  author={Park, Jeong Min and Cao, Yuan and Xia, Li-Qiao and Sun, Shuwen and Watanabe, Kenji and Taniguchi, Takashi and Jarillo-Herrero, Pablo},
  journal={Nature Materials},
  volume={21},
  number={8},
  pages={877--883},
  year={2022},
  publisher={Nature Publishing Group UK London},
  url={https://doi.org/10.1038/s41563-022-01287-1}
}

@article{zhang2022promotion,
  title={Promotion of superconductivity in magic-angle graphene multilayers},
  author={Zhang, Yiran and Polski, Robert and Lewandowski, Cyprian and Thomson, Alex and Peng, Yang and Choi, Youngjoon and Kim, Hyunjin and Watanabe, Kenji and Taniguchi, Takashi and Alicea, Jason and others},
  journal={Science},
  volume={377},
  number={6614},
  pages={1538--1543},
  year={2022},
  publisher={American Association for the Advancement of Science},
  url={https://www.science.org/doi/10.1126/science.abn8585}
}

@article{mukherjee2025superconducting,
  title={Superconducting magic-angle twisted trilayer graphene with competing magnetic order and moir{\'e} inhomogeneities},
  author={Mukherjee, Ayshi and Layek, Surat and Sinha, Subhajit and Kundu, Ritajit and Marchawala, Alisha H and Hingankar, Mahesh and Sarkar, Joydip and Sangani, LD Varma and Agarwal, Heena and Ghosh, Sanat and others},
  journal={Nature Materials},
  pages={1--7},
  year={2025},
  publisher={Nature Publishing Group UK London},
  url={https://doi.org/10.1038/s41563-025-02252-4}
}

@article{khalaf2019magic,
  title={Magic angle hierarchy in twisted graphene multilayers},
  author={Khalaf, Eslam and Kruchkov, Alex J and Tarnopolsky, Grigory and Vishwanath, Ashvin},
  journal={Physical Review B},
  volume={100},
  number={8},
  pages={085109},
  year={2019},
  publisher={APS},
  url = {https://link.aps.org/doi/10.1103/PhysRevB.100.085109}
}

@article{pierce2025tunable,
  title={Tunable interplay between light and heavy electrons in twisted trilayer graphene},
  author={Pierce, Andrew T and Xie, Yonglong and Park, Jeong Min and Cai, Zhuozhen and Watanabe, Kenji and Taniguchi, Takashi and Jarillo-Herrero, Pablo and Yacoby, Amir},
  journal={Nature Physics},
  pages={1--6},
  year={2025},
  publisher={Nature Publishing Group UK London},
  url={https://doi.org/10.1038/s41567-025-02956-z}
}

@article{christos2022correlated,
  title={Correlated insulators, semimetals, and superconductivity in twisted trilayer graphene},
  author={Christos, Maine and Sachdev, Subir and Scheurer, Mathias S},
  journal={Physical Review X},
  volume={12},
  number={2},
  pages={021018},
  year={2022},
  publisher={APS},
  url = {https://link.aps.org/doi/10.1103/PhysRevX.12.021018}
}

@article{fischer2022unconventional,
  title={Unconventional superconductivity in magic-angle twisted trilayer graphene},
  author={Fischer, Ammon and Goodwin, Zachary AH and Mostofi, Arash A and Lischner, Johannes and Kennes, Dante M and Klebl, Lennart},
  journal={npj Quantum Materials},
  volume={7},
  number={1},
  pages={5},
  year={2022},
  publisher={Nature Publishing Group UK London},
  url={https://doi.org/10.1038/s41535-021-00410-w}
}

@article{yu2023magic,
  title={Magic-angle twisted symmetric trilayer graphene as a topological heavy-fermion problem},
  author={Yu, Jiabin and Xie, Ming and Bernevig, B Andrei and Das Sarma, Sankar},
  journal={Physical Review B},
  volume={108},
  number={3},
  pages={035129},
  year={2023},
  publisher={APS},
  url = {https://link.aps.org/doi/10.1103/PhysRevB.108.035129}
}

@article{song2022magic,
  title={Magic-angle twisted bilayer graphene as a topological heavy fermion problem},
  author={Song, Zhi-Da and Bernevig, B Andrei},
  journal={Physical review letters},
  volume={129},
  number={4},
  pages={047601},
  year={2022},
  publisher={APS},
  url = {https://link.aps.org/doi/10.1103/PhysRevLett.129.047601}
}

@article{zhang2025electrically,
  title={Electrically tunable heavy fermion and quantum criticality in magic-angle twisted trilayer graphene},
  author={Zhang, Le and Zhou, Wenqiang and Fang, Xinjie and Zhan, Zhen and Watanabe, Kenji and Taniguchi, Takashi and Yang, Yi-feng and Xu, Shuigang},
  journal={arXiv preprint arXiv:2507.12254},
  year={2025},
  url={https://arxiv.org/abs/2507.12254}
}

@article{zhao2025mixed,
  title={Mixed valence Mott insulator and composite excitation in twisted bilayer graphene},
  author={Zhao, Jing-Yu and Zhou, Boran and Zhang, Ya-Hui},
  journal={arXiv preprint arXiv:2507.00139},
  year={2025},
  url = {https://link.aps.org/doi/10.1103/kts3-81nk}
}

@article{shackleton2024emergent,
  title={Emergent polaronic correlations in doped spin liquids},
  author={Shackleton, Leyna and Zhang, Shiwei},
  journal={arXiv preprint arXiv:2408.02190},
  year={2024},
  url={https://arxiv.org/abs/2408.02190}
}

@article{muller2025polaronic,
  title={Polaronic correlations from optimized ancilla wave functions for the Fermi--Hubbard model},
  author={M{\"u}ller, Tobias and Thomale, Ronny and Sachdev, Subir and Iqbal, Yasir},
  journal={Proceedings of the National Academy of Sciences},
  volume={122},
  number={20},
  pages={e2504261122},
  year={2025},
  publisher={National Academy of Sciences},
  url={https://doi.org/10.1073/pnas.2504261122}
}

@article{zhao2025ancilla,
  title={Ancilla theory of twisted bilayer graphene I: topological Mott localization and pseudogap metal in twisted bilayer graphene},
  author={Zhao, Jing-Yu and Zhou, Boran and Zhang, Ya-Hui},
  journal={arXiv preprint arXiv:2502.17430},
  year={2025},
  url={https://arxiv.org/abs/2502.17430}
}

@article{zhao2025resonating,
  title={Resonating-valence-bond superconductor from small Fermi surface in twisted bilayer graphene},
  author={Zhao, Jing-Yu and Zhang, Ya-Hui},
  journal={arXiv preprint arXiv:2510.26801},
  year={2025},
  url={https://arxiv.org/abs/2510.26801}
}

@article{mascot2022electronic,
  title={Electronic spectra with paramagnon fractionalization in the single-band Hubbard model},
  author={Mascot, Eric and Nikolaenko, Alexander and Tikhanovskaya, Maria and Zhang, Ya-Hui and Morr, Dirk K and Sachdev, Subir},
  journal={Physical Review B},
  volume={105},
  number={7},
  pages={075146},
  year={2022},
  publisher={APS},
  url = {https://link.aps.org/doi/10.1103/PhysRevB.105.075146}}

@article{zhang2020pseudogap,
  title={From the pseudogap metal to the Fermi liquid using ancilla qubits},
  author={Zhang, Ya-Hui and Sachdev, Subir},
  journal={Physical Review Research},
  volume={2},
  number={2},
  pages={023172},
  year={2020},
  publisher={APS},
  url = {https://link.aps.org/doi/10.1103/PhysRevResearch.2.023172}
}

@article{zhou2025variational,
  title={Variational wavefunction for a Mott insulator at finite U using ancilla qubits},
  author={Zhou, Boran and Jin, Hui-Ke and Zhang, Ya-Hui},
  journal={Physical Review B},
  volume={112},
  number={11},
  pages={115159},
  year={2025},
  publisher={APS},
  url = {https://link.aps.org/doi/10.1103/8knn-mr5x}
}

@article{wang2025electron,
  title={Electron-phonon coupling in the topological heavy fermion model of twisted bilayer graphene},
  author={Wang, Yi-Jie and Zhou, Geng-Dong and Lian, Biao and Song, Zhi-Da},
  journal={Physical Review B},
  volume={111},
  number={3},
  pages={035110},
  year={2025},
  publisher={APS},
  url = {https://link.aps.org/doi/10.1103/PhysRevB.111.035110}
}

@article{lau2025topological,
  title={Topological mixed valence model for twisted bilayer graphene},
  author={Lau, Liam LH and Coleman, Piers},
  journal={Physical Review X},
  volume={15},
  number={2},
  pages={021028},
  year={2025},
  publisher={APS},
  url = {https://link.aps.org/doi/10.1103/PhysRevX.15.021028}

}

@article{wong2020cascade,
  title={Cascade of electronic transitions in magic-angle twisted bilayer graphene},
  author={Wong, Dillon and Nuckolls, Kevin P and Oh, Myungchul and Lian, Biao and Xie, Yonglong and Jeon, Sangjun and Watanabe, Kenji and Taniguchi, Takashi and Bernevig, B Andrei and Yazdani, Ali},
  journal={Nature},
  volume={582},
  number={7811},
  pages={198--202},
  year={2020},
  publisher={Nature Publishing Group UK London},
  url={https://doi.org/10.1038/s41586-020-2339-0}
}

@article{andrei2020graphene,
  title={Graphene bilayers with a twist},
  author={Andrei, Eva Y and MacDonald, Allan H},
  journal={Nature materials},
  volume={19},
  number={12},
  pages={1265--1275},
  year={2020},
  publisher={Nature Publishing Group UK London},
  url={https://doi.org/10.1038/s41563-020-00840-0}
}

@article{andrei2021marvels,
  title={The marvels of moir{\'e} materials},
  author={Andrei, Eva Y and Efetov, Dmitri K and Jarillo-Herrero, Pablo and MacDonald, Allan H and Mak, Kin Fai and Senthil, T and Tutuc, Emanuel and Yazdani, Ali and Young, Andrea F},
  journal={Nature Reviews Materials},
  volume={6},
  number={3},
  pages={201--206},
  year={2021},
  publisher={Nature Publishing Group UK London},
  url={https://doi.org/10.1038/s41578-021-00284-1}
}

@article{nuckolls2024microscopic,
  title={A microscopic perspective on moir{\'e} materials},
  author={Nuckolls, Kevin P and Yazdani, Ali},
  journal={Nature Reviews Materials},
  volume={9},
  number={7},
  pages={460--480},
  year={2024},
  publisher={Nature Publishing Group UK London},
  url={https://doi.org/10.1038/s41578-024-00682-1}
}

@article{kim2022evidence,
  title={Evidence for unconventional superconductivity in twisted trilayer graphene},
  author={Kim, Hyunjin and Choi, Youngjoon and Lewandowski, Cyprian and Thomson, Alex and Zhang, Yiran and Polski, Robert and Watanabe, Kenji and Taniguchi, Takashi and Alicea, Jason and Nadj-Perge, Stevan},
  journal={Nature},
  volume={606},
  number={7914},
  pages={494--500},
  year={2022},
  publisher={Nature Publishing Group UK London},
  url={https://doi.org/10.1038/s41586-022-04715-z}
}

@article{shen2023dirac,
  title={Dirac spectroscopy of strongly correlated phases in twisted trilayer graphene},
  author={Shen, Cheng and Ledwith, Patrick J and Watanabe, Kenji and Taniguchi, Takashi and Khalaf, Eslam and Vishwanath, Ashvin and Efetov, Dmitri K},
  journal={Nature Materials},
  volume={22},
  number={3},
  pages={316--321},
  year={2023},
  publisher={Nature Publishing Group UK London},
  url={https://doi.org/10.1038/s41563-022-01428-6}
}

@article{oh2021evidence,
  title={Evidence for unconventional superconductivity in twisted bilayer graphene},
  author={Oh, Myungchul and Nuckolls, Kevin P and Wong, Dillon and Lee, Ryan L and Liu, Xiaomeng and Watanabe, Kenji and Taniguchi, Takashi and Yazdani, Ali},
  journal={Nature},
  volume={600},
  number={7888},
  pages={240--245},
  year={2021},
  publisher={Nature Publishing Group UK London},
  url={https://doi.org/10.1038/s41586-021-04121-x}
}

@article{park2026experimental,
  title={Experimental evidence for nodal superconducting gap in moir{\'e} graphene},
  author={Park, Jeong Min and Sun, Shuwen and Watanabe, Kenji and Taniguchi, Takashi and Jarillo-Herrero, Pablo},
  journal={Science},
  volume={391},
  number={6780},
  pages={79--83},
  year={2026},
  publisher={American Association for the Advancement of Science},
  url={https://doi.org/10.1038/s41586-021-04121-x}
}

@article{kim2026resolving,
  title={Resolving intervalley gaps and many-body resonances in moir{\'e} superconductors},
  author={Kim, Hyunjin and Rai, Gautam and Crippa, Lorenzo and C{\u{a}}lug{\u{a}}ru, Dumitru and Hu, Haoyu and Choi, Youngjoon and Kong, Lingyuan and Baum, Eli and Zhang, Yiran and Holleis, Ludwig and others},
  journal={Nature},
  pages={1--7},
  year={2026},
  publisher={Nature Publishing Group UK London},
  url={https://doi.org/10.1038/s41586-025-10067-1}
}

@article{chen2024strong,
  title={Strong electron--phonon coupling in magic-angle twisted bilayer graphene},
  author={Chen, Cheng and Nuckolls, Kevin P and Ding, Shuhan and Miao, Wangqian and Wong, Dillon and Oh, Myungchul and Lee, Ryan L and He, Shanmei and Peng, Cheng and Pei, Ding and others},
  journal={Nature},
  volume={636},
  number={8042},
  pages={342--347},
  year={2024},
  publisher={Nature Publishing Group UK London},
  url={https://doi.org/10.1038/s41586-024-08227-w}
}

@article{wang2024molecular,
  title={Molecular pairing in twisted bilayer graphene superconductivity},
  author={Wang, Yi-Jie and Zhou, Geng-Dong and Peng, Shi-Yu and Lian, Biao and Song, Zhi-Da},
  journal={Physical review letters},
  volume={133},
  number={14},
  pages={146001},
  year={2024},
  publisher={APS},
  url = {https://link.aps.org/doi/10.1103/PhysRevLett.133.146001}
}

@article{rozen2021entropic,
  title={Entropic evidence for a Pomeranchuk effect in magic-angle graphene},
  author={Rozen, Asaf and Park, Jeong Min and Zondiner, Uri and Cao, Yuan and Rodan-Legrain, Daniel and Taniguchi, Takashi and Watanabe, Kenji and Oreg, Yuval and Stern, Ady and Berg, Erez and others},
  journal={Nature},
  volume={592},
  number={7853},
  pages={214--219},
  year={2021},
  publisher={Nature Publishing Group UK London},
  url={https://doi.org/10.1038/s41586-021-03319-3}
}

@article{saito2021isospin,
  title={Isospin Pomeranchuk effect in twisted bilayer graphene},
  author={Saito, Yu and Yang, Fangyuan and Ge, Jingyuan and Liu, Xiaoxue and Taniguchi, Takashi and Watanabe, Kenji and Li, JIA and Berg, Erez and Young, Andrea F},
  journal={Nature},
  volume={592},
  number={7853},
  pages={220--224},
  year={2021},
  publisher={Nature Publishing Group UK London},
  url={https://doi.org/10.1038/s41586-021-03409-2}
}

@article{nuckolls2023quantum,
  title={Quantum textures of the many-body wavefunctions in magic-angle graphene},
  author={Nuckolls, Kevin P and Lee, Ryan L and Oh, Myungchul and Wong, Dillon and Soejima, Tomohiro and Hong, Jung Pyo and C{\u{a}}lug{\u{a}}ru, Dumitru and Herzog-Arbeitman, Jonah and Bernevig, B Andrei and Watanabe, Kenji and others},
  journal={Nature},
  volume={620},
  number={7974},
  pages={525--532},
  year={2023},
  publisher={Nature Publishing Group UK London},
  url={https://doi.org/10.1038/s41586-023-06226-x}
}

@article{hu2023kondo,
  title={Kondo lattice model of magic-angle twisted-bilayer graphene: Hund’s rule, local-moment fluctuations, and low-energy effective theory},
  author={Hu, Haoyu and Bernevig, B Andrei and Tsvelik, Alexei M},
  journal={Physical review letters},
  volume={131},
  number={2},
  pages={026502},
  year={2023},
  publisher={APS},
  url = {https://link.aps.org/doi/10.1103/PhysRevLett.131.026502}
}

@article{chou2023kondo,
  title={Kondo lattice model in magic-angle twisted bilayer graphene},
  author={Chou, Yang-Zhi and Das Sarma, Sankar},
  journal={Physical Review Letters},
  volume={131},
  number={2},
  pages={026501},
  year={2023},
  publisher={APS},
  url = {https://link.aps.org/doi/10.1103/PhysRevLett.131.026501}
}

@article{kang2023pseudomagnetic,
  title = {Pseudomagnetic fields, particle-hole asymmetry, and microscopic effective continuum Hamiltonians of twisted bilayer graphene},
  author = {Kang, Jian and Vafek, Oskar},
  journal = {Phys. Rev. B},
  volume = {107},
  issue = {7},
  pages = {075408},
  numpages = {34},
  year = {2023},
  month = {Feb},
  publisher = {American Physical Society},
  doi = {10.1103/PhysRevB.107.075408},
  url = {https://link.aps.org/doi/10.1103/PhysRevB.107.075408}
}

@article{kang2025analytical,
  title = {Analytical solution for the relaxed atomic configuration of twisted bilayer graphene including heterostrain},
  author = {Kang, Jian and Vafek, Oskar},
  journal = {Phys. Rev. B},
  volume = {112},
  issue = {12},
  pages = {125138},
  numpages = {20},
  year = {2025},
  month = {Sep},
  publisher = {American Physical Society},
  doi = {10.1103/s3s7-513d},
  url = {https://link.aps.org/doi/10.1103/s3s7-513d}
}

@article{herzog2024topological,
  title={Topological heavy fermion principle for flat (narrow) bands with concentrated quantum geometry},
  author={Herzog-Arbeitman, Jonah and Yu, Jiabin and C{\u{a}}lug{\u{a}}ru, Dumitru and Hu, Haoyu and Regnault, Nicolas and Liu, Chaoxing and Vafek, Oskar and Coleman, Piers and Tsvelik, Alexei and Song, Zhi-da and others},
  journal={arXiv preprint arXiv:2404.07253},
  year={2024},
  url={https://arxiv.org/abs/2404.07253}
}

@article{rai2024dynamical,
  title={Dynamical correlations and order in magic-angle twisted bilayer graphene},
  author={Rai, Gautam and Crippa, Lorenzo and C{\u{a}}lug{\u{a}}ru, Dumitru and Hu, Haoyu and Paoletti, Francesca and De’Medici, Luca and Georges, Antoine and Bernevig, B Andrei and Valent{\'\i}, Roser and Sangiovanni, Giorgio and others},
  journal={Physical Review X},
  volume={14},
  number={3},
  pages={031045},
  year={2024},
  publisher={APS},
  url = {https://link.aps.org/doi/10.1103/PhysRevX.14.031045}

}

@article{youn2024hundness,
  title={Hundness in twisted bilayer graphene: Correlated gaps and pairing},
  author={Youn, Seongyeon and Goh, Beomjoon and Zhou, Geng-Dong and Song, Zhi-Da and Lee, Seung-Sup B},
  journal={arXiv preprint arXiv:2412.03108},
  year={2024},
  url={https://arxiv.org/abs/2412.03108}
}

@article{xiao2026imaging,
  title={Imaging the flat bands of magic-angle graphene reshaped by interactions},
  author={Xiao, J and Inbar, A and Birkbeck, J and Gershon, N and Zamir, Y and Vituri, Y and Taniguchi, T and Watanabe, K and Berg, E and Ilani, S},
  journal={Nature},
  volume={653},
  number={8113},
  pages={68--75},
  year={2026},
  publisher={Nature Publishing Group UK London},
  url={https://doi.org/10.1038/s41586-026-10378-x}
}

@article{zhao2026pseudogap,
  title={Pseudogap and Non-Fermi-liquid criticality in double Kondo model for bilayer nickelates},
  author={Zhao, Jing-Yu and Zhang, Ya-Hui},
  journal={arXiv preprint arXiv:2603.25742},
  year={2026},
  url={https://arxiv.org/abs/2603.25742}
}

@article{zhang2020deconfined,
  title={Deconfined criticality and ghost Fermi surfaces at the onset of antiferromagnetism in a metal},
  author={Zhang, Ya-Hui and Sachdev, Subir},
  journal={Physical Review B},
  volume={102},
  number={15},
  pages={155124},
  year={2020},
  publisher={APS},
  url = {https://link.aps.org/doi/10.1103/PhysRevB.102.155124}
}

@article{song2019all,
  title={All magic angles in twisted bilayer graphene are topological},
  author={Song, Zhida and Wang, Zhijun and Shi, Wujun and Li, Gang and Fang, Chen and Bernevig, B Andrei},
  journal={Physical review letters},
  volume={123},
  number={3},
  pages={036401},
  year={2019},
  publisher={APS},
  url = {https://link.aps.org/doi/10.1103/PhysRevLett.123.036401}
}

@article{kang2018symmetry,
  title={Symmetry, maximally localized Wannier states, and a low-energy model for twisted bilayer graphene narrow bands},
  author={Kang, Jian and Vafek, Oskar},
  journal={Physical Review X},
  volume={8},
  number={3},
  pages={031088},
  year={2018},
  publisher={APS},
  url = {https://link.aps.org/doi/10.1103/PhysRevX.8.031088}
}

@article{bultinck2020ground,
  title={Ground state and hidden symmetry of magic-angle graphene at even integer filling},
  author={Bultinck, Nick and Khalaf, Eslam and Liu, Shang and Chatterjee, Shubhayu and Vishwanath, Ashvin and Zaletel, Michael P},
  journal={Physical Review X},
  volume={10},
  number={3},
  pages={031034},
  year={2020},
  publisher={APS}
}

@article{parker2021strain,
  title={Strain-induced quantum phase transitions in magic-angle graphene},
  author={Parker, Daniel E and Soejima, Tomohiro and Hauschild, Johannes and Zaletel, Michael P and Bultinck, Nick},
  journal={Physical review letters},
  volume={127},
  number={2},
  pages={027601},
  year={2021},
  publisher={APS},
   url = {https://link.aps.org/doi/10.1103/PhysRevX.10.031034}
}

@article{kwan2021kekule,
  title={Kekul{\'e} spiral order at all nonzero integer fillings in twisted bilayer graphene},
  author={Kwan, Yves H and Wagner, Glenn and Soejima, Tomohiro and Zaletel, Michael P and Simon, Steven H and Parameswaran, Siddharth A and Bultinck, Nick},
  journal={Physical Review X},
  volume={11},
  number={4},
  pages={041063},
  year={2021},
  publisher={APS},
   url = {https://link.aps.org/doi/10.1103/PhysRevX.11.041063}
}

@article{wagner2022global,
  title={Global phase diagram of the normal state of twisted bilayer graphene},
  author={Wagner, Glenn and Kwan, Yves H and Bultinck, Nick and Simon, Steven H and Parameswaran, SA},
  journal={Physical review letters},
  volume={128},
  number={15},
  pages={156401},
  year={2022},
  publisher={APS},
  url = {https://link.aps.org/doi/10.1103/PhysRevLett.128.156401}
}

@article{ghosh2025thermopower,
  title={Thermopower probes of emergent local moments in magic-angle twisted bilayer graphene},
  author={Ghosh, Ayan and Chakraborty, Souvik and Dutta, Ranit and Agarwala, Adhip and Watanabe, K and Taniguchi, T and Banerjee, Sumilan and Trivedi, Nandini and Mukerjee, Subroto and Das, Anindya},
  journal={Nature Physics},
  volume={21},
  number={5},
  pages={732--739},
  year={2025},
  publisher={Nature Publishing Group UK London},
  url={https://doi.org/10.1038/s41567-025-02849-1}
}

@article{cualuguaru2021twisted,
  title={Twisted symmetric trilayer graphene: Single-particle and many-body Hamiltonians and hidden nonlocal symmetries of trilayer moir{\'e} systems with and without displacement field},
  author={C{\u{a}}lug{\u{a}}ru, Dumitru and Xie, Fang and Song, Zhi-Da and Lian, Biao and Regnault, Nicolas and Bernevig, B Andrei},
  journal={Physical Review B},
  volume={103},
  number={19},
  pages={195411},
  year={2021},
  publisher={APS},
  url = {https://link.aps.org/doi/10.1103/PhysRevB.103.195411}
}

@article{lei2021mirror,
  title={Mirror symmetry breaking and lateral stacking shifts in twisted trilayer graphene},
  author={Lei, Chao and Linhart, Lukas and Qin, Wei and Libisch, Florian and MacDonald, Allan H},
  journal={Physical Review B},
  volume={104},
  number={3},
  pages={035139},
  year={2021},
  publisher={APS},
  url = {https://link.aps.org/doi/10.1103/PhysRevB.104.115167}
}

@article{wu2021lattice,
  title={Lattice relaxation, mirror symmetry and magnetic field effects on ultraflat bands in twisted trilayer graphene},
  author={Wu, Zewen and Zhan, Zhen and Yuan, Shengjun},
  journal={Science China Physics, Mechanics \& Astronomy},
  volume={64},
  number={6},
  pages={267811},
  year={2021},
  publisher={Springer},
  url={https://doi.org/10.1007/s11433-020-1690-4}
}

@article{xie2021twisted,
  title={Twisted symmetric trilayer graphene. II. Projected Hartree-Fock study},
  author={Xie, Fang and Regnault, Nicolas and C{\u{a}}lug{\u{a}}ru, Dumitru and Bernevig, B Andrei and Lian, Biao},
  journal={Physical Review B},
  volume={104},
  number={11},
  pages={115167},
  year={2021},
  publisher={APS},
  url = {https://link.aps.org/doi/10.1103/PhysRevB.104.115167}
}

\appendix
\onecolumngrid

\section{Review: BM model of TTG}
\label{appendix:section to the model of TTG}
The typical twisted trilayer graphene(TTG) system is constructed from a AAA-stacking graphen, with the top layer($l=1$) and the bottom layer($l=3$) twisted by $\theta/2$ and the middle layer($l=2$) twisted by $-\theta/2$, where $\theta$ is the twist angle and $l=1,2,3$ is the layer index. The (extended) BM Hamiltonian of TTG with perpendicular electric field is 
\begin{equation}
H_{TTG}=\sum_{\eta s}\int d^2 r\left(\phi_{\bm{r}, 1,\eta s}^{\dagger} \quad \phi_{\bm{r}, 2,\eta s}^{\dagger} \quad \phi_{\bm{r}, 3,\eta s}^{\dagger}\right)\left(\begin{array}{ccc}
-\mathrm{i} v_0 \bm{\sigma}_{\theta/2} \cdot \bm{\nabla}-\frac{D}{2} & T(\bm{r}) & \\
T^{\dagger}(\bm{r}) & -\mathrm{i} v_0 \bm{\sigma}_{-\theta/2} \cdot \bm{\nabla} & T^{\dagger}(\bm{r}) \\
& T(\bm{r}) & -\mathrm{i} v_0 \bm{\sigma}_{\theta/2} \cdot \bm{\nabla}+\frac{D}{2}
\end{array}\right) \left(\begin{array}{c}
\phi_{\bm{r}, 1,\eta s} \\
\phi_{\bm{r}, 2,\eta s} \\
\phi_{\bm{r}, 3,\eta s}
\end{array}\right),
\label{full BM}
\end{equation}
where $\phi^{\dagger}_{\bm{r},l,\eta s}=(\phi^{\dagger}_{\bm{r},l,\eta s,A} \quad \phi^{\dagger}_{\bm{r},l,\eta s,B})$, $\eta=\pm$ is the valley index, $s=\pm$ is the spin index and  $A,B$ denote the two sublattices. $v_0$ is the Fermi velocity, $\bm{\sigma}_{\theta / 2}=e^{-(i \theta / 4) \sigma_z}\left(\sigma_x, \sigma_y\right) e^{(i \theta / 4) \sigma_z}$ and $D$ is the energy difference generated by the electric field.  $T(\bm{r})=\sum_{j=1,2,3} T_j e^{i \bm{q}_j \cdot \bm{r} }$ is the hopping matrix between the neighboring layers, where
$\bm{q}_1=k_{\theta}(0,-1)$, and $\bm{q}_{2,3}=k_{\theta}(\pm \frac{\sqrt{3}}{2},\frac{1}{2})$. Here $k_{{\theta/2}}=\frac{8\pi}{3a}\sin({\frac{\theta}{2}})$ is the Dirac momentum, and 

\begin{equation}
\begin{aligned}
T_j= & w_0 \sigma_0+w_1\left[\cos \left(\frac{2 \pi}{3}(j-1)\right) \sigma_x\right. \\
& \left.+\sin \left(\frac{2 \pi}{3}(j-1)\right) \sigma_y\right] .
\end{aligned}
\end{equation}

where $w_{0}=110\,\mathrm{meV}$ and $w_1=0.8w_0$ are the AA and AB tunneling.

When $D=0$, based on the mirror symmetry, equation (\ref{full BM}) can be decomposed into a twisted bilayer grahene and a monolayer graphene. To be specific, we can define the symmetric part
\begin{equation}
\begin{aligned} 
& \tilde{\phi}_{ \bm{r}, t,\eta s}^{\dagger}=\frac{1}{\sqrt{2}}\left(\phi_{\bm{r}, 3, \eta s}^{\dagger}+\phi_{\bm{r}, 1, \eta s}^{\dagger}\right) \\ & \tilde{\phi}_{\bm{r}, b,\eta s}^{\dagger}=\phi_{\bm{r}, 2, \eta s}^{\dagger}
\end{aligned}
\end{equation}
and the antisymmetric part
\begin{equation}
d_{r,\eta s}^{\dagger}=\frac{1}{\sqrt{2}}\left(\phi_{ \bm{r}, 3,\eta s}^{\dagger}-\phi_{\bm{r}, 1,\eta s}^{\dagger}\right)
\end{equation}
$H_{BM}$ can be divided into three parts
\begin{equation}
H_{TTG}=H_{\mathrm{TBG}}+H_{Dirac}+H_{D},
\end{equation}

where
\begin{equation}
H_{TBG}=\sum_{\eta s}\int d^2 r \tilde{\phi}_{\bm{r},\eta s}^{\dagger}\left(\begin{array}{cc}
-\mathrm{i} \bm{\sigma}_{\theta/2} \cdot \nabla & \sqrt{2} T(\bm{r}) \\
\sqrt{2} T^{\dagger}(\bm{r}) & -\mathrm{i} \bm{\sigma}_{-\theta/2} \cdot \nabla
\end{array}\right) \tilde{\phi}_{\bm{r},{\eta s}}
\end{equation}
is equivalent to BM model of TBG with $w_0=\sqrt{2}w_0$ and $w_1=\sqrt{2}w_1$.
\begin{equation}
H_{Dirac}=\sum_{\eta s}\int d^2 r d_{\bm{r},\eta s}^{\dagger}(-\mathrm{i} v_0) \bm{\sigma}_{\theta/2} \cdot \nabla d_{\bm{r},\eta s}
\end{equation}

is the Hamiltonian of a monolayer graphene. The displacement field term is 
\begin{equation}
H_{D}=\sum_{\eta s}\int d^2 r \frac{D}{2} \tilde{\phi}_{\bm{r}, t,\eta s}^{\dagger} d_{\bm{r},\eta s}+\text { H.c. }
\end{equation}

When the displacement field $D=0$, due to the equivalence between TTG and TBG + Dirac cone, there is also a magic angle when the flat band is perfectly flat. The angle is $\sqrt{2}$ times the magic angle in TBG, which is 

\begin{equation}
\theta_{TTG} \approx 1.5 ^\circ
\end{equation}

\begin{figure}
    \centering
    \includegraphics[width=0.6\columnwidth]{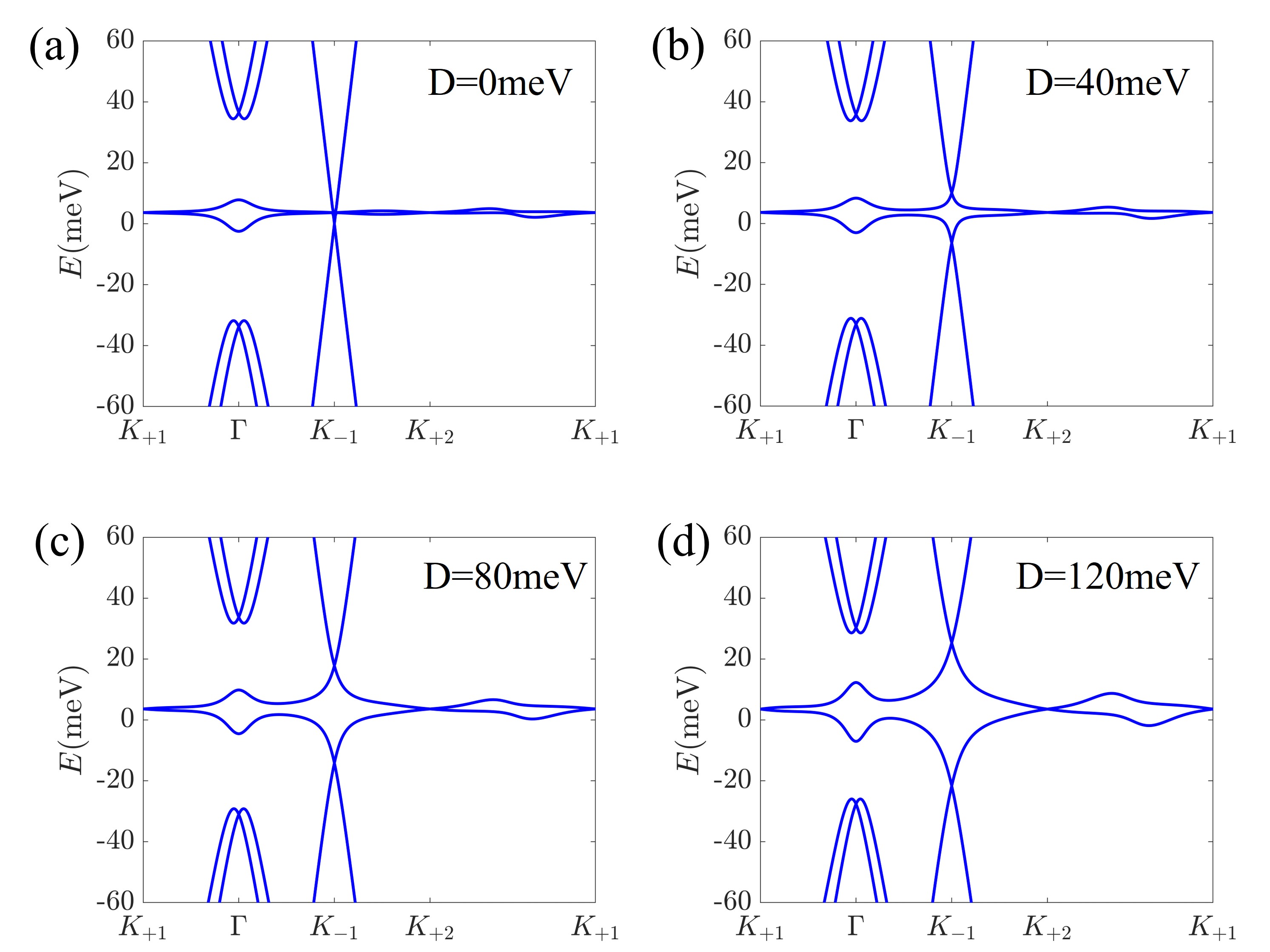}
    
    \caption{ The bare band result of TTG calculated from the BM model at different displacement field $D$, $\theta=1.48^\circ$.
    }
    
    \label{figure_bare_TTG}	
\end{figure}

We choose $\theta=1.48^\circ$ in this paper. In Fig~\ref{figure_bare_TTG}, we show the bare band results of TTG at different fillings. We see that the Dirac cone splits into two and these two Dirac cone keep shifting away from each other with increasing $D$.

\section{Details of the $f$-$c$-$d$ model}
\label{appendix:details of fcd}
We employ the $f$-$c$-$d$ model framework \cite{yu2023magic}, which describes TTG in a manner analogous to the topological heavy fermion model (THFM) for twisted bilayer graphene (TBG) \cite{song2022magic}. The full Hamiltonian is given by
\begin{equation}
H_0 = H_0^{d} + H_0^{c} + H_0^{fd} + H_0^{fc} - \mu(N_d + N_f + N_c) +wN_d+ H_{\text{int}}^{(f)}.
\label{original_Hamiltonian_app}
\end{equation}
The Hamiltonian consists of two dispersive conduction bands ($c_1$ and $c_2$) governed by $H_0^{c}$, a Dirac band ($d$) governed by $H_0^{d}$, and localized $f$ electrons. The $f$ electrons hybridize with the $c$ and $d$ bands as follows:
\begin{equation}
H_{0}^{fd} = \sum_{\bm{k} \in \text{MBZ}}\sum_{\bm{p}}^{|\bm{p}| < \Lambda_d} f_{\bm{k}}^{\dagger} M_1 D\tilde{h}(\bm{k},{\bm{p}})d_{\bm{p}} + \text{h.c.},
\end{equation}
\begin{equation}
H_{0}^{fc} = \sum_{\bm{k}, \bm{G}} f_{\bm{k}}^{\dagger} \gamma(\bm{k}+\bm{G})c_{\bm{k}+\bm{G}} + \text{h.c.},
\end{equation}
where $f_{\bm{k}} = \{f_{\bm{k};\alpha}\}$, $c_{\bm{k}} = \{c_{\bm{k};\alpha}\}$, and $d_{\bm{k}} = \{d_{\bm{k};\alpha}\}$. Here, $\alpha = (a, \eta, s)$ is a composite index encompassing the orbital $a$, valley $\eta \in \{K,K'\}$, and spin $s \in \{\uparrow, \downarrow\}$. The $f$, $c$, and $d$ fermions possess 2, 4, and 2 orbital degrees of freedom, respectively. An important point here is that the orbital indices $a=\pm$ here represents the state with angular momentum $L=\pm1$ for $c_1$ and $f$ electron, which is different from the sublattice basis usually used in \cite{song2022magic,yu2023magic}. 
Therefore, we use a modified form of the THFM. In the sublattice basis, $H_0^c$, $H_0^d$ and the hybridization matrices  
 $\tilde{h}(\bm{k},{\bm{p}})$ and $\gamma(\bm{k})$ are defined as:
 \begin{equation}
H_0^c=v_{\star} \sum_{\mathbf{k}}\left(c_{1, \mathbf{k}}^{\dagger} \left( k_x \eta_z\sigma_0+\mathrm{i} k_y \eta_0 \sigma_z\right)s_0 c_{2, \mathbf{k}}+\text { h.c. }\right)+\sum_{\mathbf{k}} c_{2, \mathbf{k}}^{\dagger} M \eta_0\sigma_x s_0 c_{2, \mathbf{k}},
\end{equation}

\begin{equation}
H_{0}^d=\sum_p^{|\boldsymbol{p}| \leqslant \Lambda_d} d_{\eta, \boldsymbol{p}}^{\dagger}v_0\left(p_x \eta_z \sigma_x+p_y \eta_0\sigma_y\right)s_0 d_{\eta, \boldsymbol{p}},
\end{equation}

\begin{equation}
\tilde{h}(\bm{k},{\bm{p}}) = \sum_{\bm{G}} e^{-\frac{|\bm{p}|^2 \lambda^2}{2}} (\eta_0\sigma_0 + \mathrm{i} \eta_z \sigma_z)s_0 \delta_{\eta_z \bm{K}_M + \bm{p}, \bm{k}+\bm{G}},
\end{equation}
\begin{equation}
\gamma(\bm{k}) = e^{-\frac{|\bm{k}|^2 \lambda^2}{2}} 
\left( \gamma \eta_0 \sigma_0 + v_{\star}^{\prime} (k_x\eta_z \sigma_x + k_y \eta_0 \sigma_y), \right. \\
\left. \vphantom{\gamma \eta_0 \sigma_0} v_{\star}^{\prime \prime} (k_x \eta_z\sigma_x - k_y \eta_0 \sigma_y) \right) s_0,
\end{equation}

where $D$ represents the displacement field, and $\sigma_i$, $\eta_i$, and $s_i$ are the Pauli matrices acting on the orbital, valley, and spin spaces, respectively. 
The angular momentum of the $c_1$ and $f$ electron in orbitals $a=+,-$ in the valley $\eta$ are $\eta,-\eta$, respectively. While in the orbital basis we used in our paper, the angular momentum of the $c_1$ and $f$ electron in orbitals $a=+,-$ in the valley $\eta$ are $1,-1$, respectively. To recover this convention, we interchange the two orbitals in the $\eta=-$ valley in the sublattice basis, which can be achieved by adding a $\sigma_x$ transformation to the original Hamiltonian in the $\eta=-$ valley, which gives the final form of the Hamiltonian:

$\tilde{h}(\bm{k},{\bm{p}})$ and $\gamma(\bm{k})$ are defined as:
 \begin{equation}
H_0^c=v_{\star} \sum_{\mathbf{k}}\left(c_{1, \mathbf{k}}^{\dagger} \left( k_x \eta_z\sigma_0+\mathrm{i} k_y \eta_z \sigma_z\right)s_0 c_{2, \mathbf{k}}+\text { h.c. }\right)+\sum_{\mathbf{k}} c_{2, \mathbf{k}}^{\dagger} M \eta_0\sigma_x s_0 c_{2, \mathbf{k}},
\end{equation}

\begin{equation}
H_{0}^d=\sum_p^{|\boldsymbol{p}| \leqslant \Lambda_d} d_{\eta, \boldsymbol{p}}^{\dagger}v_0\left(p_x \eta_z \sigma_x+p_y \eta_z\sigma_y\right)s_0 d_{\eta, \boldsymbol{p}},
\end{equation}

\begin{equation}
\tilde{h}(\bm{k},{\bm{p}}) = \sum_{\bm{G}} e^{-\frac{|\bm{p}|^2 \lambda^2}{2}} M_1(\eta_0\sigma_0 + \mathrm{i} \eta_0 \sigma_z)s_0 \delta_{\eta_z \bm{K}_M + \bm{p}, \bm{k}+\bm{G}},
\end{equation}
\begin{equation}
\gamma(\bm{k}) = e^{-\frac{|\bm{k}|^2 \lambda^2}{2}} 
\left( \gamma \eta_0 \sigma_0 + v_{\star}^{\prime} (k_x\eta_z \sigma_x + k_y \eta_z \sigma_y), \right. \\
\left. \vphantom{\gamma \eta_0 \sigma_0} v_{\star}^{\prime \prime} (k_x \eta_z\sigma_x - k_y \eta_z \sigma_y) \right) s_0,
\end{equation}

The parameter choices are as follows. We choose $\theta=1.48^\circ$ and set $k_\theta=\frac{8\pi}{3a}\sin{\frac{\theta}{2}}=1$, and an energy scale $E_0=269.967 \,\mathrm{meV}$. Under this basis, $v_{\star}=-0.7176$, $\gamma=-0.1265$, $v_{\star}'=0.2711$, $v_{\star}^{\prime \prime}=0.005768$, $M=-0.0194$, $v_0=1.9703$ and $M_1=-0.1397$. We can see that the parameter $v_{\star}^{\prime \prime}$ is negligibly small compared to $\gamma$ and $v_{\star}^{\prime}$. 

Furthermore, we introduce the on-site interaction for the $f$ electrons:
\begin{equation}
H_{\text{int}}^{(f)} = \frac{U}{2} \sum_i \left(n_{i ; f} - 4 - \kappa \nu\right)^2 + \sum_i h_{i ; J}^{(f)}.
\end{equation}
This term includes the spin interaction $h_{i ; J}^{(f)}$, which originates from both electron-electron and electron-phonon couplings\cite{wang2025electron}. As discussed in \cite{zhao2025resonating}, $h_{i ; J}^{(f)}$ favors inter-valley spin-singlet pairing. To account for the residual repulsive interactions between the $f$ electrons and other itinerant electrons at the Hartree level, we introduce a phenomenological parameter $\kappa$. This modification yields better agreement with experimental observations\cite{wong2020cascade,lau2025topological}. Throughout this work, we set $\kappa=0.8$.
\section{ancilla fermion theory of TTG}
\label{appendix:section_for_ancilla}

Ancilla fermion theory was first developed to understand the pseudogap phase in early works\cite{zhang2020pseudogap,mascot2022electronic}. Later it is proved to describe the Mott insulator state of spin-$1/2$ Hubbard model at half filling both analytically and numerically\cite{zhou2025variational}. Variational Monte Carlo study also demonstrated that the ancilla wave function can effectively capture the polaronic correlations in doped Mott insulators\cite{shackleton2024emergent, muller2025polaronic}. Also in \cite{zhao2025ancilla}, the ancilla fermion theory is established in TBG system, to capture the topological Mott localization, which shows good agreement to the slave particle model results at $\nu=-2$\cite{zhao2025mixed,zhao2025resonating}. Readers may refer to the above papers, especially \cite{zhao2025ancilla}, for detailed discussion of the ancilla fermion theory. 

In MATTG system, because the interaction is only dominant in the flat band, to capture the Mott physics, we introduce a couple of ancilla fermions $\psi$ and ${\psi^{\prime}}$  and let it couple with the TBG-sector of the TTG. To be clear, we replace the TBG part annihilation operator $\tilde{\phi}$ by $c$ in the following. $c$ and  $\psi$ couple to open the charge gap, and the spin sector $\psi^{\prime}$ represents the local moment. In this spirit, we fix the number of $\psi$ to be $4-\nu_c$ and the number of $\psi^{\prime}$ to be $4+\nu_c$, where $\nu_c \in[-4,4]$ is the filling factor of TBG part. The total number 8 comes from the two spin indexes, two valleys and two orbitals. Due to the large dispersion of the Dirac cone compared to our energy scale, the change in the fillings of the monolayer part of TTG is small, so in this section we omit the change of fillings in the monolayer part and set $\nu_c=\nu$, where $\nu$ is the total filling factor of electrons. The general mean field Hamiltonian allowed by symmetry is as follows.

\begin{equation}
H^{(c \psi \psi')}=H_{TTG}+H_{\mathrm{hyb}}^{(c \psi)}+H_{\mathrm{hyb}}^{(c\psi')}-\mu\left(N_e+N_\psi+N_{\psi'}\right)+\Delta_1\left(N_\psi+N_{\psi'}-N_c\right) / 2+\Delta_2(N_{\psi'}-N_\psi)/2
\label{full ancilla function}
\end{equation}

where $H_{\mathrm{hyb}}^{(c \psi)}$ and $H_{\mathrm{hyb}}^{(c \psi')}$ are defined as follows:

\begin{equation}
H_{\mathrm{hyb}}^{(c \psi)}=\Phi_0 \sum_{l a \sigma s} \sum_{\bm{k} \in \mathrm{MBZ}, \bm{G}_M} c_{0, \bm{k}+\bm{G}_M ; l a \sigma s}^{\dagger} e^{i \frac{\pi}{4} l a} \psi_{\bm{k} ; a \sigma s}+\text { h.c. }
\end{equation}

\begin{equation}
H_{\mathrm{hyb}}^{(c \psi')}=\Phi_1 \sum_{l a \eta s} \sum_{\bm{k} \in \mathrm{MBZ}, \bm{G}_M} c_{0, \bm{k}+\bm{G}_M ; l a \eta s}^{\dagger} e^{i \frac{\pi}{4} l a} \psi'_{\bm{k} ; a \eta s}+\text { h.c. }
\end{equation}

Here $l=1,-1$ corresponds to the two layers in the TBG model, $a=1,-1$ represents two orbitals in the TBG model, $\eta$ and $s$ are the valley and spin index. $H_{\mathrm{hybd}}^{(\psi \psi')}$ is 

\begin{equation}
H_{\mathrm{hyb}}^{(\psi \psi')}=\Phi_2 \sum_{a \eta s} \psi^{\dagger} \psi'+\text{h.c.}
\end{equation}

$N_e$, $N_\psi$ and $N_{\psi'}$ are the particle numbers of physical electron, ancilla fermion $\psi$ and $\psi'$. The chemical potential parameters $\mu$, $\Delta_1$ and $\Delta_2$ are introduced to fix the filling factor of physical electron, $\psi$,$\psi'$ to be $\nu$, $-\nu$, $\nu$.

As discussed earlier in \cite{zhang2020deconfined}, ancilla fermion theory provides a description of the phase transitions in Kondo lattice. When the charge sector denoted by $c$ and $\psi$ is decoupled from the spin sector denoted by $\psi'$($\Phi_1=0,\Phi_2=0$), local moments are isolated, forming spin orders dominated by RKKY interaction and other ferromagnetic interactions. When $\Phi_1$ or $\Phi_2$ $\neq 0$, electrons are coupled to local moments, resulting in the formation of the heavy fermion liquid. Next, we would show and discuss the band structures of these two cases.

\subsection{$\Phi_1=0$, $\Phi_2=0$}
\begin{figure*}[t]
    \centering
    \includegraphics[width=\textwidth]{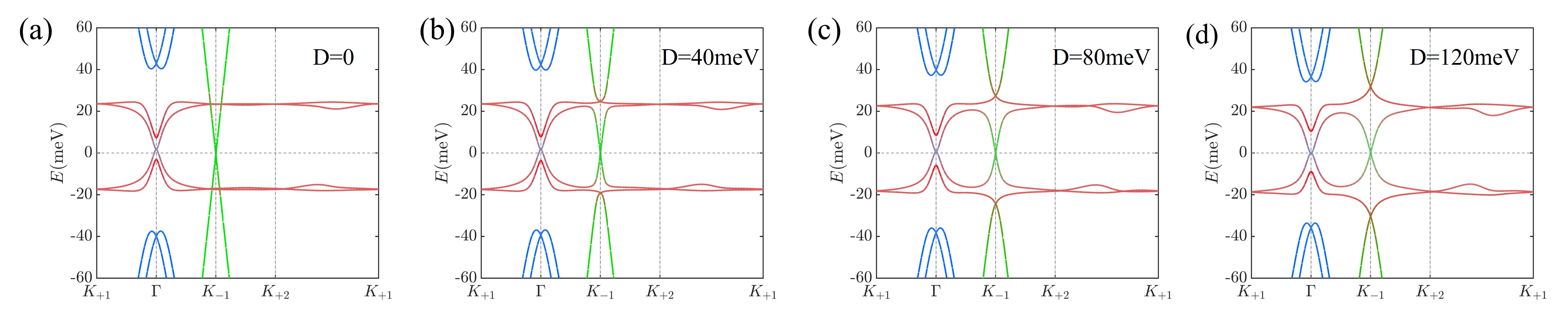}
    
    \caption{ Mott band calculated at $\nu=0$, when the charge sector is decoupled from the local moment, with only ancilla fermion $\psi$. The red, green, blue and gray colors of the dispersion line represent the weight of flat bands, Dirac cone, remote bands and ancilla fermion $\psi$.The gap at $K_+$ point is $40$meV. }
    
    \label{figure_for_ancilla_fermion_niu1}	
\end{figure*}

\begin{figure*}[t]
    \centering
    \includegraphics[width=\textwidth]{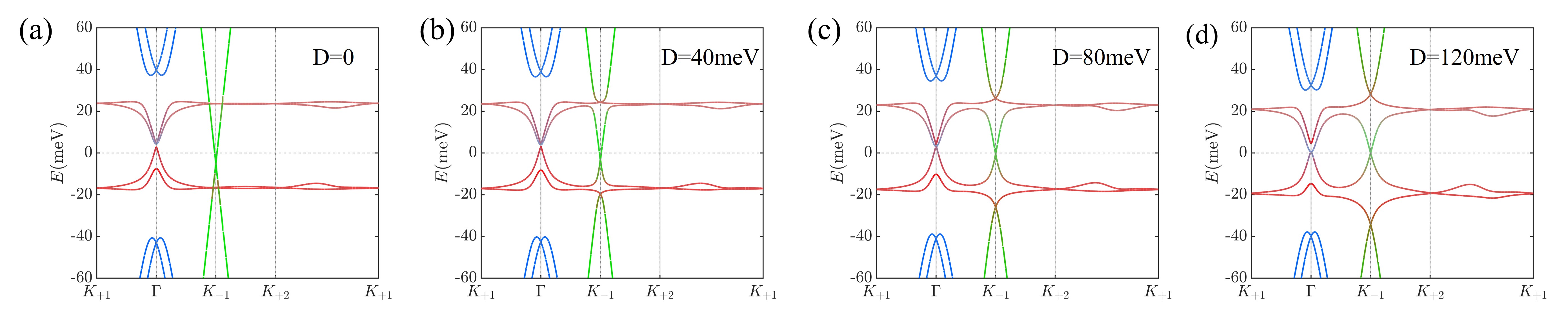}
    
    \caption{Mott band calculated at $\nu=1$, when the charge sector is decoupled from the local moment, with only ancilla fermion $\psi$. The red, green, blue and gray colors of the dispersion line represent the weight of flat bands, Dirac cone, remote bands and ancilla fermion $\psi$.The gap at $K_+$ point is $40$meV.}
    
    \label{figure_for_ancilla_fermion_niu2}	
\end{figure*}

\begin{figure*}[t]
    \centering
    \includegraphics[width=\textwidth]{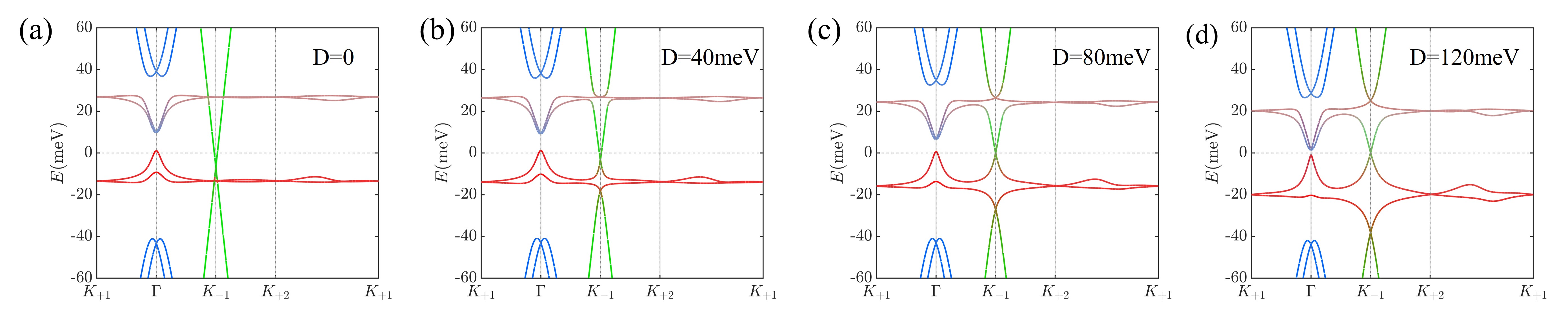}
    
    \caption{Mott band calculated at $\nu=2$, when the charge sector is decoupled from the local moment, with only ancilla fermion $\psi$. The red, green, blue and gray colors of the dispersion line represent the weight of flat bands, Dirac cone, remote bands and ancilla fermion $\psi$. The gap at $K_+$ point is $40$meV. }
    
    \label{figure_for_ancilla_fermion_niu3}	
\end{figure*}

\begin{figure*}[t]
    \centering
    \includegraphics[width=\textwidth]{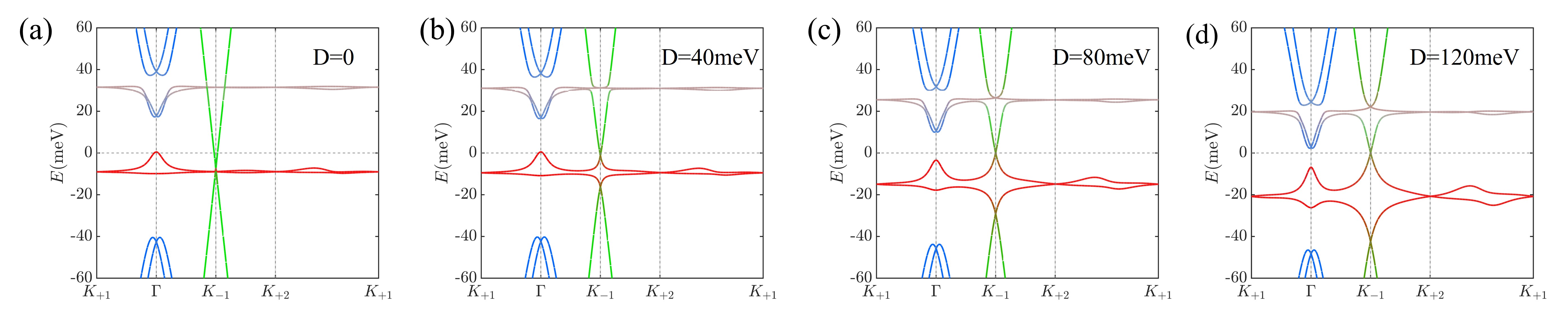}
    
    \caption{Mott band calculated at $\nu=3$, when the charge sector is decoupled from the local moment, with only ancilla fermion $\psi$. The red, green, blue and gray colors of the dispersion line represent the weight of flat bands, Dirac cone, remote bands and ancilla fermion $\psi$. The gap at $K_+$ point is $40$meV. }
    
    \label{figure_for_ancilla_fermion_niu4}	
\end{figure*}

In this section, we would discuss the the band result when the charge sector and the spin sector are decoupled($\Phi_1=0$, $\Phi_2=0$). According to the experimental result, the Hubbard repulsion parameter $U$ in TBG is about $20 \sim 30$meV. Given that the energy scale is $\sqrt{2}$ times larger in TTG system, the Hubbard repulsion parameter $U$ in TTG is about $30 \sim 45$ meV. Because the Hubbard repulsion term is only the most dominant at $K_+$ point, so in this section, we fix the Hubbard gap at $K_+$ point to be $40$ meV. 

In Fig.~\ref{figure_for_ancilla_fermion_niu1}, \ref{figure_for_ancilla_fermion_niu2}, \ref{figure_for_ancilla_fermion_niu3}, \ref{figure_for_ancilla_fermion_niu4}, we show the band result for filling factor $\nu=1,2,3$ and its relationship with $D$. With increasing $\nu$, the bandgap at $\Gamma$ point becomes larger. At all fillings($\nu=1,2,3$), we can see that when $D$ is small, the Dirac cone is slightly electron doped and the lower Hubbard band around $\Gamma$ point is slightly hole doped. When $\nu=1$, as illustrated in Fig.~\ref{figure_for_ancilla_fermion_niu1}, the bandgap at $\Gamma$ point is small. As the bare band result we disucssed before, with $D$ increasing, the Dirac cone splits into two and they keep moving away from each other. At medium $D$, the upper Dirac cone lifts up and lies within the bandgap. When $D$ further increases, the Dirac cone gets higher than the upper Hubbard band at the $\Gamma$ point, which leads to the Dirac cone getting hole doped and the upper Hubbard band gets slightly electron doped. 

When $\nu=2$, as demonstrated in Fig.~\ref{figure_for_ancilla_fermion_niu2}, due to the larger gap at $\Gamma$, the systems remains semimetal (the upper Dirac cone lies in the bandgap at $\Gamma$) for a large range of $D$. At fairly large $D$, the upper Dirac cone would transcend the upper Hubbard band and gets hole doped.  

When $\nu=3$, as shown in Fig.~\ref{figure_for_ancilla_fermion_niu3}, the gap becomes even larger. Like the case of $\nu=2$, the system remains in the semimetal state after the Dirac cone exceeds the lower Hubbard band.

\subsection{$\Phi_1$ or $\Phi_2 \neq 0$ }
\begin{figure}[t]
    \centering
    \includegraphics[width=\textwidth]{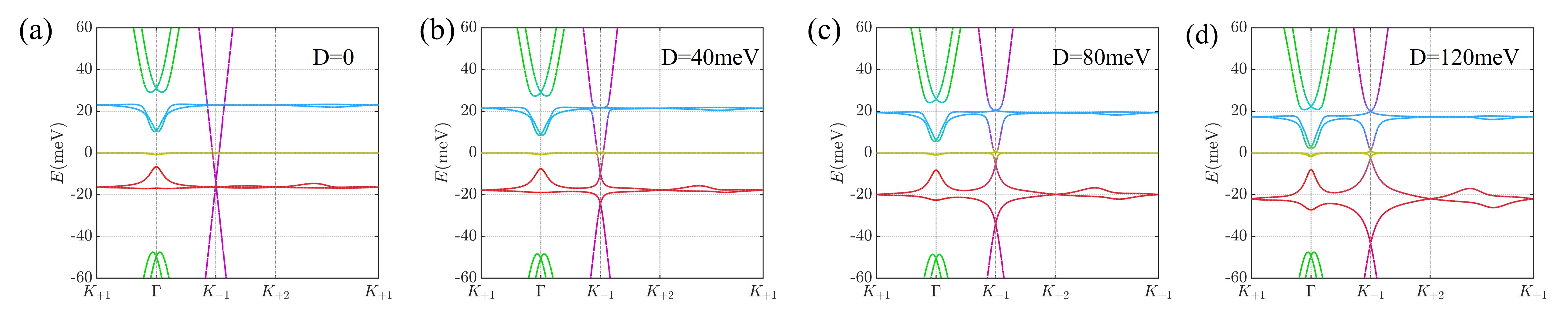}
    
    \caption{ Mott band calculated at $\nu=3$, when the charge sector is coupled to the local moment, with ancilla fermion $\psi$ and $\psi'$.  The red, green, blue and gray colors of the dispersion line represent the weight of flat bands, Dirac cone, remote bands and ancilla fermion $\psi$. $\Phi_0=4meV$, $\Phi_1=2meV$, $\Phi_2=0meV$. }
    
    \label{figure_for_new_ancilla1}	
\end{figure}

\begin{figure*}[t]
    \centering
    \includegraphics[width=\textwidth]{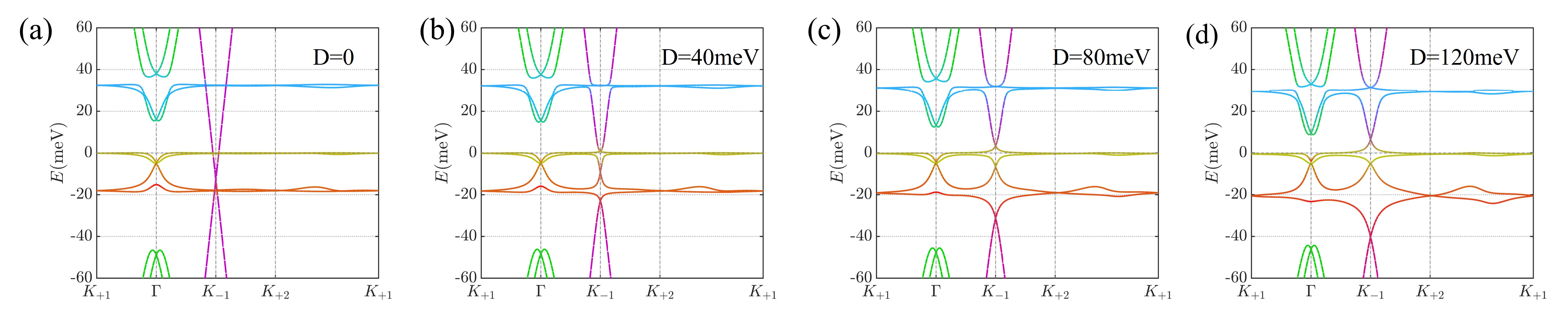}
    
    \caption{ Mott band calculated at $\nu=3$, when the charge sector is coupled to the local moment, with ancilla fermion $\psi$ and $\psi'$. The red, purple, green, blue and yellow colors of the dispersion line represent the weight of flat bands, Dirac cone, remote bands, ancilla fermion $\psi$ and ancilla fermion $\psi'$. $\Phi_0=4meV$, $\Phi_1=6meV$, $\Phi_2=4meV$.}
    
    \label{figure_for_new_ancilla2}	
\end{figure*}

In this section, we consider the coupling between the charge sector and the spin sector. That is, set $\Phi_1$ or $\Phi_2 \neq 0$. In Fig.~\ref{figure_for_new_ancilla1} and \ref{figure_for_new_ancilla2}, we show the Mott band for two different parameter choices, corresponding to two different cases: the coupling between the local moment and the charge sector is weak, the coupling is strong.
We see that the local moment $\psi'$ bands are pinned at zero energy, which is consistent with the results of the model in the main text.

\section{Details of the slave particle model}
\label{section of slave particle model}

\subsection{Spin interaction and local-moment pairing}
\label{sec:spin_interaction}

In this subsection we discuss the on-site spin interaction used in our paper, the main discussion follows the Appendix of \cite{zhao2025resonating}. Since the localized $f$ orbitals are centered on the AA sites and carry orbital, valley, and spin indices, the on-site Hilbert space contains a large multiplet structure. The role of the spin interaction is to lift this multiplet degeneracy and select a spin-singlet local-moment state in the doubly occupied sector.

The local interaction of the $f$ orbitals is written as
\begin{equation}
    H_{\rm int}^{(f)}
    =
    \frac{U}{2}
    \sum_i
    \left(n_{i;f}-4-\kappa\nu\right)^2
    +
    \sum_i h_{i;J}^{(f)} .
    \label{eq:Hint_f_spin}
\end{equation}
The first term controls the charge valence of the local $f$ orbital. The
second term acts within the local spin-orbital Hilbert space and determines
the structure of the local moment. Following the notation
$\alpha=(a,\eta,s)$, where $a=\pm$ is the orbital index, $\eta=K,K'$ is the
valley index, and $s=\uparrow,\downarrow$ is the spin index, we decompose
\begin{equation}
    h_{i;J}^{(f)}
    =
    h_{i;J_A}^{(f)}
    +
    h_{i;J_H}^{(f)} .
    \label{eq:hJ_decomposition}
\end{equation}
Here $h_{i;J_A}^{(f)}$ denotes an anti-Hund interaction, which can be
generated by electron-phonon coupling, while $h_{i;J_H}^{(f)}$ denotes a
Hund-type interaction originating from Coulomb interactions. Explicitly, we
take
\begin{align}
h_{i;J_A}^{(f)}
=&
-\frac{J_A}{2}
\sum_{a,\eta,b,s,s'}
f_{i;b\eta\bar s}^{\dagger}
f_{i;a\eta s'}^{\dagger}
f_{i;b\eta\bar s'}f_{i;a\eta s}
\nonumber\\
&-
\frac{J_A}{2}
\sum_{a,\eta,s,s'}
f_{i;a\eta\bar s}^{\dagger}
f_{i;\bar a\eta s'}^{\dagger}
f_{i;\bar a\eta\bar s'}f_{i;a\eta s},
\label{eq:anti_Hund}
\end{align}
and
\begin{align}
h_{i;J_H}^{(f)}
=&
\sum_{a,s,s'}
\sum_{\eta_1,\eta_2,\eta_3,\eta_4}
\delta_{\eta_1+\eta_2,\eta_3+\eta_4}
\frac{J_H}{2}
f_{i;(a\eta_1)\eta_1 s}^{\dagger}
f_{i;(a\eta_2)\eta_2 s'}^{\dagger}
f_{i;(a\eta_3)\eta_3 s'}
f_{i;(a\eta_4)\eta_4 s}
\nonumber\\
&+
\sum_{a,s,s'}
\sum_{\eta_1,\eta_2}
\bigg[
\frac{J_H'}{2}
f_{i;(a\eta_1)\eta_1 s}^{\dagger}
f_{i;(a\bar\eta_1)\bar\eta_1 s'}^{\dagger}
f_{i;(\bar a\bar\eta_2)\bar\eta_2 s'}
f_{i;(\bar a\eta_2)\eta_2 s}
\nonumber\\
&\hspace{2.2cm}
+
\frac{J_H'}{2}
f_{i;(a\eta_1)\eta_1 s}^{\dagger}
f_{i;(\bar a\eta_2)\eta_2 s'}^{\dagger}
f_{i;\bar a\eta_2 s'}
f_{i;(a\eta_1)\eta_1 s}
\nonumber\\
&\hspace{2.2cm}
+
\frac{J_H'}{2}
f_{i;(a\eta_1)\eta_1 s}^{\dagger}
f_{i;(\bar a\eta_2)\eta_2 s'}^{\dagger}
f_{i;(a\eta_1)\eta_1 s'}
f_{i;(\bar a\eta_2)\eta_2 s}
\bigg].
\label{eq:Hund}
\end{align}
The anti-Hund term energetically favors inter-valley spin singlets, whereas the Hund term distinguishes different orbital and angular-momentum channels.
In the present work we do not attempt to determine the microscopic values of $J_A,J_H,J_H'$ from first principles. Instead, we use this local interaction
as an effective low-energy parametrization of the spin sector.

The spin interaction can be diagonalized independently in the singly, doubly, and triply occupied sectors of the local $f$ orbital. The singly occupied sector does not gain energy from the intra-site spin interaction.
The triply occupied sector is included in the parton construction through the triplon fields, but its internal spin splitting is not essential for the
phase-transition mechanism discussed in the main text. The most important sector is the doubly occupied sector, which forms the parent local moment near $\nu=-2$. In this sector, the competition between
$h_{i;J_A}^{(f)}$ and $h_{i;J_H}^{(f)}$ selects an inter-valley spin-singlet state.

For the parameter choice $J_H'=J_H/3$, one possible lowest-energy state is an angular-momentum-zero singlet,
\begin{align}
|\Delta_{i;s}\rangle
=
\frac{1}{2}
\big(
&
f_{i;+K\uparrow}^{\dagger}
f_{i;-K'\downarrow}^{\dagger}
-
f_{i;+K\downarrow}^{\dagger}
f_{i;-K'\uparrow}^{\dagger}
\nonumber\\
&
+
f_{i;-K\uparrow}^{\dagger}
f_{i;+K'\downarrow}^{\dagger}
-
f_{i;-K\downarrow}^{\dagger}
f_{i;+K'\uparrow}^{\dagger}
\big)|0\rangle .
\label{eq:local_s_state}
\end{align}
This state is favored when the anti-Hund interaction dominates the relevant Hund splitting, for example in the regime $J_H<J_A/2$ in the parametrization above. It is an inter-valley spin singlet with total angular momentum
$L_z=0$.

In another parameter regime, the lowest-energy manifold is twofold degenerate and consists of two inter-valley spin singlets with angular
momenta $L_z=\pm2$:
\begin{align}
|\Delta_{i;d1}\rangle
&=
\frac{1}{\sqrt{2}}
\left(
f_{i;+K\uparrow}^{\dagger}
f_{i;+K'\downarrow}^{\dagger}
-
f_{i;+K\downarrow}^{\dagger}
f_{i;+K'\uparrow}^{\dagger}
\right)|0\rangle,
\nonumber\\
|\Delta_{i;d2}\rangle
&=
\frac{1}{\sqrt{2}}
\left(
f_{i;-K\uparrow}^{\dagger}
f_{i;-K'\downarrow}^{\dagger}
-
f_{i;-K\downarrow}^{\dagger}
f_{i;-K'\uparrow}^{\dagger}
\right)|0\rangle .
\label{eq:local_d_states}
\end{align}
For example, this occurs in the regime
$J_A<J_H<3J_A/2$ within the same parametrization. In our mean-field
calculation we focus on the time-reversal-invariant linear combination
\begin{equation}
|\Delta_{i;d}\rangle
=
\frac{1}{\sqrt{2}}
\left(
|\Delta_{i;d1}\rangle
+
|\Delta_{i;d2}\rangle
\right),
\label{eq:local_d_TR}
\end{equation}
which preserves $C_{2z}T$ but breaks the threefold rotation symmetry. Such a
state can be selected by a weak heterostrain or by other small symmetry
breaking perturbations. This choice gives the local $d$-wave ansatz used in
the main text.

In the slave-particle representation, the doubly occupied sector is written in terms of the neutral local spinor $\psi'$. The local pairing selected by
the spin interaction is therefore encoded as
\begin{equation}
    \Delta
    =
    \frac{1}{N}
    \sum_{i,a,\eta,s}
    s\,
    \left\langle
    \psi'_{i;a\eta\bar s}
    \psi'_{i;a\eta s}
    \right\rangle ,
    \label{eq:Delta_spinon}
\end{equation}
where $s=\pm1$ for spin up and down. The corresponding mean-field
decoupling of the spin interaction is
\begin{equation}
    H_{J,\mathrm{MF}}
    =
    -2J
    \sum_{i,a,\eta,s}
    \left[
    \Delta^{\ast}
    s\,
    \psi'_{i;a\eta\bar s}
    \psi'_{i;a\eta s}
    +
    \mathrm{h.c.}
    \right].
    \label{eq:HJ_MF_spin}
\end{equation}
Here $J$ is the effective coupling in the selected local pairing channel. In
the notation of the microscopic interaction, $J$ is a linear combination of
the parameters $J_A,J_H,J_H'$, after projecting to the relevant two-electron
local ground-state manifold.

Several comments are useful. First, $\Delta$ describes on-site pairing of the neutral local spinors $\psi'$ and therefore represents local-moment pairing, not directly superconducting order of physical electrons. In the decoupled phase, where the hybridization between $\psi'$ and the charge-carrying physical electrons vanishes, a nonzero $\Delta$ only produces a local
spin-singlet background. A physical superconducting gap appears only after the hybridization order parameters become nonzero, which transfers the local-moment pairing into the physical-electron sector.

Second, the choice between the $L_z=0$ and $L_z=\pm2$ local singlet channels does not strongly affect the charge-sector mechanism of the displacement-field-driven transition. The transition is mainly controlled by the band alignment between the neutral local-moment level and the relevant Hubbard band. The spin interaction fixes the internal symmetry of the local pair and sets the scale of the pseudogap, but the existence of the transition relies primarily on the onset of hybridization.

Finally, in realistic TTG samples additional spin interactions may be generated by RKKY-type processes. Such interactions may depend on site, valley, and orbital indices and need not take exactly the local form in Eqs.~\eqref{eq:anti_Hund} and (\ref{eq:Hund}). In the present work we use Eqs.~(\ref{eq:anti_Hund})--(\ref{eq:HJ_MF_spin}) as a minimal local ansatz
that captures the essential tendency toward inter-valley spin-singlet formation. This ansatz is sufficient to capture the essential physics: it provides a preformed local-moment pairing scale, which is subsequently transferred to the physical electrons once the displacement field induces the hybridization transition.

\subsection{Details of the mean-field construction}

\label{sec:MF_decoupling}

In this subsection we give the explicit mean-field decoupling used in the slave-particle calculation. The construction follows the parton mean-field theory of Ref.~\cite{zhao2025resonating}, with the only essential
extension that the physical-electron sector contains both the TBG-like $c$ fermions and the monolayer Dirac $d$ fermions.

As discussed in the main text, in the restricted Hilbert space near $\nu=-2$ the
projected physical $f$ operator is represented as
\begin{equation}
P_G f_{i;\alpha} P_G
=
\sum_{\beta}
s_{i;\beta}^{\dagger}
\psi'_{i;\beta}\psi'_{i;\alpha}
+
\sum_{\beta<\gamma}
t_{i;\alpha\beta\gamma}
\psi_{i;\beta}^{\prime\dagger}
\psi_{i;\gamma}^{\prime\dagger},
\label{eq:PG_f_parton}
\end{equation}
where $\alpha=(a,\eta,s)$ labels orbital, valley, and spin. We denote by
$\bar\alpha=(a,\eta,\bar s)$ the spin-singlet partner of $\alpha$, and use
$\alpha=\pm 1$ to denote the spin sign when $\alpha$ is not used as a
flavor index. The time-reversal-invariant $d$-wave local-moment pairing gives 
\begin{equation}
\left\langle \psi'_{i;\beta} \psi'_{i;\alpha} \right\rangle
= \alpha \Delta \delta_{\alpha\bar{\beta}}
\label{pairing instablity}
\end{equation}

and the corresponding mean-field spin-interaction term is
\begin{equation}
H_{J,\mathrm{MF}}
=
-2J
\sum_{i,\alpha}
\left[
\Delta^{\ast}\alpha
\psi'_{i;\bar\alpha}\psi'_{i;\alpha}
+\mathrm{h.c.}
\right].
\label{eq:HJ_MF_SM}
\end{equation}

Readers can find Ref.~\cite{zhao2025resonating} for more details about the spin interaction. In short, the spin interaction comes from the combination of the Hund and anti-Hund coupling.

Next we decouple the hybridization terms in the pairing channel. In this section, the matrices $\gamma(\bm{k})$ and $\tilde{h}(\bm{k},\bm{p})$ are taken to be the Hermitian conjugates of their counterparts defined in Appendix~\ref{appendix:details of fcd}.
For the $c$-electron hybridization, the pairing instability defined in Eq.~\eqref{pairing instablity} could lead to the following channel:
\begin{align}
P_G c_{k+G;\lambda}^{\dagger}
\gamma_{\lambda\alpha}(k+G)
f_{k;\alpha} P_G
&=
c_{k+G;\lambda}^{\dagger}
\gamma_{\lambda\alpha}(k+G)
\left(
\sum_{\beta}
s_{k;\beta}^{\dagger}
\psi'_{k;\beta}\psi'_{k;\alpha}
+
\sum_{\beta<\gamma}
t_{k;\alpha\beta\gamma}
\psi_{k;\beta}^{\prime\dagger}
\psi_{k;\gamma}^{\prime\dagger}
\right)
\nonumber\\
&\simeq
c_{k+G;\lambda}^{\dagger}
\gamma_{\lambda\alpha}(k+G)
\left(
\alpha\Delta\,s_{-k;\bar\alpha}^{\dagger}
+
\Delta^{\ast}
\sum_{\beta<\bar\beta}
\beta\,t_{k;\alpha\beta\bar\beta}
\right).
\label{eq:c_pairing_decoupling}
\end{align}
Similarly, the $f$-$d$ hybridization gives
\begin{align}
P_G d_{p;\lambda}^{\dagger}
\widetilde h_{\lambda\alpha}^{\dagger}(p,k)
f_{k;\alpha} P_G
&=
d_{p;\lambda}^{\dagger}
\widetilde h_{\lambda\alpha}^{\dagger}(p,k)
\left(
\sum_{\beta}
s_{k;\beta}^{\dagger}
\psi'_{k;\beta}\psi'_{k;\alpha}
+
\sum_{\beta<\gamma}
t_{k;\alpha\beta\gamma}
\psi_{k;\beta}^{\prime\dagger}
\psi_{k;\gamma}^{\prime\dagger}
\right)
\nonumber\\
&\simeq
d_{p;\lambda}^{\dagger}
\widetilde h_{\lambda\alpha}^{\dagger}(p,k)
\left(
\alpha\Delta\,s_{-k;\bar\alpha}^{\dagger}
+
\Delta^{\ast}
\sum_{\beta<\bar\beta}
\beta\,t_{k;\alpha\beta\bar\beta}
\right),
\label{eq:d_pairing_decoupling}
\end{align}
where
\(\widetilde h^{\dagger}_{\lambda\alpha}(p,k)
=[\widetilde h_{\alpha\lambda}(k,p)]^{\ast}\).
Only eight linear combinations of the original triplon fields enter at the
mean-field level. We therefore define the recombined triplon
\begin{equation}
t_{i;\alpha}
=
\frac{1}{\sqrt{3}}
\sum_{\beta<\bar\beta}
\beta\,t_{i;\alpha\beta\bar\beta}.
\label{eq:recombined_t}
\end{equation}
With this definition,
\begin{equation}
P_G f_{i;\alpha}P_G
\xrightarrow{\Delta}
\alpha\Delta\,s_{i;\bar\alpha}^{\dagger}
+
\sqrt{3}\Delta^{\ast}t_{i;\alpha}.
\label{eq:f_pairing_MF}
\end{equation}
The pairing-channel decoupling of the physical hybridization terms therefore gives
\begin{align}
H_{\Delta}^{\mathrm{MF}}
=&
\sum_{k,G,\lambda,\alpha}
c_{k+G;\lambda}^{\dagger}
\gamma_{\lambda\alpha}(k+G)
\left(
\alpha\Delta\,s_{-k;\bar\alpha}^{\dagger}
+
\sqrt{3}\Delta^{\ast}t_{k;\alpha}
\right)
+\mathrm{h.c.}
\nonumber\\
&+
\sum_{k,p,\lambda,\alpha}
d_{p;\lambda}^{\dagger}
\widetilde h_{\lambda\alpha}^{\dagger}(p,k)
\left(
\alpha\Delta\,s_{-k;\bar\alpha}^{\dagger}
+
\sqrt{3}\Delta^{\ast}t_{k;\alpha}
\right)
+\mathrm{h.c.}
\nonumber\\
&-
2J
\sum_{k,\alpha}
\left[
\Delta^{\ast}\alpha
\psi'_{-k;\bar\alpha}\psi'_{k;\alpha}
+\mathrm{h.c.}
\right].
\label{eq:HDelta_MF_full}
\end{align}

Next we decouple the same hybridization terms in the normal and anomalous
hybridization channels between the neutral spinon $\psi'$ and the charged
fermions. For the $c$ channel,
\begin{align}
&
\sum_{\lambda,\alpha}
P_G c_{k+G;\lambda}^{\dagger}
\gamma_{\lambda\alpha}(k+G)
f_{k;\alpha}P_G
\nonumber\\
&\simeq
\sum_{\alpha,\beta,\lambda}
\left\langle
c_{k+G;\lambda}^{\dagger}
\gamma_{\lambda\alpha}(k+G)
\psi'_{k;\alpha}
\right\rangle
s_{k;\beta}^{\dagger}\psi'_{k;\beta}
+
\sum_{\alpha,\beta,\lambda}
c_{k+G;\lambda}^{\dagger}
\gamma_{\lambda\alpha}(k+G)
\psi'_{k;\alpha}
\left\langle
s_{k;\beta}^{\dagger}\psi'_{k;\beta}
\right\rangle
\nonumber\\
&\quad
+
\frac{1}{\sqrt{3}}
\sum_{\alpha,\beta,\lambda}
\alpha
\left\langle
c_{k+G;\lambda}^{\dagger}
\gamma_{\lambda\alpha}(k+G)
\psi_{-k;\bar\alpha}^{\prime\dagger}
\right\rangle
\psi_{k;\beta}^{\prime\dagger}t_{k;\beta}
\nonumber\\
&\quad
+
\frac{1}{\sqrt{3}}
\sum_{\alpha,\beta,\lambda}
\alpha
c_{k+G;\lambda}^{\dagger}
\gamma_{\lambda\alpha}(k+G)
\psi_{-k;\bar\alpha}^{\prime\dagger}
\left\langle
\psi_{k;\beta}^{\prime\dagger}t_{k;\beta}
\right\rangle .
\label{eq:c_B_decoupling}
\end{align}
The corresponding $d$-electron decoupling is
\begin{align}
&
\sum_{\lambda,\alpha}
P_G d_{p;\lambda}^{\dagger}
\widetilde h_{\lambda\alpha}^{\dagger}(p,k)
f_{k;\alpha}P_G
\nonumber\\
&\simeq
\sum_{\alpha,\beta,\lambda}
\left\langle
d_{p;\lambda}^{\dagger}
\widetilde h_{\lambda\alpha}^{\dagger}(p,k)
\psi'_{k;\alpha}
\right\rangle
s_{k;\beta}^{\dagger}\psi'_{k;\beta}
+
\sum_{\alpha,\beta,\lambda}
d_{p;\lambda}^{\dagger}
\widetilde h_{\lambda\alpha}^{\dagger}(p,k)
\psi'_{k;\alpha}
\left\langle
s_{k;\beta}^{\dagger}\psi'_{k;\beta}
\right\rangle
\nonumber\\
&\quad
+
\frac{1}{\sqrt{3}}
\sum_{\alpha,\beta,\lambda}
\alpha
\left\langle
d_{p;\lambda}^{\dagger}
\widetilde h_{\lambda\alpha}^{\dagger}(p,k)
\psi_{-k;\bar\alpha}^{\prime\dagger}
\right\rangle
\psi_{k;\beta}^{\prime\dagger}t_{k;\beta}
\nonumber\\
&\quad
+
\frac{1}{\sqrt{3}}
\sum_{\alpha,\beta,\lambda}
\alpha
d_{p;\lambda}^{\dagger}
\widetilde h_{\lambda\alpha}^{\dagger}(p,k)
\psi_{-k;\bar\alpha}^{\prime\dagger}
\left\langle
\psi_{k;\beta}^{\prime\dagger}t_{k;\beta}
\right\rangle .
\label{eq:d_B_decoupling}
\end{align}

The corresponding order parameters are given by
\begin{equation}
\begin{aligned}
\Delta
&=
\frac{1}{8N_k}
\sum_{k}
\alpha
\left\langle
\psi'_{k;\bar\alpha}\psi'_{-k;\alpha}
\right\rangle,
\\
B_s
&=
7\alpha_B
\frac{1}{8N_k}
\sum_{k,\alpha}
\left\langle
s_{k;\alpha}^{\dagger}\psi'_{k;\alpha}
\right\rangle,
\\
B_t
&=
2\sqrt{3}\alpha_B
\frac{1}{8N_k}
\sum_{k,\alpha}
\left\langle
t_{k;\alpha}^{\dagger}\psi'_{k;\alpha}
\right\rangle,
\\
B_c'
&=
7\alpha_B
\frac{1}{8N_k}
\sum_{k,G,\lambda,\alpha}
\left\langle
c_{k+G;\lambda}^{\dagger}
\gamma_{\lambda\alpha}(k+G)
\psi'_{k;\alpha}
\right\rangle,
\\
B_d'
&=
7\alpha_B
\frac{1}{8N_k}
\sum_{k,p,\lambda,\alpha}
\left\langle
d_{p;\lambda}^{\dagger}
\widetilde h_{\lambda\alpha}^{\dagger}(p,k)
\psi'_{k;\alpha}
\right\rangle,
\\
\Delta_c'
&=
2\sqrt{3}\alpha_B
\frac{1}{8N_k}
\sum_{k,G,\lambda,\alpha}
\alpha
\left\langle
c_{k+G;\lambda}^{\dagger}
\gamma_{\lambda\alpha}(k+G)
\psi_{-k;\bar\alpha}^{\prime\dagger}
\right\rangle,
\\
\Delta_d'
&=
2\sqrt{3}\alpha_B
\frac{1}{8N_k}
\sum_{k,p,\lambda,\alpha}
\alpha
\left\langle
d_{p;\lambda}^{\dagger}
\widetilde h_{\lambda\alpha}^{\dagger}(p,k)
\psi_{-k;\bar\alpha}^{\prime\dagger}
\right\rangle.
\end{aligned}
\label{eq:Deltad_def}
\end{equation}
Here all flavor indices are summed over the total eight orbitals. $\alpha_B$ is the same phenomenological reduction factor introduced in
Ref.~\cite{zhao2025resonating}. It is introduced to account for the effect that in the real systems, not all electrons would interact in the mean-field channel. It is also vital for achieving numerical convergence. In this paper, we choose $\alpha_B$ to be $\frac{1}{2}$. The numerical coefficients in Eq.~\eqref{eq:Deltad_def} follow from flavor counting in the restricted  Hilbert space and from the normalization of the recombined triplon in Eq.~\eqref{eq:recombined_t}.

With these definitions, the hybridization-channel mean-field Hamiltonian is
\begin{align}
H_{B}^{\mathrm{MF}}
=&
\sum_{k,G,\lambda,\alpha}
c_{k+G;\lambda}^{\dagger}
\gamma_{\lambda\alpha}(k+G)
\left(
B_s\psi'_{k;\alpha}
+
\alpha B_t
\psi_{-k;\bar\alpha}^{\prime\dagger}
\right)
+\mathrm{h.c.}
\nonumber\\
&+
\sum_{k,p,\lambda,\alpha}
d_{p;\lambda}^{\dagger}
\widetilde h_{\lambda\alpha}^{\dagger}(p,k)
\left(
B_s\psi'_{k;\alpha}
+
\alpha B_t
\psi_{-k;\bar\alpha}^{\prime\dagger}
\right)
+\mathrm{h.c.}
\nonumber\\
&+
\sum_{k,\alpha}
\left(
B_c' s_{k;\alpha}^{\dagger}\psi'_{k;\alpha}
+
\Delta_c' t_{k;\alpha}^{\dagger}\psi'_{k;\alpha}
\right)
+\mathrm{h.c.}
\nonumber\\
&+
\sum_{k,\alpha}
\left(
B_d' s_{k;\alpha}^{\dagger}\psi'_{k;\alpha}
+
\Delta_d' t_{k;\alpha}^{\dagger}\psi'_{k;\alpha}
\right)
+\mathrm{h.c.}.
\label{eq:HB_MF_full}
\end{align}

Combining Eqs.~(\ref{eq:HDelta_MF_full}) and
(\ref{eq:HB_MF_full}), the full mean-field Hamiltonian is
\begin{align}
H_{\mathrm{MF}}
=&
H_c^0
+
H_d^0
+
wN_d
-\mu(N_c+N_d)
+
H_{\Delta}^{\mathrm{MF}}
+
H_B^{\mathrm{MF}}
\nonumber\\
&+
(E_s+\mu-\lambda)
\sum_{i,\alpha}
s_{i;\alpha}^{\dagger}s_{i;\alpha}
-
\frac{\lambda}{2}
\sum_{i,\alpha}
\psi_{i;\alpha}^{\prime\dagger}\psi'_{i;\alpha}
+
(E_t-\mu-\lambda)
\sum_{i,\alpha}
t_{i;\alpha}^{\dagger}t_{i;\alpha}
+
\mathrm{const.}
\label{eq:HMF_full}
\end{align}
The Lagrange multiplier $\lambda$ enforces the local constraint on average,
and the chemical potential $\mu$ fixes the total density. The order
parameters in Eq.~\eqref{eq:Deltad_def} are solved self-consistently by
diagonalizing Eq.~\eqref{eq:HMF_full} in Nambu space.

Finally, the projected physical $f$ operator used to compute the spectral
function is obtained by keeping both the pairing and hybridization
mean-field contractions:
\begin{equation}
P_G f_{i;\alpha}P_G
\simeq
\alpha\Delta\,s_{i;\bar\alpha}^{\dagger}
+
\sqrt{3}\Delta^{\ast}t_{i;\alpha}
+
B_s\psi'_{i;\alpha}
+
\sqrt{3}\alpha B_t^{\ast}
\psi_{i;\bar\alpha}^{\prime\dagger}.
\label{eq:f_spectral_full}
\end{equation}
 The physical STM spectrum is then obtained from the physical $c$ and $d$ components of
the BdG eigenvectors together with the projected $f$-electron weight in
Eq.~\eqref{eq:f_spectral_full}.

\section{Equivalence between the ancilla theory and the slave-particle model}
\label{appendix:euqivelence between ancilla and slave particle}
As discussed in Ref.~\cite{zhao2025mixed}, when the local moments are decoupled from the physical electrons, i.e., when all order parameters other
than $\Delta$ vanish and $\Delta$ is real, Eq.~\eqref{eq:f_spectral_full} reduces to
\begin{equation}
P_G f_{i;\alpha}P_G
\simeq
\alpha\Delta\,s_{i;\bar\alpha}^{\dagger}
+
\sqrt{3}\Delta t_{i;\alpha}.
\label{eq:f_spectral_decoupled}
\end{equation}
One can then introduce an orthogonal fermionic operator,
\begin{equation}
\psi_{i;\alpha}^{\dagger}
=-\frac{\sqrt{3}}{2}\alpha s_{i;\bar\alpha}
+\frac{1}{2}t_{i;\alpha}^{\dagger}.
\label{eq:orthogonal_psi}
\end{equation}
In terms of this orthogonal fermion, the Hamiltonian can be rewritten in the ancilla-theory form
\begin{equation}
H_{\rm ancilla}
=H_0^{(c_1,c_2)}
+H_0^{(c_1,f)}
+\Phi \sum_{\mathbf k} \left(
f_{\mathbf k}^{\dagger}\psi_{\mathbf k}
+
{\rm H.c.}
\right)
-\mu N_c-\mu_f N_f-\mu_\psi N_\psi ,
\label{eq:ancilla_equivalent}
\end{equation}
where $\mu_\psi$ is introduced to enforce the constraint
$\langle n_{i;\psi}\rangle=6$. This is precisely the form of Eq.~\eqref{full ancilla function} in the limit $\Phi_1=0$ and $\Phi_2=0$. Therefore, the ancilla bands provide a faithful description of
the Mott bands and serve as a natural starting point for describing the transition into the superconducting state.

We emphasize, however, that after including the coupling to the local moment
$\psi^\prime$, Eq.~\eqref{eq:ancilla_equivalent} does not yet contain the
additional hybridization channels between $\psi^\prime$ and the $c,d$ electrons. These couplings can be added artificially to the ancilla Hamiltonian. Nevertheless, the band structure obtained from Eq.~\eqref{eq:ancilla_equivalent} already
captures the essential Mott-band structure and provides a useful starting point for understanding the interplay between the on-site interaction and
the hybridization with local moments.

\section{Minimal model near the $K_{-}$ point and the $D$-tuned band repulsion}
\label{appendix:minimal model}

\begin{figure*}[t]
    \centering
    \includegraphics[width=\textwidth]{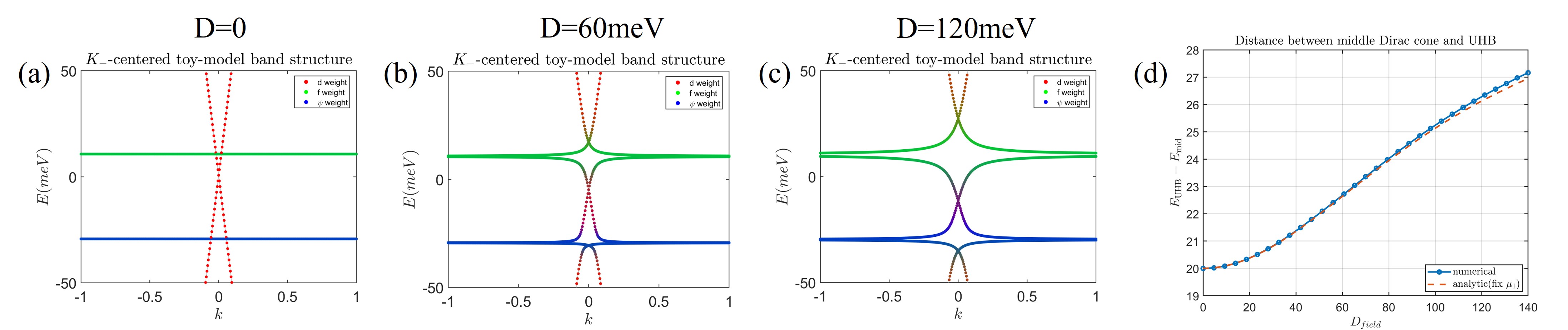}
    
    \caption{
    Results of the toy-model calculation. In the numerical calculation, the
    constant $M_1$ in the $d$-$f$ hybridization matrix $V_{df}$ is restored,
    such that the effective hybridization is proportional to $M_1D$.
    The parameters are $\nu=-2$, $U=40\,\mathrm{meV}$.
    (a)--(c) Band structures for representative values of the displacement
    field $D$. (d) Comparison between the analytic expression and the
    numerical result for the energy separation between the middle Dirac point
    and the UHB.
    }

    \label{fig:toy_model}	
\end{figure*}

In this section we introduce a minimal three-component model to illustrate
the displacement-field-induced level repulsion near the $K_-$ point. The
model retains a Dirac-like physical-electron sector, denoted by $d$, a flat
band sector, denoted by $f$, and an ancilla fermion $\psi$. The ancilla
fermion accounts for the on-site interaction by hybridizing with the $f$
sector and producing the lower and upper Hubbard bands. This model is
essentially the charge-sector ancilla theory introduced in
Sec.~\ref{appendix:section_for_ancilla}, with the coupling to the local moment
$\psi^\prime$ turned off and with the $c$ electrons neglected. We also neglect the flavor index. In the basis
\[
(d_1,d_2,f_1,f_2,\psi_1,\psi_2),
\]
the Hamiltonian is written as
\begin{equation}
H(\mathbf{k})=
\begin{pmatrix}
h_d(\mathbf{k}) & V_{df} & 0 \\
V_{df}^{\dagger} & \frac{\mu_1}{2} \sigma_0 & V_{f\psi} \\
0 & V_{f\psi}^{\dagger} & -\frac{\mu_1}{2} \sigma_0
\end{pmatrix}
-\mu I_6 ,
\label{eq:toy_full_H}
\end{equation}
where
\begin{equation}
h_d(\mathbf{k})
=
v k_x \sigma_x+v k_y \sigma_y,
\qquad
V_{df}
=
D(\sigma_0+i\sigma_z),
\qquad
V_{f\psi}
=
\phi_1 \sigma_0 .
\label{eq:toy_blocks}
\end{equation}
Here $\mathbf{k}$ is measured from the Dirac point, which is the $K_{-}$
point in our original model in the main text. For the analytic derivation
below, we omit the constant $M_1$ in the $V_{df}$ term for simplicity; it is
restored in the numerical calculation shown in
Fig.~\ref{fig:toy_model}. The parameter $\phi_1$ controls the
$f$-$\psi$ hybridization, while $\mu_1$ and $\mu$ are chosen to satisfy the
constraints
\begin{equation}
\begin{aligned}
n_d+n_f &= 2+\nu/4,
\\
n_\psi &= 1-\nu/4.
\end{aligned}
\label{equation_constraint_on_toy}
\end{equation}
The resulting band structures are shown in
Fig.~\ref{fig:toy_model}, using the same parameter regime as in
the TTG calculation. We now analyze how the displacement field shifts the
middle Dirac point through band repulsion.

We first consider the band structure at $\mathbf{k}=0$. At this point,
$h_d(\mathbf{k}=0)=0$. Since
\begin{equation}
D(\sigma_0+i\sigma_z)
=
D
\begin{pmatrix}
1+i & 0 \\
0 & 1-i
\end{pmatrix},
\label{eq:toy_D_matrix}
\end{equation}
the two internal sectors decouple and are related by complex conjugation.
For each sector, the phase of the $d$-$f$ hybridization can be removed by a
unitary transformation, leaving the effective hybridization amplitude
\begin{equation}
\sqrt{2}D .
\label{eq:t_eff}
\end{equation}
Thus the $\mathbf{k}=0$ Hamiltonian reduces to two degenerate copies of the
following $3\times 3$ problem:
\begin{equation}
H_{K_-}^{(0)}
=
\begin{pmatrix}
w & \sqrt{2}D & 0 \\
\sqrt{2}D & \frac{\mu_1}{2} & \phi_1 \\
0 & \phi_1 & -\frac{\mu_1}{2}
\end{pmatrix}.
\label{eq:H_Gamma_3by3}
\end{equation}
The global shift $-\mu I_3$ has been omitted because it does not affect
energy separations.

\subsection{Perturbation theory based on diagonal band basis}

We first set $D=0$ and then diagonalize the $f$-$\psi$ block,
\begin{equation}
H_{f\psi}
=
\begin{pmatrix}
\frac{\mu_1}{2} & \phi_1 \\
\phi_1 & -\frac{\mu_1}{2}
\end{pmatrix}.
\label{eq:SM_H_fpsi}
\end{equation}
Its two eigenvalues are
\begin{equation}
\epsilon_{\pm}
=
\pm \Lambda,
\qquad
\Lambda
=
\sqrt{\phi_1^2+\frac{\mu_1^2}{4}}
=
\frac{1}{2}\sqrt{\mu_1^2+4\phi_1^2}.
\label{eq:SM_eps_pm}
\end{equation}
These two bands correspond to the upper and lower Hubbard bands in this minimal model. The corresponding normalized eigenstates of these Hubbard bands can be written as
\begin{equation}
|\pm\rangle
=
u_\pm |f\rangle
+
v_\pm |\psi\rangle ,
\label{eq:SM_eigenstate_pm}
\end{equation}
where, for real $\phi_1$,
\begin{equation}
u_\pm
=
\mathcal{N}_\pm ,
\qquad
v_\pm
=
\frac{\epsilon_\pm-\mu_1/2}{\phi_1}\mathcal{N}_\pm 
\qquad
\text{with} \
\mathcal{N}_\pm
=
\left[
1+
\left(
\frac{\epsilon_\pm-\mu_1/2}{\phi_1}
\right)^2
\right]^{-1/2}.
\label{eq:SM_uv_pm_raw}
\end{equation}

Therefore 
\begin{equation}
Z_+
\equiv
|u_+|^2
=
\frac{1}{2}
\left(
1+\frac{\mu_1}{2\Lambda}
\right),
\qquad
Z_-
\equiv
|u_-|^2
=
\frac{1}{2}
\left(
1-\frac{\mu_1}{2\Lambda}
\right).
\label{eq:SM_Z_pm}
\end{equation}
Here $Z_\pm$ are the $f$ weights of the two $f$-$\psi$ eigenstates. Since the
Dirac state couples only to the $f$ component, these weights directly control the strength of the level repulsion. From the constraint in Eq.~\eqref{equation_constraint_on_toy} and the fact that the Mott gap is equal to $U$. We can get 
\begin{equation}
\mu_1=-\frac{\nu}{4}U, \quad \Phi=\frac{\sqrt{1-\frac{\nu^2}{16}}}{2} \Delta_{\mathrm{Mott}},
\end{equation}

In the basis
\begin{equation}
|d\rangle,\quad |+\rangle,\quad |-\rangle,
\end{equation}
the Hamiltonian becomes
\begin{equation}
H_{\Gamma}
=
\begin{pmatrix}
0 & V_{d,+} & V_{d,-} \\
V_{d,+}^\ast & \epsilon_+ & 0 \\
V_{d,-}^\ast & 0 & \epsilon_-
\end{pmatrix}.
\label{eq:SM_H_diag_basis}
\end{equation}
Here $V_{d,+}=t u_+$, $V_{d,-}=t u_- $
and $t=\sqrt{2}D$
is the effective $d$-$f$ hybridization at $\mathbf{k}=0$, after removing the
phase of $D(\sigma_0+i\sigma_z)$ by a unitary transformation. 

At small $D$ when the detuning between the bare Dirac level and the two $f$-$\psi$ levels
is large compared with the hybridization, namely
\begin{equation}
|t u_\pm|\ll |\epsilon_d-\epsilon_\pm|,
\label{eq:SM_perturbative_condition}
\end{equation}
the energy shift of the Dirac level is given by second-order perturbation
theory:
\begin{equation}
\delta E_d
=
-\frac{|V_{d,+}|^2}{\epsilon_+}
-
\frac{|V_{d,-}|^2}{\epsilon_-}
=
-D^2
\frac{\mu_1}
{\phi_1^2+\mu_1^2/4}.
\label{eq:SM_second_order_general}
\end{equation}
The separation between the upper Hubbard branch and the middle Dirac point is
therefore
\begin{equation}
\Delta_{\rm UHB-mid}(D)
=
\epsilon_+-\delta E_d
=
U/2-\frac{\nu D^2}{U}
\label{eq:SM_Delta_w0}
\end{equation}
Thus, for $\mu_1>0$, the middle Dirac point is pushed downward by the
hybridization, and its distance from the upper Hubbard branch increases quadratically with $D$.
We see that the sign of the energy shift is controlled by $\mu_1$. For $\nu<0$, increasing $D$ pushes the middle Dirac point downward. By contrast, for $\nu>0$, one has $\mu_1<0$, and the same mechanism shifts the
middle Dirac point upward. This is consistent with the band evolution discussed in the main text. 

\subsection{Full eigenvalue calculation}

Let $E$ denote an eigenvalue of Eq.~\eqref{eq:H_Gamma_3by3}. The
characteristic equation is obtained from
\begin{equation}
\det\left[
E-H_{K_-}^{(0)}
\right]
=
E^3
-
\left(
2D^2+\phi_1^2+\frac{\mu_1^2}{4}
\right)E
-
\mu_1D^2
=0.
\label{eq:det_equation}
\end{equation}
At $D=0$, the three eigenvalues are
\begin{equation}
E_0=0,
\qquad
E_{\pm}
=
\pm
\sqrt{\phi_1^2+\frac{\mu_1^2}{4}}
=
\pm
\frac{1}{2}
\sqrt{\mu_1^2+4\phi_1^2}.
\label{eq:D0_eigenvalues}
\end{equation}
The eigenvalue $E_0=0$ corresponds to the original Dirac point. When $D$ is
turned on, the middle Dirac point is identified as the root of
Eq.~\eqref{eq:det_equation} that is continuously connected to $E_0=0$. We denote
this root by $E_{K_-}(D)$.

For completeness, we can give the full expression of the roots for Eq.~\eqref{eq:det_equation}, which gives
\begin{equation}
E_j(D)
=
2\sqrt{-\frac{p}{3}}
\cos\left[
\frac{1}{3}
\cos^{-1}
\left(
\frac{3q}{2p}\sqrt{-\frac{3}{p}}
\right)
-
\frac{2\pi j}{3}
\right],
\qquad
j=0,1,2 .
\label{eq:cardano_roots}
\end{equation}
The physical middle Dirac point $E_{K_-}(D)$ is the branch satisfying
$E_{K_-}(0)=0$.

We next define the reference energy of the UHB in this minimal model. In the
limit $|\mathbf{k}|\rightarrow\infty$, the $d$-like Dirac states move to
high energies of order $\pm v|\mathbf{k}|$, while the finite-energy sector
is described by the residual $f$-$\psi$ Hamiltonian
\begin{equation}
H_{f\psi}^{(\infty)}
=
\begin{pmatrix}
\frac{\mu_1}{2} & \phi_1 \\
\phi_1 & -\frac{\mu_1}{2}
\end{pmatrix}.
\label{eq:fpsi_infty}
\end{equation}
The upper finite branch is
\begin{equation}
E_{\rm UHB}
=
\sqrt{\phi_1^2+\frac{\mu_1^2}{4}}
=
\frac{1}{2}
\sqrt{\mu_1^2+4\phi_1^2}.
\label{eq:EUHB}
\end{equation}
This gives the asymptotic position of the upper $f$-derived branch in the
toy model. 
The energy separation between the middle Dirac point and the UHB is then
\begin{equation}
\Delta_{\rm UHB-K_-}(D)
=
E_{\rm UHB}-E_{K_-}(D).
\label{eq:Delta_exact}
\end{equation}

We can get the perturbative results in the last subsection by solving Eq.~\eqref{eq:det_equation} order by order
\begin{equation}
E_{K_-}(D)
=
-\frac{\mu_1}{\phi_1^2+\mu_1^2/4}D^2
+
\frac{2\mu_1}{\left(\phi_1^2+\mu_1^2/4\right)^2}D^4
+
O(D^6).
\label{eq:EGamma_expansion_explicit}
\end{equation}

We can also get the large-$D$ limiting behavior from Eq.~\eqref{eq:det_equation}. For $D^2\gg \phi_1^2+\mu_1^2/4$, the finite-energy root satisfies
\begin{equation}
E_{K_-}(D)
=
-\frac{\mu_1}{2}
+
\frac{\mu_1\phi_1^2}{4D^2}
+
O(D^{-4}).
\label{eq:large_D_EK}
\end{equation}
Thus the energy separation saturates to
\begin{equation}
\Delta_{\rm UHB-K_-}(D\rightarrow\infty)
=
E_{\rm UHB}
+
\frac{\mu_1}{2}
=
\sqrt{\phi_1^2+\frac{\mu_1^2}{4}}
+
\frac{\mu_1}{2},
\label{eq:large_D_Delta}
\end{equation}
up to corrections of order $D^{-2}$. If the Hubbard-band splitting is fixed
as
\begin{equation}
E_{\rm UHB}-E_{\rm LHB}=U,
\end{equation}
then
\begin{equation}
E_{\rm UHB}=\frac{U}{2}.
\end{equation}
Furthermore, if the $D=0$ filling constraint is used to estimate
\begin{equation}
\mu_1\simeq -\frac{\nu U}{4},
\label{eq:mu1_nu_estimate}
\end{equation}
the large-$D$ saturation value becomes
\begin{equation}
\Delta_{\rm UHB-K_-}(D\rightarrow\infty)
\simeq
\frac{U}{2}
-
\frac{\nu U}{8}
=
\frac{4-\nu}{8}U .
\label{eq:large_D_Delta_nu}
\end{equation}
This provides a simple interpretation of the saturation scale in terms of
the Hubbard-band splitting and the filling-dependent offset $\mu_1$.

\section{Results away from $|\nu|=2$}

\begin{figure*}[t]
    \centering
    \includegraphics[width=0.6\textwidth]{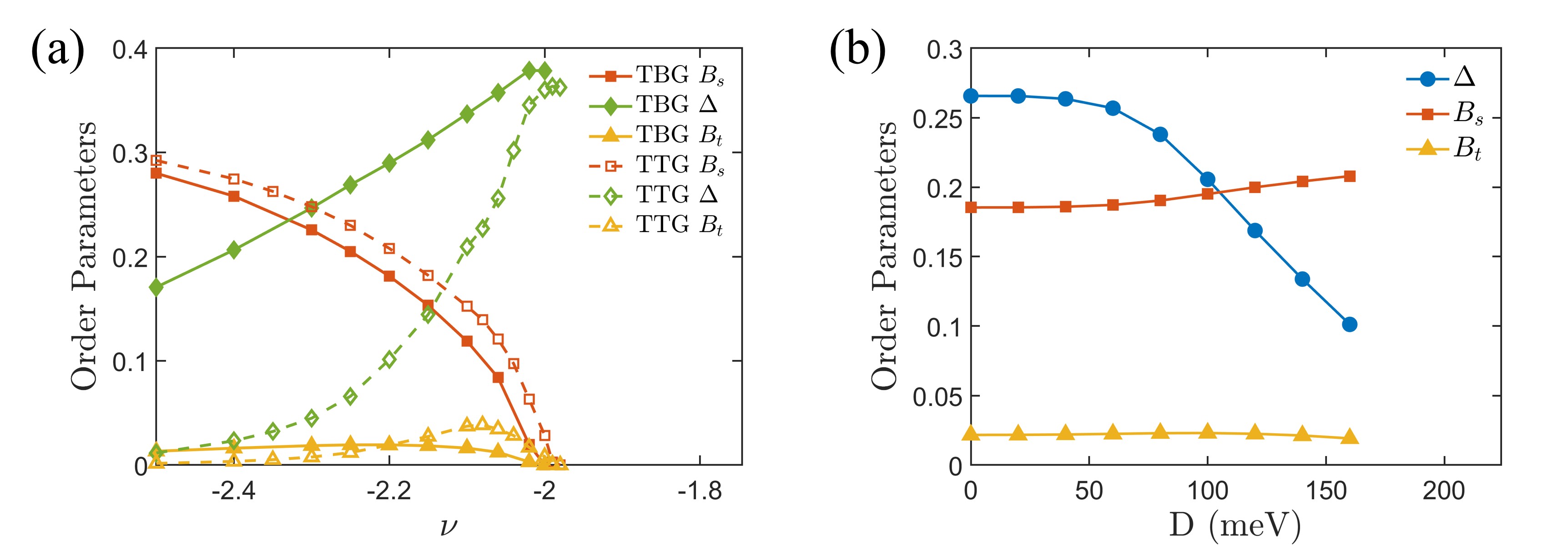}
    \caption{
    Superconducting order parameters away from $|\nu|=2$.
    (a) Comparison of the mean-field order parameters between the TBG reference and TTG with $D=160\mathrm{meV}$ and $w=-15\mathrm{meV}$.
    Away from $\nu=-2$, the spinon pairing amplitude $\Delta$ is strongly suppressed in TTG.
    (b) Evolution of the order parameters as a function of displacement field $D$ at a fixed filling $\nu=-2.2$.
    At large $D$, $\Delta$ decreases substantially, indicating the suppression of superconductivity away from the integer filling.
    }
    \label{fig:results_at_finite_filling}
\end{figure*}

We now present additional results for fillings farther away from $|\nu|=2$.
Figure~\ref{fig:results_at_finite_filling}(a) compares the mean-field order parameters of TTG at large displacement field with those of the TBG reference.
When the filling is moved farther away from $\nu=-2$, the spinon pairing amplitude $\Delta$ is strongly suppressed in TTG, resulting in a smaller superconducting gap.
We further track the evolution of the order parameters with displacement field at fixed filling $\nu=-2.2$, as shown in Fig.~\ref{fig:results_at_finite_filling}(b).
Although the hybridization order parameter $B_s$ is enhanced by increasing $D$, the spinon pairing amplitude $\Delta$ is substantially reduced at large $D$.
As a result, superconductivity is suppressed away from the integer filling.
This behavior is consistent with experiments showing that, in the intermediate filling regime between $\nu=-2$ and $\nu=-3$, the superconducting critical temperature is strongly suppressed at large displacement field~\cite{park2021tunable,hao2021electric,zhang2022promotion}.

\section{Linear response theory for the phase transition}
\subsection{Construction of the linear response matrix}

\label{appendix:linear response theory}

To systematically investigate the phase transition driven by the energy offset $w$ and the displacement field $D$, we employ a rigorous linear response formalism. This approach is naturally connected with the method of expanding the free energy, while it directly evaluates how the quantum fluctuations within the parent state self-consistently amplify the spontaneous order parameters, providing a straightforward understanding to the numerical matrix implementations.

Within the parton construction, the macroscopic phase transition is driven by the condensation of slave bosons. We perform the linear-response analysis around the parent state, in which all order parameters vanish except for the local-moment pairing $\Delta$. The linear-response theory then determines when this parent state becomes unstable against slave-boson condensation, thereby identifying the onset of the phase transition.

The momentum-resolved fluctuation operators are
\begin{equation}
\begin{aligned}
    \hat{O}_s(\bm{k})
    &=
    \sum_{\alpha}
    s^\dagger_{\bm{k},\alpha}
    \psi'_{\bm{k},\alpha},
    \quad
    \hat{O}_{t}(\bm{k})
    =
    \sum_{\alpha}
    t^\dagger_{\bm{k},\alpha}
    \psi'_{\bm{k},\alpha},
    \\
    \hat{O}_{c}(\bm{k})
    &=
    \sum_{\bm{G},\lambda,\alpha}
    c^\dagger_{\bm{k}+\bm{G},\lambda}
    \gamma_{\lambda\alpha}(\bm{k}+\bm{G})
    \psi'_{\bm{k},\alpha},
    \\
    \hat{O}_{d}(\bm{k})
    &=
    \sum_{\bm{p},\lambda,\alpha}
    d^\dagger_{\bm{p},\lambda}
    \tilde h^\dagger_{\lambda\alpha}(\bm{p},\bm{k})
    \psi'_{\bm{k},\alpha},
    \\
    \hat{O}_{c}^{\prime}(\bm{k})
    &=
    \sum_{\bm{G},\lambda,\alpha}
    \zeta_\alpha
    c^\dagger_{\bm{k}+\bm{G},\lambda}
    \gamma_{\lambda\alpha}(\bm{k}+\bm{G})
    \psi^{\prime\dagger}_{-\bm{k},\bar\alpha},
    \\
    \hat{O}_{d}^{\prime}(\bm{k})
    &=
    \sum_{\bm{p},\lambda,\alpha}
    \zeta_\alpha
    d^\dagger_{\bm{p},\lambda}
    \tilde h^\dagger_{\lambda\alpha}(\bm{p},\bm{k})
    \psi^{\prime\dagger}_{-\bm{k},\bar\alpha}.
\end{aligned}
\label{mean field operator}
\end{equation}

The uniform order parameters correspond to the zero-momentum components of these bilinears. Explicitly, denoting the numerical integration weight of each momentum point by $w_{\bm{k}}$, we define
\begin{equation}
    \langle \hat{O}_{\mu}\rangle
    \equiv
    \frac{1}{N_k}\sum_{\bm{k}} 
    \langle \hat{O}_{\mu}(\bm{k})\rangle .
\end{equation}
The emergence of the ordered phase is characterized by non-zero expectation values of these uniform bilinears. According to the mean-field ansatz, the effective order parameters are dictated by the self-consistency conditions
\begin{equation}
\begin{aligned}
    B_s &= g_1 \langle \hat{O}_s \rangle,
    \quad
    B_t = g_2 \langle \hat{O}_t \rangle,
    \\
    B'_c &= g_1 \langle \hat{O}_c \rangle,
    \quad
    \Delta_c' = g_2 \langle \hat{O}_c^\prime \rangle,
    \\
    B'_d &= g_1 \langle \hat{O}_d \rangle,
    \quad
    \Delta_d' = g_2 \langle \hat{O}_d^\prime \rangle .
\end{aligned}
\end{equation}
From the last subsection, we see that the value of $g_1$ and $g_2$ is $\frac{7\alpha_B}{8}$ and $\frac{\sqrt{3} \alpha_B}{4}$. In the numerical implementation used below, a fourfold flavor degeneracy is absorbed into the prefactors. Therefore the coefficients appearing in the code are equivalently $4g_1=7\alpha_B/2$ and $4g_2=\sqrt{3}\alpha_B$.

Before constructing the real linear response function, we need to first clarify which operators could be coupled together following the constraint of the gauge charge. Within the parton construction, the physical electron is fractionalized, introducing an internal $U(1)$ gauge redundancy subject to the local constraint $n_{i;s} + n_{i;\psi'}/2 + n_{i;t} = 1$. Under a local gauge transformation $s \rightarrow s e^{2i\varphi}$, $t \rightarrow t e^{2i\varphi}$ and $\psi' \rightarrow \psi' e^{i\varphi}$, the mean-field operators transform distinctively:
\begin{equation}
\begin{aligned}
    B_s
    &\propto
    \sum_{\bm{k},\alpha}
    \langle
    s^\dagger_{\bm{k},\alpha}
    \psi'_{\bm{k},\alpha}
    \rangle
    \quad
    (\text{Gauge charge: } -1),
    \nonumber
    \\
    B'_c
    &\propto
    \sum_{\bm{k},\bm{G},\lambda,\alpha}
    \left\langle
    c^\dagger_{\bm{k}+\bm{G},\lambda}
    \gamma_{\lambda\alpha}(\bm{k}+\bm{G})
    \psi'_{\bm{k},\alpha}
    \right\rangle
    \quad
    (\text{Gauge charge: } +1),
    \nonumber
    \\
    B'_d
    &\propto
    \sum_{\bm{k},\bm{p},\lambda,\alpha}
    \left\langle
    d^\dagger_{\bm{p},\lambda}
    \tilde h^\dagger_{\lambda\alpha}(\bm{p},\bm{k})
    \psi'_{\bm{k},\alpha}
    \right\rangle
    \quad
    (\text{Gauge charge: } +1),
    \nonumber
    \\
    B_t
    &\propto
    \sum_{\bm{k},\alpha}
    \langle
    t^\dagger_{\bm{k},\alpha}
    \psi'_{\bm{k},\alpha}
    \rangle
    \quad
    (\text{Gauge charge: } -1),
    \nonumber
    \\
    \Delta'_c
    &\propto
    \sum_{\bm{k},\bm{G},\lambda,\alpha}
    \zeta_\alpha
    \left\langle
    c^\dagger_{\bm{k}+\bm{G},\lambda}
    \gamma_{\lambda\alpha}(\bm{k}+\bm{G})
    \psi^{\prime\dagger}_{-\bm{k},\bar\alpha}
    \right\rangle
    \quad
    (\text{Gauge charge: } -1),
    \\
    \Delta'_d
    &\propto
    \sum_{\bm{k},\bm{p},\lambda,\alpha}
    \zeta_\alpha
    \left\langle
    d^\dagger_{\bm{p},\lambda}
    \tilde h^\dagger_{\lambda\alpha}(\bm{p},\bm{k})
    \psi^{\prime\dagger}_{-\bm{k},\bar\alpha}
    \right\rangle
    \quad
    (\text{Gauge charge: } -1).
\end{aligned}
\end{equation}

Therefore, the full collection of the order parameters should be
\begin{equation}
    \bm{\Phi}
    =
    \left(
    B_c^{\prime *},
    B_d^{\prime *},
    B_s,
    \Delta_c^\prime,
    \Delta_d^{\prime},
    B_t
    \right)^{T}.
\end{equation}
Here $B_c^{\prime *}$ and $B_d^{\prime *}$ mean taking the complex conjugate of the related normal hybridization order parameters. Correspondingly, the first two components of the response vector are associated with $\hat O_c^\dagger$ and $\hat O_d^\dagger$, rather than $\hat O_c$ and $\hat O_d$.

In the linear response regime, the expectation value of an operator induced by the spontaneous emergence of the order parameters $\Phi_j$ is captured by the susceptibility matrix. To make the source-response structure explicit, we write
\begin{equation}
    \langle \hat{R}_i \rangle
    =
    \sum_j \chi_{ij}\Phi_j,
\end{equation}
where
\begin{equation}
    \hat{\bm R}
    =
    \left(
    \hat O_c^\dagger,
    \hat O_d^\dagger,
    \hat O_s,
    \hat O_c',
    \hat O_d',
    \hat O_t
    \right)^T .
\end{equation}
The imaginary-time susceptibility is defined as
\begin{equation}
    \chi_{ij}(\bm{q},\tau)
    =
    -
    \left\langle
    T_\tau
    \hat R_i(\bm{q},\tau)
    \hat V_j^\dagger(\bm{q},0)
    \right\rangle_0 .
\end{equation}
Here $\hat V_j$ denotes the bilinear operator that is multiplied by the source field $\Phi_j$ in the mean-field Hamiltonian, and the expectation value is taken in the parent sFL state, where all hybridization order parameters vanish while $\Delta\neq0$. The Matsubara-frequency response is
\begin{equation}
    \chi_{ij}(\bm{q},i\Omega_\ell)
    =
    \int_0^\beta d\tau\,
    e^{i\Omega_\ell\tau}
    \chi_{ij}(\bm{q},\tau),
    \qquad
    \Omega_\ell=\frac{2\pi \ell}{\beta}.
\end{equation}
The instability discussed below is controlled by the static uniform limit
\begin{equation}
    \chi_{ij}
    \equiv
    \chi_{ij}(\bm{q}=0,i\Omega_\ell=0).
\end{equation}
Substituting the linear response relation back into the self-consistency conditions yields a closed set of linearized coupled equations:
\begin{equation}
    \bm{M}_{LR} \bm{\Phi}=0 ,
    \label{linearresponse}
\end{equation}
where $\bm{M}_{LR}=(\bm{I} - \bm{g}\bm{\chi})$, $\bm{I}$ is the identity matrix, and
\begin{equation}
    \bm{g}
    =
    \text{diag}(g_1,g_1,g_1,g_2,g_2,g_2).
\end{equation}

The full form of the matrix is
\begin{equation}
\bm{M_{LR}}=
\left(
\begin{array}{cccccc}
1-g_1 \chi_{sc}
&
-g_1\chi_{sc}
&
-g_1 (\chi_{cc}+\chi_{cd})
&
-g_1 \chi_{ct}
&
-g_1\chi_{ct}
&
-g_1(\chi_{paircc}+\chi_{paircd})
\\
-g_1 \chi_{sd}
&
1-g_1 \chi_{sd}
&
-g_1(\chi_{dd}+\chi_{cd})
&
-g_1 \chi_{dt}
&
-g_1 \chi_{dt}
&
-g_1(\chi_{pairdd}+\chi_{paircd})
\\
-g_1 \chi_{ss}
&
-g_1 \chi_{ss}
&
1-g_1(\chi_{sc}+\chi_{sd})
&
-g_1 \chi_{st}
&
-g_1 \chi_{st}
&
-g_1(\chi_{paircs}+\chi_{pairds})
\\
-g_2 \chi_{paircs}
&
-g_2 \chi_{pairds}
&
-g_2(\chi_{paircc}+\chi_{paircd})
&
1-g_2 \chi_{pairtc}
&
-g_2\chi_{pairtc}
&
-g_2 (\chi_{cc}'+\chi_{cd}')
\\
-g_2 \chi_{paircs}
&
-g_2 \chi_{pairds}
&
-g_2(\chi_{pairdd}+\chi_{paircd})
&
-g_2 \chi_{pairtd}
&
1-g_2\chi_{pairtd}
&
-g_2 (\chi_{dd}'+\chi_{cd}')
\\
-g_2 \chi_{st}
&
-g_2 \chi_{st}
&
-g_2(\chi_{ct}+\chi_{dt})
&
-g_2 \chi_{tt}
&
-g_2\chi_{tt}
&
1-g_2 (\chi_{pairtc}+\chi_{pairtd})
\end{array}
\right).
\label{details of the linear response matrix}
\end{equation}

In the following, we use a compact notation for the dressed $c$ and $d$ operators:
\begin{equation}
\begin{aligned}
C_{\bm{k},\alpha}
&=\sum_{\bm{G},\lambda}
\gamma_{\alpha\lambda}(\bm{k}+\bm{G})c_{\bm{k}+\bm{G},\lambda},
&
D_{\bm{k},\alpha}
&=\sum_{\bm{p},\lambda}
\tilde h_{\alpha\lambda}(\bm{k},\bm{p})d_{\bm{p},\lambda}.
\end{aligned}
\end{equation}
Thus the $\gamma$ and $\tilde h$ matrices absorbed into $C_{\bm{k},\alpha}$ and $D_{\bm{k},\alpha}$ in the notation below. We further define the normal bilinears
\begin{equation}
\begin{aligned}
X_s(\bm{k})&=\sum_\alpha s^\dagger_{\bm{k},\alpha}\psi'_{\bm{k},\alpha},
&
X_t(\bm{k})&=\sum_\alpha t^\dagger_{\bm{k},\alpha}\psi'_{\bm{k},\alpha},
\\
X_c(\bm{k})&=\sum_\alpha \psi'^\dagger_{\bm{k},\alpha} C_{\bm{k},\alpha},
&
X_d(\bm{k})&=\sum_\alpha \psi'^\dagger_{\bm{k},\alpha}D_{\bm{k},\alpha},
\end{aligned}
\end{equation}
and the anomalous bilinears
\begin{equation}
\begin{aligned}
Y_c(\bm{k})
&=\sum_\alpha
C^\dagger_{\bm{k},\alpha}\psi^{\prime\dagger}_{-\bm{k},\bar\alpha},
&
Y_d(\bm{k})
&=\sum_\alpha
D^\dagger_{\bm{k},\alpha}\psi^{\prime\dagger}_{-\bm{k},\bar\alpha}.
\end{aligned}
\end{equation}
The uniform order parameters are the zero-momentum components of these bilinears:
\begin{equation}
\begin{aligned}
B_s&=g_1{1\over N_k}\sum_{\bm{k}}\langle X_s(\bm{k})\rangle,
&
B_t&=g_2{1\over N_k}\sum_{\bm{k}}\langle X_t(\bm{k})\rangle,
\\
B_c^{\prime *}&=g_1{1\over N_k}\sum_{\bm{k}}\langle X_c(\bm{k})\rangle,
&
B_d^{\prime*}&=g_1{1\over N_k}\sum_{\bm{k}}\langle X_d(\bm{k})\rangle,
\\
\Delta'_c&=g_2{1\over N_k}\sum_{\bm{k}}\langle Y_c(\bm{k})\rangle,
&
\Delta'_d&=g_2{1\over N_k}\sum_{\bm{k}}\langle Y_d(\bm{k})\rangle .
\end{aligned}
\end{equation}

The susceptibility components entering Eq.~\eqref{details of the linear response matrix} are static uniform correlation functions evaluated in the parent state. For the normal channels, we define
\begin{equation}
\begin{aligned}
\chi_{sc}&=-{1\over N_k}\sum_{\bm{k}}\int_0^\beta d\tau\,
\langle T_\tau X_c(\bm{k},\tau)X_s^\dagger(\bm{k},0)\rangle_0,
&
\chi_{sd}&=-{1\over N_k}\sum_{\bm{k}}\int_0^\beta d\tau\,
\langle T_\tau X_d(\bm{k},\tau)X_s^\dagger(\bm{k},0)\rangle_0,
\\
\chi_{ct}&=-{1\over N_k}\sum_{\bm{k}}\int_0^\beta d\tau\,
\langle T_\tau X_c(\bm{k},\tau)X_t^\dagger(\bm{k},0)\rangle_0,
&
\chi_{dt}&=-{1\over N_k}\sum_{\bm{k}}\int_0^\beta d\tau\,
\langle T_\tau X_d(\bm{k},\tau)X_t^\dagger(\bm{k},0)\rangle_0,
\\
\chi_{cc}&=-{1\over N_k}\sum_{\bm{k}}\int_0^\beta d\tau\,
\langle T_\tau X_c(\bm{k},\tau)X_c^\dagger(\bm{k},0)\rangle_0,
&
\chi_{cd}&=-{1\over N_k}\sum_{\bm{k}}\int_0^\beta d\tau\,
\langle T_\tau X_d(\bm{k},\tau)X_c^\dagger(\bm{k},0)\rangle_0,
\\
\chi_{dd}&=-{1\over N_k}\sum_{\bm{k}}\int_0^\beta d\tau\,
\langle T_\tau X_d(\bm{k},\tau)X_d^\dagger(\bm{k},0)\rangle_0,
&
\chi_{ss}&=-{1\over N_k}\sum_{\bm{k}}\int_0^\beta d\tau\,
\langle T_\tau X_s(\bm{k},\tau)X_s^\dagger(\bm{k},0)\rangle_0,
\\
\chi_{st}&=-{1\over N_k}\sum_{\bm{k}}\int_0^\beta d\tau\,
\langle T_\tau X_s(\bm{k},\tau)X_t^\dagger(\bm{k},0)\rangle_0,
&
\chi_{tt}&=-{1\over N_k}\sum_{\bm{k}}\int_0^\beta d\tau\,
\langle T_\tau X_t(\bm{k},\tau)X_t^\dagger(\bm{k},0)\rangle_0.
\end{aligned}
\end{equation}
The anomalous channels are
\begin{equation}
\begin{aligned}
\chi_{paircc}&=-{1\over N_k}\sum_{\bm{k}}\int_0^\beta d\tau\,
\langle T_\tau X_c(\bm{k},\tau)Y_c(\bm{k},0)\rangle_0,
&
\chi_{paircd}&=-{1\over N_k}\sum_{\bm{k}}\int_0^\beta d\tau\,
\langle T_\tau X_c(\bm{k},\tau)Y_d(\bm{k},0)\rangle_0,
\\
\chi_{pairdd}&=-{1\over N_k}\sum_{\bm{k}}\int_0^\beta d\tau\,
\langle T_\tau X_d(\bm{k},\tau)Y_d(\bm{k},0)\rangle_0,
&
\chi_{pairsc}&=-{1\over N_k}\sum_{\bm{k}}\int_0^\beta d\tau\,
\langle T_\tau Y_c(\bm{k},\tau)X_s^\dagger(\bm{k},0)\rangle_0,
\\
\chi_{pairsd}&=-{1\over N_k}\sum_{\bm{k}}\int_0^\beta d\tau\,
\langle T_\tau Y_d(\bm{k},\tau)X_s^\dagger(\bm{k},0)\rangle_0,
&
\chi_{pairtc}&=-{1\over N_k}\sum_{\bm{k}}\int_0^\beta d\tau\,
\langle T_\tau Y_c(\bm{k},\tau)X_t^\dagger(\bm{k},0)\rangle_0,
\\
\chi_{pairtd}&=-{1\over N_k}\sum_{\bm{k}}\int_0^\beta d\tau\,
\langle T_\tau Y_d(\bm{k},\tau)X_t^\dagger(\bm{k},0)\rangle_0.
\end{aligned}
\end{equation}
Finally, the primed anomalous susceptibilities are
\begin{equation}
\begin{aligned}
\chi_{cc}'&=-{1\over N_k}\sum_{\bm{k}}\int_0^\beta d\tau\,
\langle T_\tau Y_c(\bm{k},\tau)Y_c^\dagger(\bm{k},0)\rangle_0,
&
\chi_{cd}'&=-{1\over N_k}\sum_{\bm{k}}\int_0^\beta d\tau\,
\langle T_\tau Y_c(\bm{k},\tau)Y_d^\dagger(\bm{k},0)\rangle_0,
\\
\chi_{dd}'&=-{1\over N_k}\sum_{\bm{k}}\int_0^\beta d\tau\,
\langle T_\tau Y_d(\bm{k},\tau)Y_d^\dagger(\bm{k},0)\rangle_0.
\end{aligned}
\end{equation}

In evaluating the susceptibility components, we use Wick decomposition. Since all hybridization order parameters vanish in the parent state, the mixed propagators between the neutral spinon $\psi'$ and the charged fields $s,t,c,d$ vanish. Therefore, only contractions within the charged sector and within the $\psi'$ sector are retained. For example, the component $\chi_{cs}$ should be understood as a zero-total-momentum bubble,
\begin{equation}
\begin{aligned}
\chi_{cs}
&=
-{1\over N_k}
\sum_{\bm{k},\alpha}
\int_0^\beta d\tau\,
\left\langle
T_\tau
\left[
\psi_{\bm{k},\alpha}^{\prime\dagger}
C_{\bm{k},\alpha}
\right](\tau)
\left[
\psi_{-\bm{k},\bar\alpha}^{\prime\dagger}
s_{-\bm{k},\bar\alpha}
\right](0)
\right\rangle_0
\\
&\simeq
-{1\over N_k}
\sum_{\bm{k},\alpha}
\int_0^\beta d\tau\,
\left\langle
T_\tau
C_{\bm{k},\alpha}(\tau)
s_{-\bm{k},\bar\alpha}(0)
\right\rangle_0
\left\langle
T_\tau
\psi_{\bm{k},\alpha}^{\prime\dagger}(\tau)
\psi_{-\bm{k},\bar\alpha}^{\prime\dagger}(0)
\right\rangle_0 .
\end{aligned}
\end{equation}
Here $\bar\alpha$ denotes the spin-singlet partner of $\alpha$. 
The omitted Wick contractions contain mixed propagators such as
$\langle T_\tau s(\tau)\psi^{\prime\dagger}(0)\rangle_0$,
$\langle T_\tau t(\tau)\psi^{\prime\dagger}(0)\rangle_0$,
$\langle T_\tau C(\tau)\psi^{\prime\dagger}(0)\rangle_0$, and
$\langle T_\tau D(\tau)\psi^{\prime\dagger}(0)\rangle_0$, which vanish in the parent state. Thus the retained terms are precisely the zero-total-momentum anomalous contractions between the charged sector and the paired $\psi'$ sector.

The linear response matrix Eq.~\eqref{linearresponse} demonstrates that the order parameters sustain themselves through a feedback loop. The instability threshold of the parent state is mathematically equivalent to the condition where the smallest eigenvalue of $\bm{M_{LR}}$ becomes negative:
\begin{equation}
    \lambda_{min}<0 .
\end{equation}
This condition is equivalent to the emergence of a negative eigenvalue in the Hessian matrix of the free energy, which means the system spontaneously amplifies specific fluctuations, thereby destabilizing the phase with vanishing hybridization order parameters.

\subsection{Explicit calculation of the susceptibility matrix}

Evaluating the susceptibility matrix $\bm{\chi}$ involves calculating the fermionic bubble diagrams using the unperturbed Nambu-Gor'kov Green's functions of the parent state, where
\begin{equation}
    B_s=B_t=B_c'=\Delta_c'=B_d'=\Delta_d'=0,
    \qquad
    \Delta\neq0 .
\end{equation}

By exactly diagonalizing the unperturbed Hamiltonian 
\begin{equation}
    \mathcal{H}_{\text{BdG}}(\bm{k}) |u_{m}(\bm{k})\rangle = E_{m}(\bm{k}) |u_{m}(\bm{k})\rangle,
\end{equation}
we obtain the complete set of quasiparticle eigenenergies $E_{m}(\bm{k})$ and their corresponding multi-component Bloch spinors $|u_{m}(\bm{k})\rangle$, where the global band index $m$ encompasses both particle and hole branches across all flavors.

Using this complete spectral basis, the Matsubara Green's function is exactly expanded as
\begin{equation}
    \hat{\mathcal{G}}_0(\bm{k}, i\omega_n)
    =
    \sum_{m}
    \frac{
    |u_{m}(\bm{k})\rangle \langle u_{m}(\bm{k})|
    }{
    i\omega_n - E_m(\bm{k})
    },
    \qquad
    \omega_n=\frac{(2n+1)\pi}{\beta}.
\end{equation}
The susceptibility can first be written in the standard bubble form
\begin{equation}
    \chi_{\mu\nu}(\bm{q},i\Omega_\ell)
    =
    -\sum_{\bm{k}}
    \frac{1}{\beta}\sum_{i\omega_n}
    \mathrm{Tr}
    \left[
    R_\mu(\bm{k}+\bm{q},\bm{k})
    \hat{\mathcal{G}}_0(\bm{k},i\omega_n)
    V_\nu^\dagger(\bm{k},\bm{k}+\bm{q})
    \hat{\mathcal{G}}_0(\bm{k}+\bm{q},i\omega_n+i\Omega_\ell)
    \right].
\end{equation}
Here $R_\mu$ and $V_\nu$ are the vertex matrices in the Nambu basis corresponding to the response and source bilinears defined above. The static uniform susceptibility used in the instability analysis is obtained by taking $\bm{q}=0$ and $i\Omega_\ell=0$.

Substituting the spectral representation into the bubble, the trace over the internal Hilbert spaces transforms into transition amplitudes between the BdG eigenstates. The generic matrix element linking the fluctuation channels $\mu$ and $\nu$ is expressed as
\begin{equation}
    \chi_{\mu\nu}
    =
    \sum_{\bm{k}}
    \sum_{m,n}
    W_{\mu\nu}^{mn}(\bm{k})
    \Pi(E_m(\bm{k}), E_n(\bm{k}), T).
\end{equation}
Here, $W_{\mu\nu}^{mn}(\bm{k})$ represents the generalized coherence form factor between states $m$ and $n$. Numerically, it is constructed by projecting the full eigenvectors onto the relevant physical subspaces, such as itinerant electrons $c$, Dirac electrons $d$, singlons $s$, triplons $t$, and spinons $\psi'$, together with the appropriate $\gamma$ and $\tilde h$ vertex matrices:
\begin{equation}
    W_{\mu\nu}^{mn}(\bm{k})
    \equiv
    \langle u_{n}(\bm{k})| R_\mu(\bm{k}) | u_{m}(\bm{k}) \rangle
    \langle u_{m}(\bm{k})| V_\nu^\dagger(\bm{k}) | u_{n}(\bm{k}) \rangle .
\end{equation}
This formulation naturally captures both the normal polarization processes, such as standard particle-hole bubbles like $\chi_{cc}$, and the anomalous cross-polarizations, such as Andreev-like bubbles involving $\chi_{paircc}$ or $\chi_{paircs}$, within the same Nambu-space calculation.

The summation over the fermionic Matsubara frequencies is performed analytically. The Matsubara kernel before taking the static limit is
\begin{equation}
    \Pi_{mn}(\bm{k},i\Omega_\ell)
    =
    -\frac{1}{\beta}
    \sum_{i\omega_n}
    \frac{1}{i\omega_n-E_m(\bm{k})}
    \frac{1}{i\omega_n+i\Omega_\ell-E_n(\bm{k})}.
\end{equation}
After carrying out the Matsubara summation, this gives
\begin{equation}
    \Pi_{mn}(\bm{k},i\Omega_\ell)
    =
    \frac{
    n_F(E_m(\bm{k}))-n_F(E_n(\bm{k}))
    }{
    E_n(\bm{k})-E_m(\bm{k})-i\Omega_\ell
    }.
\end{equation}
For the static response, this reduces to
\begin{equation}
    \Pi(E_m,E_n,T)
    =
    \begin{cases}
    \dfrac{n_F(E_{m}) - n_F(E_{n})}{E_{n} - E_{m}},
    & E_{m} \neq E_{n},
    \\[8pt]
    -\dfrac{\partial n_F(E)}{\partial E}\bigg|_{E = E_m}
    =
    \dfrac{\beta e^{\beta E_m}}{(1 + e^{\beta E_m})^2}
    =
    \dfrac{\beta}{4\cosh^2(\beta E_m/2)},
    & E_{m} = E_{n}.
    \end{cases}
    \label{Matsubara summation}
\end{equation}
where $n_F(E)$ is the Fermi-Dirac distribution. 

From equation (\ref{Matsubara summation}), we can see that when the temperature $T$ is small, a band touching or a near degeneracy close to the Fermi level strongly amplifies the susceptibility component. 

\subsection{Results and analysis}
\begin{figure}[htbp]
    \centering
    \includegraphics[width=0.45\columnwidth]{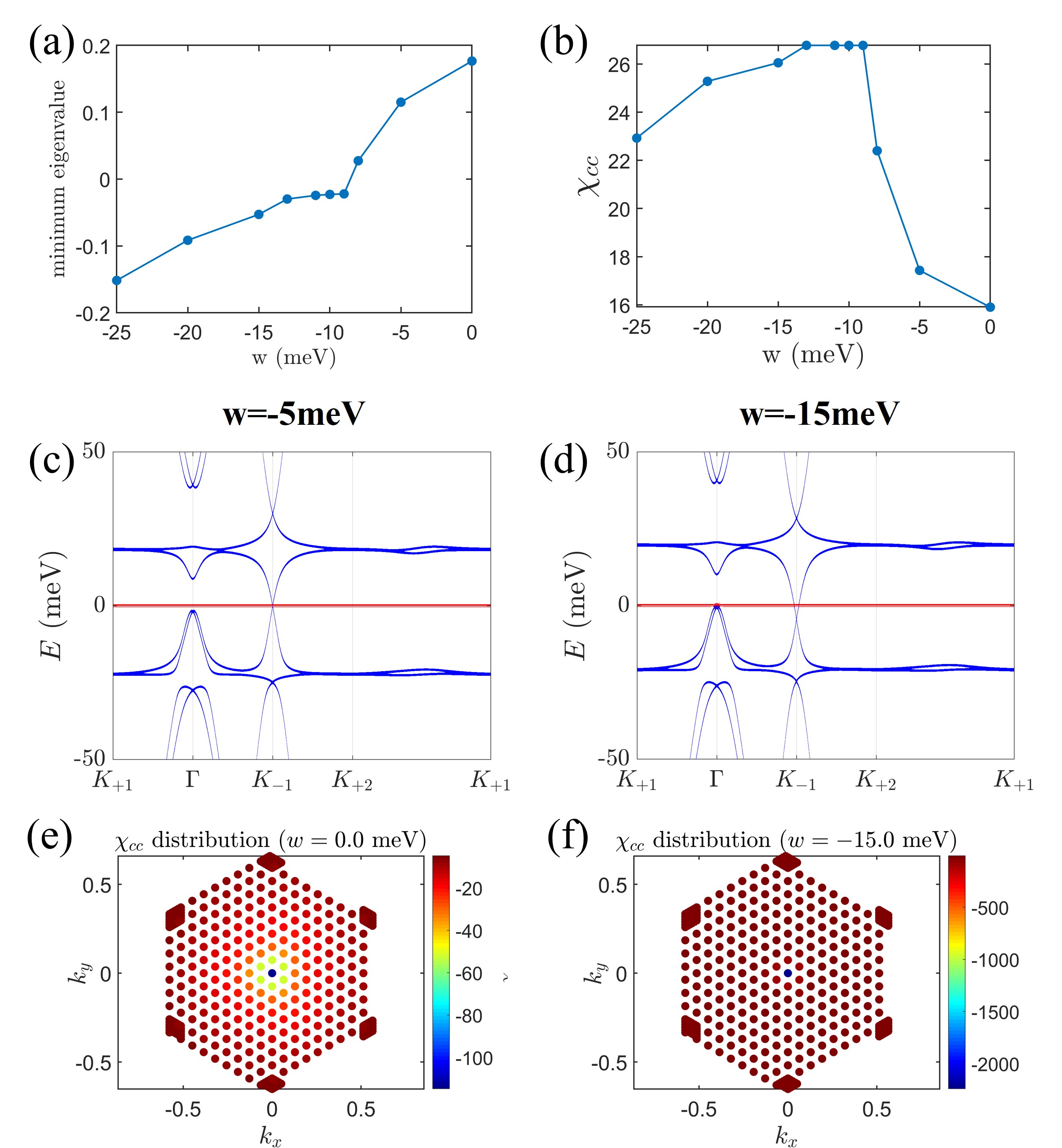}
    \caption{Evolution of the system properties driven by the energy offset $w$ when fixing $D$ at $\nu=-2$. (a) The minimum eigenvalue $\lambda_{\text{min}}$ and (b) the susceptibility component $\chi_{cc}$ as a function of $w$. (c,d) Spectral intensity showcasing the band structure evolution at different $w$. The $\psi'$ band is observed to approach the lower Hubbard band as $w$ decreases. 
    (e)-(f) Momentum-space distribution of $\chi_{cc}$ at $w = 0$~meV and $w = -15$~meV, respectively, highlighting a strong localized enhancement around the $\Gamma$ point upon crossing the phase transition.The parameters are $D=120\,\mathrm{meV}$, $J=0.3\,\mathrm{meV}$, $\kappa=0.8$, $U=40\,\mathrm{meV}$.}
    \label{linear_response1.jpg}	
    
\end{figure}

\begin{figure}[htbp]
    \centering
    \includegraphics[width=0.45\columnwidth]{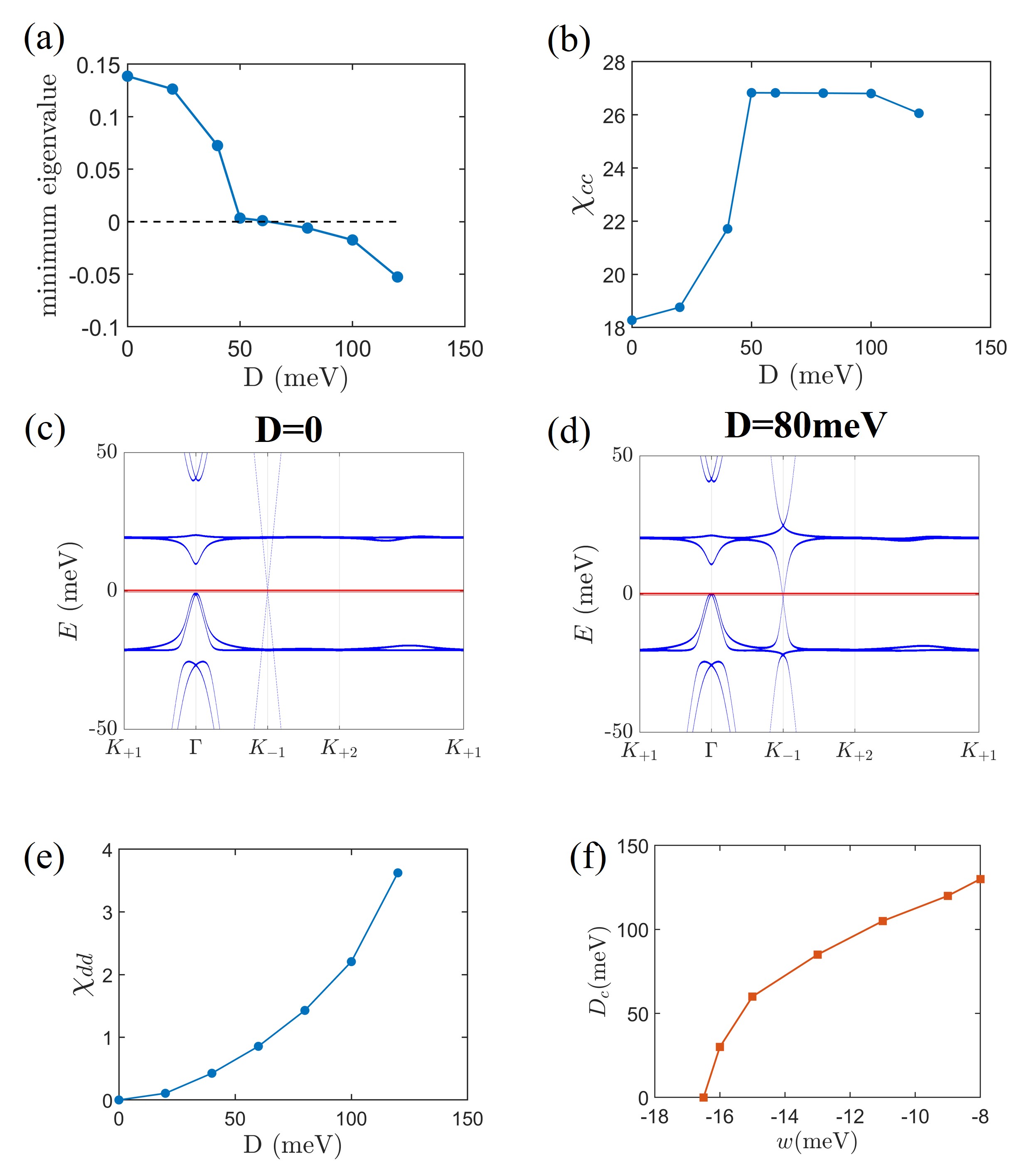}
    \caption{Evolution of the system properties driven by the displacement field $D$ when fixing $w$ at $\nu=-2$. (a) The minimum eigenvalue $\lambda_{\text{min}}$ and (b) the susceptibility component $\chi_{cc}$ as a function of $D$. (c,d) Spectral intensity demonstrating the evolution of the electron spectrum under different displacement fields. (e)The susceptibility component $\chi_{dd}$ as a function of $D$. (f) The critical displacement field $D_c$ of the phase transition as a function of $w$. The parameters are $w=-15\,\mathrm{meV}$, $J=0.3\,\mathrm{meV}$, $\kappa=0.8$, $U=40\,\mathrm{meV}$.}
    \label{linear_response2.jpg}	
    
\end{figure}

In Fig~\ref{linear_response1.jpg}(a) and \ref{linear_response2.jpg}(a), we present the minimum eigenvalue, $\lambda_{\text{min}}$, as a function of the energy shift $w$ and the displacement field $D$, respectively. At the critical point of the phase transition, $\lambda_{\text{min}}$ exhibits a sharp sign reversal from positive to negative. This instability is fundamentally driven by discontinuous enhancements in the dominant components of the bare susceptibility matrix. For instance, Fig~\ref{linear_response1.jpg}(b) and \ref{linear_response2.jpg}(b) illustrate the pronounced jump in the static susceptibility component $\chi_{cc}$ as $w$ and $D$ are varied across the phase boundary. Furthermore, the momentum-resolved susceptibility $\chi_{cc}(\mathbf{k})$ reveals a dramatic enhancement sharply localized around the $\Gamma$ point across the transition, as depicted in Fig~\ref{linear_response1.jpg}(e) and \ref{linear_response2.jpg}(f).

The microscopic origin of this susceptibility divergence can be elucidated by analyzing the evolution of the band structure. As shown in Fig~\ref{linear_response1.jpg}(c) and \ref{linear_response1.jpg}(d), decreasing $w$ drives the $\psi'$ band toward the lower Hubbard band, where the spectral weight of the $c_1$ electrons is predominantly localized. As shown by Eq.~\eqref{Matsubara summation}, the energetic proximity and eventual crossing of the $\psi'$ band with the lower Hubbard band significantly amplify the thermal response kernel. This induces a discontinuous jump in the susceptibility, ultimately triggering the phase transition.

An analogous mechanism governs the $D$-driven transition at a fixed $w$. Microscopically, as illustrated in Figs.~\ref{linear_response2.jpg}(c) and (d), the displacement field hybridizes the Dirac cones of the TBG and monolayer graphene subsystems at the $K$ point, inducing a pronounced band repulsion. At $\nu=-2$, where electrons predominantly occupy the upper Hubbard band, increasing $D$ effectively lowers the lower Dirac cone(circled in Fig ~\ref{linear_response2.jpg}(c)), thereby driving the localized $\psi^\prime$ band toward the lower Hubbard band.This evolution abruptly enhances $\chi_{cc}$ and drives $\lambda_{\text{min}}$ negative [Fig~\ref{linear_response2.jpg}(b) and (a)]. Further corroborating this physical picture, the critical displacement field $D_c$ increases monotonically with $w$ [Fig~\ref{linear_response2.jpg}(f)]. While $D$ concurrently modifies $\chi_{dd}$(Fig~\ref{linear_response2.jpg}(e)), we demonstrate in the appendix that the transition is mostly dominated by a susceptibility jump originating from a band touching at the $\Gamma$ point. 

The linear response theory provides more direct evidence for the enhanced hybridization upon hole doping in LHB. The divergence of the susceptibility matrix led by the band alignment marks the phase transition.

\subsection{Distribution of different components}
\begin{figure}[htbp]
    \centering
    \includegraphics[width=0.5\columnwidth]{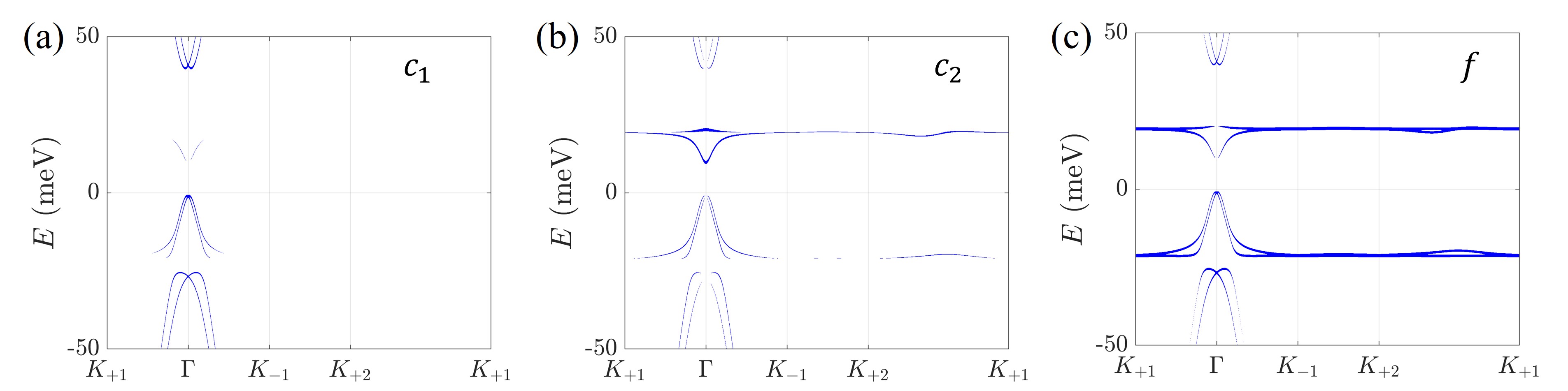}
    \caption{Distribution of different components. The parameters are $D=0$,$w=-15\,\mathrm{meV}$, $U=40\,\mathrm{meV}$, $J=0.3\,\mathrm{meV}$,$\kappa=0.8$. }
    \label{fig:distribution}	
    
\end{figure}

Figure~\ref{fig:distribution} shows the component-resolved distribution at $\nu=-2$. As discussed in the main text, the $c_1$-electron weight is concentrated predominantly in the LHB near the $\Gamma$ point, with only a weak residual contribution in the UHB. This orbital-selective distribution implies that the phase transition is primarily driven by the hybridization between the local-moment level and the LHB. Consequently, the transition can occur only when this hybridization becomes sufficiently strong.

\subsection{Further evidence for the main mechanism of the $D$-tuned phase transition}

As discussed in the main text, increasing the displacement field $D$ has two main effects. First, it reconstructs the band structure by lowering the middle Dirac cone relative to the Hubbard bands. Second, it enhances the bare hybridization between the $d$ and $f$ electrons. To isolate the role of the second effect in triggering the transition, we keep the band structure fixed and artificially vary only the hybridization strength in the linear-response equations. A natural way to implement this test is to rescale
the coefficients multiplying the operators defined in
Eq.~\eqref{mean field operator}, namely $g_1$ and $g_2$. For example, for the $d$-electron channels $B_d^\prime$ and $\Delta_d^\prime$, we multiply the corresponding coefficients by a factor of two. This procedure is equivalent to doubling the effective $d$-$f$ hybridization vertex while leaving the single-particle band structure unchanged. The modified self-consistency
relations are then
\begin{equation}
\begin{aligned}
    B_s &= g_1 \langle \hat{O}_s \rangle, 
    \quad &B_t &= g_2 \langle \hat{O}_t \rangle,\\
    B'_c &= g_1 \langle \hat{O}_c \rangle, 
    \quad &\Delta_c' &= g_s \langle \hat{O}_c^\prime \rangle,\\
    B'_d &= 2g_1 \langle \hat{O}_d \rangle, 
    \quad &\Delta_d' &= 2g_s \langle \hat{O}_d^\prime \rangle .
\end{aligned}
\label{eq:rescaled_d_channel}
\end{equation}

\begin{figure}[htbp]
    \centering
    \includegraphics[width=0.5\columnwidth]{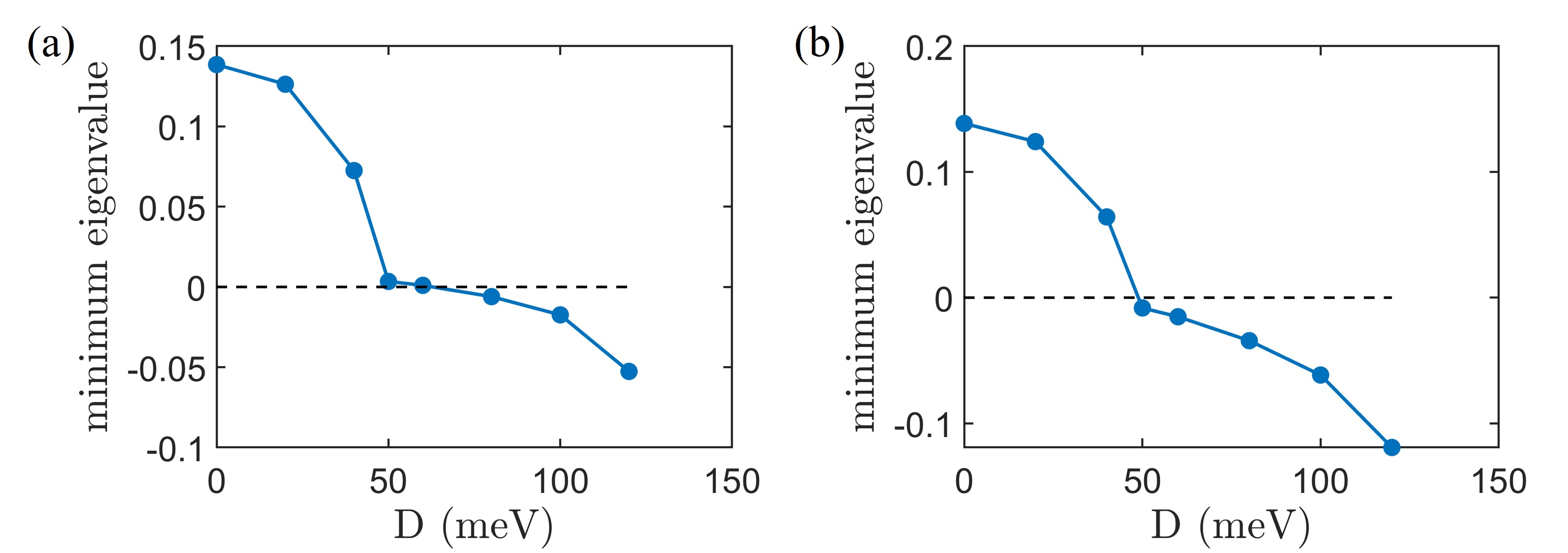}
    \caption{Comparison between the original results and the results after the coefficients in the linear response theory are multiplied with two.}
    \label{comparison_eigenvalue}	
    
\end{figure}

We then compute the smallest eigenvalue $\lambda_{\min}$ as a function of the displacement field $D$. As shown in Fig.~\ref{comparison_eigenvalue}, $\lambda_{\min}$ remains nearly unchanged even after this twofold enhancement of the $d$-electron hybridization channel, except some difference near the critical point. This result indicates that the increase of the bare $d$-$f$ hybridization is not the primary mechanism driving the transition.
Instead, the transition is mainly controlled by the $D$-induced band
reconstruction, which brings the local-moment band into resonance with the LHB and enhances the corresponding hybridization susceptibility.

\section{Additional discussion on the particle-hole asymmetry driven by the Dirac-cone position}
\label{appendix:additional discussion of ph asmmetry}

A pronounced but still poorly understood feature of TTG experiments is the
striking particle-hole asymmetry. Although particle-hole asymmetry is also
present in TBG and is often attributed to lattice relaxation~\cite{kang2023pseudomagnetic, kang2025analytical}, the
electron- and hole-doped sides of TTG exhibit qualitatively different
behaviors. The following discussion is mainly based on the low temperature results in \cite{zhang2022promotion} and the higher-temperature phase diagram in \cite{zhang2025electrically}. According to the experimental results, on the electron-doped side, especially at integer fillings
$\nu=+2$ and $+3$, the resistance is small at weak displacement field $D$,
consistent with semimetallic behavior. As $D$ is increased, the system
undergoes a transition from a semimetal to a superconducting state at low
temperature and to a heavy-fermion-liquid-like state at higher temperature.
Quantum-critical behavior is also observed near the transition, as reflected
by the enhanced resistance at intermediate $D$. In contrast, on the
hole-doped side, the resistance remains small even at integer fillings and
over a broad range of $D$, with no clear phase transition behavior. In this
section, we explain this qualitative particle-hole asymmetry by considering
the energy shift of the Dirac cone relative to the flat bands.

\begin{figure}
    \centering
    \includegraphics[width=0.5\columnwidth]{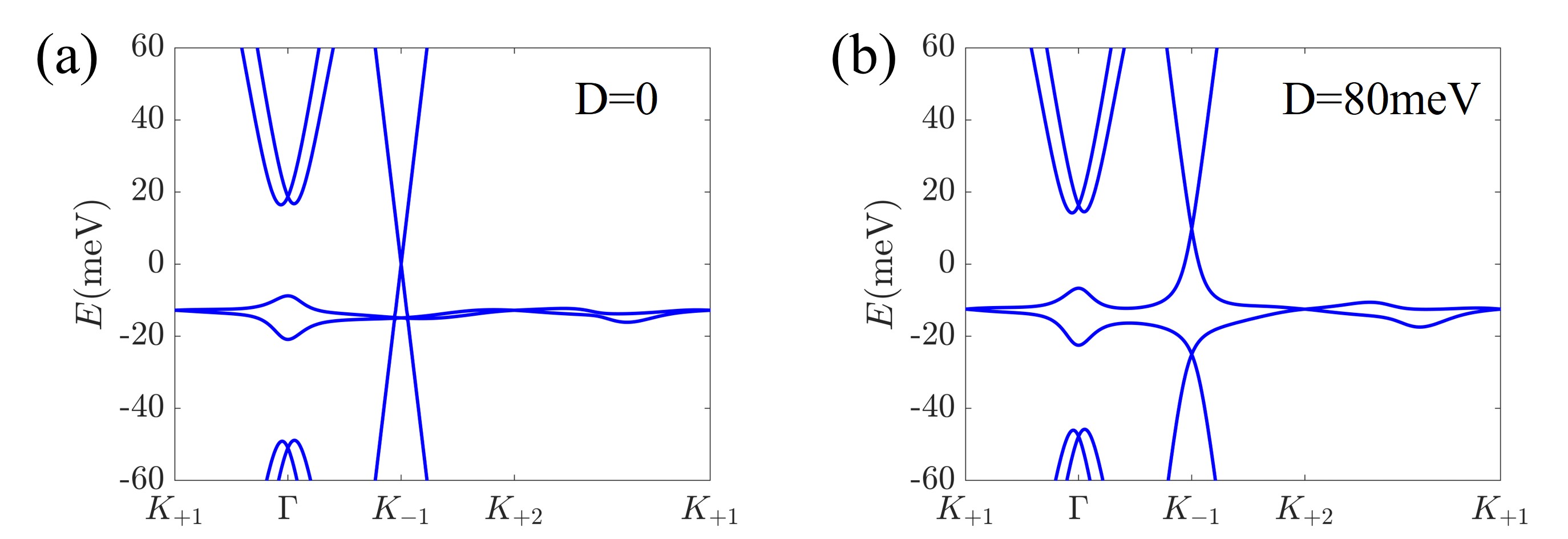}
    \caption{
    Particle-hole-asymmetric band structure in the presence of the
    middle-layer shift term. The onsite energy shift is
    $E_{sh}=-30\,{\rm meV}$.
    }
    \label{band_for_ph_asymmetry}
\end{figure}

It has been noted in DFT studies of TTG that, at charge neutrality, the
Dirac point lies above the flat bands~\cite{fischer2022unconventional}. This
energy offset originates from atomic lattice relaxation and from the
inequivalent electrostatic environments of the inner and outer layers. The
inner layer is subject to a more symmetric dielectric and structural
environment, whereas the outer layers are not. This intrinsic layer
asymmetry generates an effective layer-dependent onsite potential. In the
decoupled basis of the continuum model, we phenomenologically capture this
effect by introducing an energy offset on the middle layer of TTG:
\begin{equation}
H_{\rm shift}
=
H_{\rm TTG}
+
E_{sh}
\phi_{\bm r,2,\eta s}^{\dagger}
\sigma_0
\phi_{\bm r,2,\eta s},
\label{shift BM}
\end{equation}
where $H_{\rm TTG}$ is defined in Eq.~\eqref{full BM}. We choose
$E_{sh}=-30\,{\rm meV}$ to illustrate the essential physics, which is close to the value chosen in \cite{fischer2022unconventional}. As the band structure demonstrates in Fig.~\ref{band_for_ph_asymmetry}, this is almost equivelent to shift the Dirac cone upward by about $15\,\mathrm{meV}$, which means $w=15\,\mathrm{meV}$

In Fig.~\ref{band_for_ph_asymmetry}, we show the bare band structure in the
presence of the shift term. The Dirac cone is shifted above the flat bands,
whereas the remaining bands are nearly unchanged compared with the original
BM band structure of TBG. This indicates that the ancilla-fermion theory,
which is based on the symmetry structure of the BM model, remains applicable.
The shift term produces an effective displacement of the Dirac cone by about
$15\,{\rm meV}$, consistent with the offset introduced in
Eq.~\eqref{original_Hamiltonian}.

In the previous section, we showed that at filling $\nu=-2$, a negative
energy offset of the Dirac cone is crucial for realizing the
displacement-field-tuned phase transition. The reason is that the transition requires the LHB to become hole doped, which enhances the hybridization susceptibility. This mechanism can be extended to positive fillings by a particle-hole transformation. At $\nu=2$, realizing the transition requires a positive offset of the Dirac cone, so that the UHB becomes electron doped.

\begin{figure}[htbp]
    \centering
    \includegraphics[width=0.7\columnwidth]{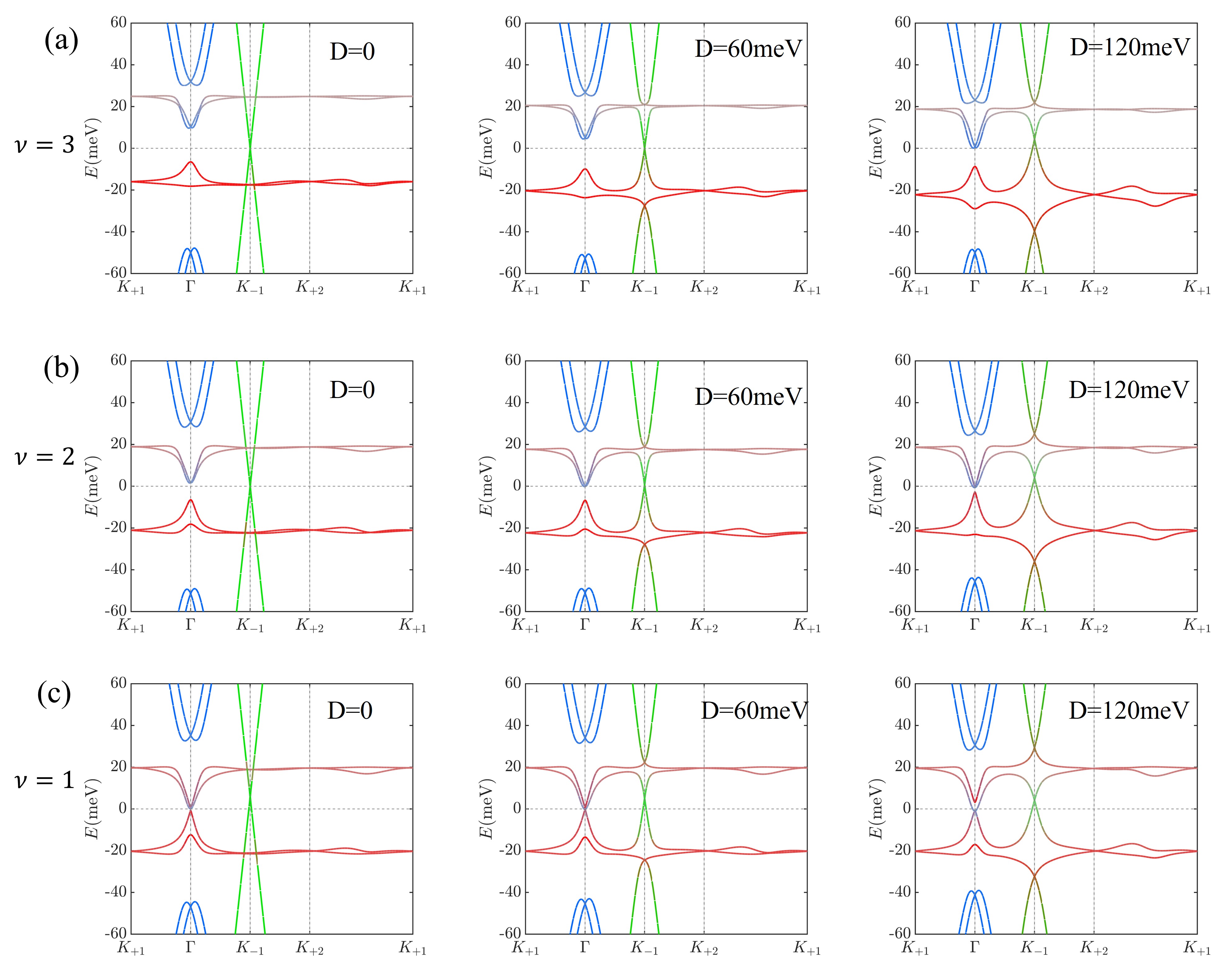}
    \caption{
    Mean-field Mott band structures including the Dirac-cone offset at
    positive fillings: (a) $\nu=3$, (b) $\nu=2$, and (c) $\nu=1$.
    The Mott gap at the $K$ point is set to $40\,{\rm meV}$, and the
    middle-layer onsite shift is $E_{sh}=-30\,{\rm meV}$.
    }
    \label{ancilla_ph_asymmetry1}
\end{figure}

\begin{figure}[htbp]
    \centering
    \includegraphics[width=0.7\columnwidth]{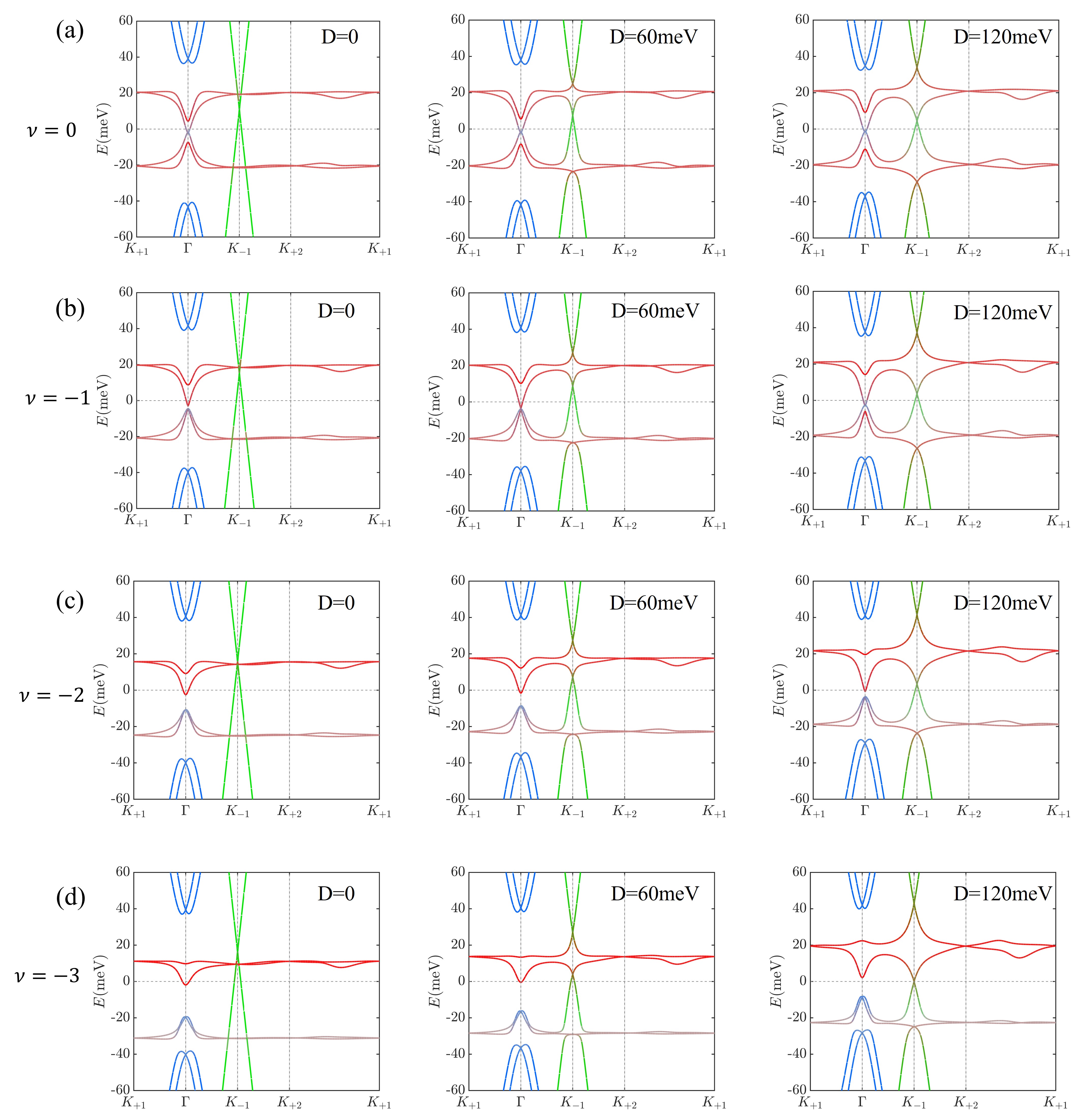}
    \caption{
    Mean-field Mott band structures including the Dirac-cone offset at
    nonpositive and negative fillings: (a) $\nu=0$, (b) $\nu=-1$,
    (c) $\nu=-2$, and (d) $\nu=-3$. The Mott gap at the $K$ point is set to
    $40\,{\rm meV}$, and the middle-layer onsite shift is
    $E_{sh}=-30\,{\rm meV}$.
    }
    \label{ancilla_ph_asymmetry2}
\end{figure}

To extend this picture to other fillings, such as $\nu=3$, we next compute the Mott band structure using the ancilla-fermion mean-field theory with the shift term in Eq.~\eqref{shift BM}. This allows us to clarify the role of the Dirac-cone shift in the phase diagram and in the particle-hole asymmetry. Following Sec.~\ref{appendix:section_for_ancilla}, we consider only the Mott band structure without the local moment $\psi^\prime$. The resulting band structures at different fillings are shown in Fig.~\ref{ancilla_ph_asymmetry1} and \ref{ancilla_ph_asymmetry2}. Comparing Fig.~\ref{ancilla_ph_asymmetry1}(b) with Fig.~\ref{linear_response2.jpg}(c) and (d), we find that the ancilla-fermion band structure at $\nu=2$ agrees well with the slave-particle spectrum at $\nu=-2$ after a particle-hole transformation. This agreement supports the validity of the ancilla-fermion description of the Mott bands.

At positive fillings, especially $\nu=2$ and $3$, the Dirac cone lies close to the UHB near the $\Gamma$ point already at $D=0$, as illustrated in Figs.~\ref{ancilla_ph_asymmetry1}(a) and \ref{ancilla_ph_asymmetry1}(b). Increasing $D$ further drives the Dirac cone toward the UHB. Combining this band evolution with the mechanism discussed above, we expect a transition from a spin-ordered semimetal to a superconducting state, consistent with experimental observations. Our results also explain why the transition at
$\nu=3$ requires a larger displacement field than at $\nu=2$: for the same parameters at $D=0$, the Dirac cone is farther from the UHB at $\nu=3$ than at $\nu=2$, so a larger $D$ is needed to achieve the required band
alignment.

At negative fillings, as shown in Fig.~\ref{ancilla_ph_asymmetry2}, the
Dirac cone is far from the LHB. According to the mechanism discussed above, this large energy separation prevents the displacement field from driving the required band alignment and therefore suppresses the phase transition. This explains why the corresponding phenomena are absent on the hole-doped side of the experimental phase diagram.

\end{document}